%% file: main.tex
\documentclass[11pt]{article}
\usepackage{amsmath,graphicx}
\usepackage{amssymb}
\usepackage{amsthm}
\usepackage{enumitem}  
\usepackage{geometry}
\usepackage{mathrsfs}
\usepackage{graphicx}
\usepackage{epstopdf}
\usepackage{hyperref}
\usepackage[applemac]{inputenc}
\usepackage{multirow}
\usepackage{color}
\usepackage{authblk}
\usepackage[export]{adjustbox}

\newcommand {\mumi} {\mu^{\mathrm{mi}}}
\newcommand {\muma} {\mu^{\mathrm{ma}}}

\newcommand {\lb} {{\langle}}
\newcommand {\rb} {{\rangle}}
\newcommand {\dd} {{\ell}}
\newcommand {\NN} {{d}}
\newcommand {\R} {{\mathbb{R}}}

\newcommand {\OO} {{{\Omega_d}}}

\newcommand {\OOey} {{{\Omega_{d,\epsilon}}}}
\newcommand {\pey} {{{p_{d, \epsilon}}}}

\newcommand {\En} {{{E}}}

\newcommand {\Z} {{\mathbb{Z}}}
\newcommand {\E} {{\mathbb{E}}}

\newcommand {\B} {{{B}}}

\newcommand {\h} {{\gamma_d}}
\newcommand {\om} {{\omega}}

\newcommand {\La} {{\Lambda}}
\newcommand {\LaN} {{\Lambda_d}}

\newcommand {\cistar} {\kern.2em\mbox{$\odot\kern-.67em\star\kern .4em$}}
\newcommand {\margdomain} {{\mathcal{I}}}
\newcommand {\domain} {{\mathcal{I}^\LaN}}
\newcommand {\domainr} {{\R^\NN}}
\newtheorem{theorem}{Theorem}[section]
\newtheorem{lemma}[theorem]{Lemma}
\newtheorem{corollary}[theorem]{Corollary}

\newtheorem{proposition}[theorem]{Proposition}
\newtheorem{definition}[theorem]{Definition}

\title{Multiscale Sparse Microcanonical Models}
\author[1,2]{Joan Bruna}
\author[3,4,5]{St\'ephane Mallat}
\affil[1]{Courant Institute of Mathematical Sciences, New York University}
\affil[2]{Center for Data Science, New York University}
\affil[3]{College de France}
\affil[4]{Ecole Normale Sup\'erieure, DIENS, PSL, Paris}
\affil[5]{Flatiron Institute, New York}

\begin{document}
%

\maketitle

\begin{abstract}
We study approximations of non-Gaussian stationary processes 
having long range correlations with microcanonical models. 
These models are conditioned by the empirical value of an
energy vector, evaluated on a single realization.
Asymptotic properties of maximum entropy microcanonical 
and macrocanonical processes and their convergence to Gibbs measures are reviewed. We show that 
the Jacobian of the energy vector controls the entropy rate of microcanonical processes. 

Sampling maximum entropy processes through MCMC algorithms
require too many operations when the number of constraints is large.
We define microcanonical gradient descent processes by
transporting a maximum entropy measure with a gradient descent algorithm
which enforces the energy conditions. Convergence and symmetries are
analyzed. 
Approximations of non-Gaussian processes with long range interactions
are defined with
multiscale energy vectors computed with wavelet and scattering transforms.
Sparsity properties are captured with $\bf l^1$ norms. 
Approximations of Gaussian, Ising and point processes are studied, as well
as image and audio texture synthesis.
\end{abstract}

%

\input{intro1.tex}

\input{microcanonical1.tex}

\input{sampling1.tex}

\input{scatt1.tex}

\input{experiments}


\input{conclusion}

{\bf Acknowledgements:} SM: This work was supported by the ERC grant InvariantClass 320959. JB: This work was partially supported by the Alfred P. Sloan Foundation, by NSF RI-1816753, and by Samsung DMC. 
We thank Zhengdao Chen and Ofer Zeitouni for valuable comments and fixes to the current manuscript, and the anonymous referees for their high-quality, valuable feedback. 

\bibliographystyle{plain}
\bibliography{references_micro}

\appendix

\input{proofs}

\end{document}

%% file: intro1.tex
\section{Introduction}

Building probabilistic models of large systems of interacting variables 
that can be efficiently estimated from data is a core problem 
in statistical physics, machine learning and signal processing. 
We consider the estimation of the probability measure of stationary processes 
$X(u)$ on the infinite grid $u \in \Z^\ell$ given a single realization
$\bar x(u)$, observed over a finite domain 
$u \in \Lambda_d \subset \Z^\ell$ of cardinality $d$.
For $\ell = 2$ and $\ell = 1$, such processes provide models
of image and audio textures. Given a piece of texture 
over $\LaN$, we may want to synthesize similar texture examples by sampling the resulting probability model.
Building probability models from a single observation is also needed in finance and in
many physical problems, such as geophysics exploration or fluid dynamics.
These estimations 
rely on the ability to build
low-dimensional approximations of the underlying stationary measure.
This paper introduces microcanonical sparse multiscale models, which can take into account non-Gaussian phenomena and long range interactions.

In his seminal paper, Jaynes \cite{jaynes1957information} interprets
statistical physics as an inference of a probability 
distribution from partial measurements, by maximizing its entropy.
In Jaynes words \cite{jaynes1957information}, maximizing the
entropy of a probability distribution 
``is maximally noncommittal with regard to missing information.''
Macrocanonical models are 
maximum entropy distributions conditioned on the expected value
of a vector of potential energies. They are used in large classes of
stochastic models \cite{geman1984stochastic}
and will thus be our departure point.

Since we only know a single realization $\bar x(u)$ of $X(u)$
in $\Lambda_d$,  the expected value of stationary energies are estimated
by the average potential energy vector $\Phi_d (\bar x)$
of $\bar x$ in the domain $\Lambda_d$ of size $d$. When $d$ is sufficiently large, weak ergodicity assumptions imply that
$\Phi_d (X)$ concentrates
near the empirical energy vector $\Phi_d (\bar x)$ with high probability.
A microcanonical model is a probability measure supported over the 
microcanonical set
of all $x$ having nearly the same energy:
$\| \Phi_d (x) - \Phi_d (\bar x)\| \leq \epsilon$. 
Maximum entropy microcanonical models have a uniform density over this
set. Under appropriate hypotheses, 
the Boltzmann equivalence principle states 
that  a maximum entropy 
microcanonical model converges to the same Gibbs measure
as the macrocanonical model, when $d$ goes to $\infty$. 
Section \ref{Gibbsconv} reviews these results.

Microcanonical models exist with mild assumptions,
even-though macrocanonical distributions may not exist, particularly for
signals $x$ having strong sparsity properties. 
We thus consider these models not as approximations of macrocanonical
models, which may not exist, but as stochastic models in their own sake. 
Section \ref{sec_algo} relates their entropy rate 
to their energy vector.
Sampling micro and macrocanonical measures is a classic problem in statistical mechanics, typically approached with MCMC algorithms or Langevin Dynamics \cite{betancourt2017conceptual,creutz1983microcanonical} or variational methods \cite{wainwright2008graphical}. 
Their numerical effectiveness on high-dimensional 
problems is hindered by the slow mixing speed of the Markov Chain \cite{creutz1983microcanonical}, which limits their applications.
To avoid this computational issue, 
we introduce an alternative class of microcanonical models where the
Markov chain is replaced by a gradient flow resulting from the microcanonical energy vector.  A microcanonical gradient descent model begins from 
a high entropy measure and
computes a progressive transport of this measure with gradient steps, 
towards the microcanonical set. Similar algorithms
have been applied to texture synthesis \cite{gatys2015texture} with deep
convolutional neural networks. Section \ref{sec_algo} studies their 
convergence to a microcanonical set. 
Although the gradient descent transport 
does not converge to a maximum entropy measure, we
prove that it preserves an important 
subset of symmetries which is specified. 

A major issue is to specify 
energy vectors $\Phi_\NN$ providing accurate microcanonical gradient descent
approximations of non-Gaussian processes with long range interactions. 
Section \ref{wavelet-sec} introduces energy vectors which take into account
long range interactions by 
separating scales with wavelet transforms. Non-Gaussian properties
are captured with $\bf l^1$ norms which measure the sparsity
of wavelet coefficients. These energy vectors are augmented
with wavelet scattering coefficients, providing information on 
the geometry of sparse wavelet coefficients
\cite{mallat2012group,bruna2013invariant}. 

Section \ref{isingsec} studies the approximation of Gaussian,
Ising and point processes, with
microcanonical gradient descent models computed with 
wavelet and scattering energy vectors. For Ising, the
wavelet scale separation 
is closely related to the Wilson
renormalization group approach \cite{Battle}. 
We show that scattering microcanonical
model can also give good 
perceptual approximations of large classes of image and audio textures.

\paragraph{Notation} We use cursive captial letters $\mathcal{A}, \mathcal{B}, \dots$ to denote sets, 
small capital letters $x, y,\dots$ to denote vectors, capitals $X, Y,Z$ to denote random processes, 
and capital letters $E, H, \Phi, \dots$ to denote operators and functions. 
$\hat{x}$ denotes the Fourier transform of $x$. $\|x\|$ denotes the Euclidean norm of $x$.

%% file: microcanonical1.tex
\section{Microcanonical and Macrocanonical Models}

We consider a stationary process $X(u)$ taking its values 
in an interval $\margdomain \subseteq \R$ for all $u \in \Z^\ell$. 
We denote by $\mu$ the probability measure of this stationary process.
We write $\E_\mu (f(x))$ 
the expected value of $f(X)$ or $\E_p (f(x))$ if $\mu$ has a density $p$.
Let $\LaN \subset \Z^\ell$ be a cube with $d$ grid points and 
$\domain$ the product domain.
Let $\bar x \in \domain$ be a realization of $X$ restricted to $\Lambda_d$.
Microcanonical models described in Section \ref{microsec} are probability densities
conditioned on a $K$-dimensional energy vector $\Phi_\NN( \bar x)$.
Section \ref{macrosec} reviews the 
properties of macrocanonical models which have 
a maximum entropy conditioned on $\E_\mu(\Phi_\NN (x))$.
We concentrate on shift-invariant
energies $\Phi_d$ introduced in Section \ref{shiftpotent}, to define
stationary maximum entropy processes. 
Section \ref{Gibbsconv} reviews the resulting
convergence properties of micro and macrocanonical models towards 
the same Gibbs measures.
In statistical physics terms, it amounts to verify the Boltzmann equivalence principle in the thermodynamical limit, for lattice gaz models. 
We shall then see that microcanonical models are also interesting
in their own sake, even in regimes where macrocanonical models do not exist.

\subsection{Maximum Entropy Microcanonical Models}
\label{microsec}

A microcanonical model is computed from 
$y = \Phi_\NN (\bar x)$. 
To estimate the measure $\mu$ of a stationary $X$ from a single
realization, we need ergodicity assumptions. We 
assume that  $\Phi_\NN (X)$ concentrates with high probability 
around $\E_\mu (\Phi_\NN (x))$
when $\NN$ goes to $\infty$:
\begin{equation}
\label{Phi-ergodic0}
\forall \epsilon >0~,~
\lim_{\NN \rightarrow \infty} {\rm Prob_\mu}(\|\Phi_\NN (X) - \E_{\mu} (\Phi_\NN (x))\| \leq \epsilon ) = 1 .
\end{equation}
If there exists $C > 0$ such that $\|\E_{\mu} (\Phi_\NN (x))\| \leq C$ then this 
convergence in probability is implied by a mean-square convergence:
\begin{equation}
\label{Phi-ergodic}
\lim_{\NN \rightarrow \infty} \E_\mu (\| \Phi_\NN (x) - \E_\mu (\Phi_\NN (x))\|^2) = 0~.
\end{equation}

The microcanonical set of width $\epsilon$ associated to 
$y = \Phi_\NN (\bar x)$ is
\[
\OOey = \{ x \in \domain~:~\|\Phi_\NN (x) - y \| \leq \epsilon \}~.
\]
The concentration property (\ref{Phi-ergodic0}) implies that 
when $d$ goes to $\infty$, $X$ belongs to microcanonical sets $\OOey$ 
of width $\epsilon=\epsilon(d)$  converging to $0$, with a probability converging to $1$.
In other words, (\ref{Phi-ergodic0}) guarantees that the support 
of the measure $\mu$ is mostly concentrated in $\OOey$ for large $d$.

The differential entropy of a probability distribution $\mu$ which admits
a density $p(x)$ relatively to the Lebesgue measure is
\begin{equation}
\label{maxentropmoment}
H (\mu) :=   - \int p(x)\, \log p(x)\, dx~.
\end{equation}
A maximum entropy microcanonical model $\mumi(\NN, \epsilon, y)$ was defined by Boltzmann as the maximum entropy distribution supported
in $\OOey$. We usually define $\Phi_\NN(x)$ so that
$\OOey$ is compact. It results the maximum entropy distribution has
a uniform density $\pey$:
\begin{equation}
\label{eqnasdf1}
\pey (x) := \frac {1_{\OOey} (x)}{\int_{\OOey} \, dx}~.
\end{equation}
Its entropy is therefore the logarithm of the volume of $\OOey$:
\begin{equation}
\label{eqnasdf198}
H(\pey) = - \int \pey(x)\, \log \pey(x)\, dx =
\log \Big(\int_{\OOey} \, dx \Big)~.
\end{equation}

We thus 
face a fundamental trade-off when constructing microcanonical models.
On the one hand, we seek representations $\Phi_\NN$ that satisfy a concentration property (\ref{Phi-ergodic0}) to ensure that typical samples from $\mu$ are included in $\OOey$ with high probability, and hence
typical for the microcanonical measure $\mumi$. On the other hand, 
the sets $\OOey$ must not be too large to avoid
having elements of $\OOey$ and hence
typical samples of $\mumi$ which are not typical for $\mu$. To obtain
an accurate microcanonical model, the energy $\Phi_\NN$ must define 
microcanonical sets of minimum
volume, while satisfying the concentration (\ref{Phi-ergodic0}).

\subsection{Macrocanonical Models}
\label{macrosec}

Since $\Phi_\NN (X)$ concentrates close to $\E_\mu (\Phi_\NN (x))$ and
$\bar x$ is a realization of $X$, one could expect 
that the maximum entropy distribution conditioned on 
$\Phi_\NN (\bar x)$ converges to the
maximum entropy distribution conditioned on $\E_\mu (\Phi_\NN (x))$ when $\NN$ goes to $\infty$. 
Section \ref{shiftpotent} studies conditions under which this 
Boltzmann equivalence principle
is verified. We begin by reviewing the properties of macrocanonical
maximum entropy models conditioned on $\E_\mu (\Phi_\NN (x)) = y$. 
Let $\mathcal{M}(\domain)$ denote the space of measures of $\domain$.

A macrocanonical measure $\muma$ with density $p_{\mathrm{ma}}$ 
has a maximum entropy conditioned on $\E_{p_{\mathrm{ma}}} (\Phi_\NN (x)) = y$:
\begin{eqnarray}
    \label{maxentropmoment2}
p_{\mathrm{ma}} &\in& \arg\max_{p \in \mathcal{A}_y} H(p) ~,\text{ with } \nonumber \\
\mathcal{A}_y &=& \{ p \in \mathcal{M}(\domain); \int_{\domain} \Phi_\NN(x)\, p(x)\,d x = y \}~.
\end{eqnarray}
The entropy is a concave function of $p$ 
whereas $\E_{p} (\Phi_\NN (x)) = y$ is a set of
linear conditions over $p$.
If $\Phi_\NN (x)$ is bounded over $\Omega_{d,\epsilon}$ 
then the set of densities
$p$ which satisfy the moment conditions is compact. 
As a consequence, there exists a 
unique macrocanonical density $p_{\mathrm{ma}}$ which maximizes $H(p)$. 
It is obtained by minimizing the following Lagrangian
\begin{equation}
\label{lagrange}
{\cal L}_\NN (p,\beta) =  - H(p) + \lb \beta , \E_p (\Phi_\NN (x)) - y \rb~,
\end{equation}
also called free energy in statistical physics.
The Lagrange multipliers
$\beta = \{\beta_k \}_{k \leq K}$ 
are adjusted so that the moment condition (\ref{maxentropmoment2}) is satisfied.
The density which minimizes (\ref{lagrange}) can be written as 
an exponential family 
\begin{equation}
\label{maxentrop}
p_{\mathrm{ma}} (x) = {\cal Z}^{-1} \exp (- \langle \beta, \Phi_\NN (x) \rangle ) ~,
\end{equation}
where $\cal Z$ guarantees that $\int p_{\mathrm{ma}}(x)\, dx = 1$ and hence
\begin{equation}
\label{maxentroppart}
{\cal Z} =  \int_{\domain} \exp(\langle \beta, \Phi_\NN (x) \rangle ) \,dx~.
\end{equation}
A direct calculation shows that the resulting maximum entropy is
\begin{equation}
\label{canonentasf}
H(p_{\mathrm{ma}}) = - \log {\cal Z} + \lb \beta , y \rb ~.
\end{equation}

If the probability measure of the restriction of $X$ to $\LaN$ has a density $p$
relatively to the Lebesgue measure, then we can also verify that the Kullback-Liebler
divergence
\[
KL(p||p_{\mathrm{ma}}) = \int_{\LaN} p(x)\, \log \frac{p_{\mathrm{ma}}(x)} {p(x)}\, dx
\]
satisfies
\begin{equation}
\label{klnonneg}
KL(p||p_{\mathrm{ma}}) = H(p_{\mathrm{ma}}) - H(p) \geq 0~.
\end{equation}
Optimizing the interaction energy
$\Phi_\NN$ thus amounts to minimizing the resulting maximum entropy
$H(p_{\mathrm{ma}})$ \cite{zhu1998filters} so that 
$H(p_{\mathrm{ma}}) = H(p)$ and hence $\muma = \mu$.

Note that it is not necessary to impose that $\Phi_\NN$ is bounded on 
$\domain$.
If there exists $\beta \in \R^K$ such that the distribution (\ref{maxentrop}) 
satisfies the moment condition (\ref{maxentropmoment2}), then 
one can verify from (\ref{klnonneg}) that $\muma$ is the unique
maximum entropy distribution.
However, if $\Phi_\NN$ is not bounded on $\domain$
then there may not exist such a
$\beta \in \R^K$. Indeed, the maximization of entropy defines
a limit distribution
over distributions which satisfy the moment constraints, but this limit
may not satisfy the moment constraints anymore. One can construct
such examples with high order moment conditions \cite{Hamburger}. 
In this case the macrocanonical model does not exist
although we may still define a microcanonical model.

\paragraph{Macrocanonical Estimation}

Given an energy vector $\Phi_\NN$, and desired 
moment constraints $y=\E_\mu[\Phi(x)]$,
fitting macrocanonical models requires 
 estimating $\E_{\muma}[\Phi_\NN(x)]$. 
This expectation can be estimated with MCMC 
algorithms such as Metropolis-Hastings, which 
sample the Gibbs distribution (\ref{maxentrop}) to estimate 
$\E_{\muma} (\Phi_\NN (x))$ and iteratively update the Lagrange multipliers 
$\beta$ until $\E_{\muma} (\Phi_\NN (x))$ converges to $y$.
However, when $\NN$ is large, this
is numerically unfeasible because sampling a high-dimensional probability
distribution is computationally dominated by the mixing 
time of the Markov Chain, which in generally has an 
exponential dependence on the data dimensionality \cite{levin2017markov}.


\subsection{Shift Equivariant and Finite Range Potentials}
\label{shiftpotent}

Microcanonical densities in (\ref{eqnasdf1}) and macrocanonical densities 
in (\ref{maxentrop}) depend on $\Phi_\NN$. These densities 
remain constant under
any transformation of $x$ which leaves $\Phi_\NN (x)$ constant.
Stationary densities are obtained with a $\Phi_\NN$ which is 
invariant to translations. It is calculated
by averaging a potential vector which is equivariant to translations.
We review simple examples with ${\bf l}^1$ and ${\bf l}^2$ norms.
It illustrates convergence issues
of micro and macrocanonical densities when $d$ goes to $\infty$, with
sparse
regimes where microcanonical models exist without macrocanonical models.

\paragraph{Equivariant Potentials}
For any $x\in \margdomain^{\Z^{\ell}}$ we define a potential $U x (u) \in \R^K$
for each $u \in \Z^\ell$.  
We write $T_\tau x (u) = x(u-\tau)$ a translation of $x$ by $\tau \in \Z^\ell$.
A potential $U$ is shift-equivariant if
\[
\forall (x,\tau) \in \margdomain^{\Z^{\dd}} \times \Z^{\ell}~,~
U T_\tau x = T_\tau U x ~.
\]
The energy $\Phi_d (x)$ is computed from the restriction of $x$ in a square
$\LaN = [a , b]^\ell$. We extend $x$ over $\Z^d$ into a signal which is 
$b-a = d^{1/\ell}$ periodic
along each of the $\ell$ generators of the grid $\Z^\ell$. With an abuse of notation
we write $U x$ the potential $U$ applied to the periodic extension of $x$ and
\begin{equation}
\label{phi-def}
\Phi_d (x) = d^{-1} \sum_{u \in \LaN} U x (u). 
\end{equation} 
Observe that $\Phi_\NN (x) \in \R^K$ is invariant to periodic translations of $x$
in $\LaN$ modulo $d^{1/\ell}$.

We say that $U x$ has a finite
range $\Delta$ if  $U x(u)$ only depends upon the values of $x(u')$
for $u-u' \in [-\Delta,\Delta]^\ell$. The resulting macrocanonical density
(\ref{maxentrop}) 
is a Markov Random Field over cliques $[u-\Delta,u+\Delta]^\ell$ around each $u$ 
\begin{equation}
\label{q-def}
p_{\mathrm{ma}} (x) = {\cal Z}^{-1} \exp (- d^{-1} \sum_{u \in \LaN}
\langle \beta , U x(u) \rangle ) ~.
\end{equation} 
To approximate random processes, we must choose $\Delta$ to be the
integral scale beyond which structures become independent. 
When there are long range interactions as in
turbulent flows, this integral scale may be very large.
Before reviewing the general
convergence properties of the resulting micro and macrocanonical
densities we consider two important examples 
obtained with ${\bf l}^r$ norms. 

\paragraph{Convergence of ${\bf l}^r$ macro and microcanonical densities}
The potential $U x(u) = |x(u)|^r$ for $u \in \Z$
defines an ${\bf l}^r$ norm energy over intervals $\LaN = [1,d] \subset \Z$:
\begin{equation}
\label{nsdfsdf}
\Phi_d (x) = d^{-1}\, \|x\|^r_r =  d^{-1} \sum_{u\in \LaN} |x(u)|^r ~.
\end{equation}

The macrocanonical measure with density $p_{\mathrm{ma}}$ defined by $\E_{p_{\mathrm{ma}}} (\Phi_d (x)) = y \geq 0$
is 
\[
p_{\mathrm{ma}} (x) = {\cal Z}^{-1} e^{- \beta\, d^{-1}\,\|x\|_r^r}
\] 
for some $\beta > 0$. It is
the density of a vector of $d$ i.i.d random variables 
$X_d (u)$ having an
exponential distribution $\propto\, e^{- \beta |z|^r }$.

A microcanonical density $p_{d,\epsilon,y}$ is uniform
over $\OOey = \{ x \in \R^d~:~| d^{-1} \|x\|_r^r - y | \leq \epsilon \}$, 
which is a thin shell around an ${\bf l}^r$ ball in $\R^d$. It is the density
of a random vector $X_{d,\epsilon} $ defined on $\LaN$. 
For a fixed $m > 0$, when $d$ goes to $\infty$ and $\epsilon$ goes 
to zero then the joint density of
$X_{d,\epsilon} (1),...,X_{d,\epsilon} (m)$ converges in total variation distance
to i.i.d random variables having an exponential distribution
$\propto e^{- \beta |z|^r }$ \cite{Barthe:05}, and $\E(|X_{d,\epsilon} (u)|^r)$ converges to $y$.
The microcanonical distribution thus 
converges to the macrocanonical distribution.
This family of results has a long history, first proved in 1906 by Borel 
\cite{Borel} for $r = 2$ and in 1987 by Diaconis and
Freeman for $r = 1$ \cite{Diaconis}. 

\paragraph{Intersections of ${\bf l}^1$ and ${\bf l}^2$ balls}
The situation becomes more complex for the two-dimensional potential 
$U x(u) = (|x(u)|^1 , |x(u)|^2)$ which defines an energy
$\Phi_d (x) = (d^{-1} \|x\|_1 , d^{-1} \|x\|_2^2)$ 
over intervals $\LaN = [1,d] \subset \Z$. We shall see that microcanonical models may exist without macrocanonical models. 
 
One can verify that there exists a unique 
maximum entropy density $p_{\mathrm{ma}}$ conditioned on $\E_{p_{\mathrm{ma}}} (\Phi_d (x)) = y$
if and only if 
\[
1 \leq \frac{y_2} {y_1^2} \leq 2 ,
\]
in which case there exists $\beta_1$ and $\beta_2$ such that
\[
p_{\mathrm{ma}} (x) = {\cal Z}^{-1} e^{-  d^{-1}\,(\beta_1 \|x\|_1 + \beta_2 \|x\|_2^2)} .
\]

The microcanonical set 
$\OOey = \{ x\,:\, \|\Phi_d (x) - y \| \leq \epsilon \}$
is a thin shell around
the intersection of the simplex $\|x\|_1 = d \,y_1$
and the sphere $\|x\|^2_2 = d \,y_2$. Since $\|x\|^2_2 \leq \|x\|^2_1 \leq d \|x\|^2_2$, 
this intersection is non-empty over a wider range defined by
\[
1 \leq \frac{y_2}{y^2_1} \leq d .
\]

When $1 < \frac{y_2} {y_1^2} \leq 2$, micro and macrocanonical densities
have the same limit when $d$ goes to $\infty$ and $\epsilon$ goes 
to zero. 
S. Chatterjee \cite{chatterjee2017note} proves that
the joint microcanonical density of
$X_{d,\epsilon} (1),...,X_{d,\epsilon} (m)$ for a fixed $m$ 
converges 
to i.i.d random variables having an exponential distribution
equal to $\alpha e^{-  \beta_1 |z| - \beta_2 |z|^2}$,
and $(\E(|X_{d,\epsilon} (u)|^1,\E(|X_{d,\epsilon} (u)|^2)$ converges to $y$. 
If $y_2/y_1^2 = 2$ then $\beta_2 = 0$.
In this regime where macrocanonical 
densities are well-defined, micro and macrocanonical measures
converge to each other so the Boltzmann equivalence principle 
is again verified.

However, when $y_2/y_1^2 > 2$ the macrocanonical density is not defined, 
so the Boltzmann equivalence principle is violated.
The microcanonical
set contains sparse signals which are not captured by exponential
distributions. In this case,
Chatterjee \cite{chatterjee2017note} proves that when $d$ goes to $\infty$ and $\epsilon$
to $0$, $X_{d,\epsilon}$ has one large coefficient randomly located at some $u_0 \in \LaN$
for which $X_{d,\epsilon}^2 (u_0) \sim d (y_2 - 2 y_1^2)$ 
with a probability which tends to $1$. 
All other coefficients have a much smaller $O(y_1)$ amplitude.
For $m$ fixed, $X_{d,\epsilon}(1),...,X_{d,\epsilon}(m)$ converge in law
to i.i.d random variables having marginals equal
to $e^{-  \beta_1 |z|}$, but there is no convergence of moments. 
This example shows that the Boltzmann equivalence principle is not necessarily satisfied, particularly when signals exhibit a strong sparsity behavior.

\subsection{Boltzmann Equivalence Principle}
\label{Gibbsconv}

Micro and macrocanonical densities are defined over configurations $x$ specified
in a finite cube $\LaN$ of dimension $\dd$.
Let $\Phi_\NN (x)$ be a shift-invariant energy vector computed
by averaging a finite range potential $U x$. 
To compute estimators which converge when $d$ goes to $\infty$, we 
need to ensure that microcanonical densities converge in the moments sense.
We consider the limit among measures defined on the configuration space $\margdomain^{\Z^\ell}$,
with the product topology of Borel fields on the interval $\margdomain \subset \R$.  
The asymptotic equivalence between micro and macrocanonical measures 
is called the Boltzmann Equivalence Principle \cite{gallagher2012newton}.
Their convergence to the same 
Gibbs measures was first proved by 
Landford \cite{lanford1975time}. 
It is the center of a large body of work, rooted in the theory of large 
deviations \cite{ellis}. 
We review results obtained
when $\margdomain$ is a bounded interval and for Gaussian processes.

\paragraph{Macrocanonical Convergence}
When $\margdomain$ is a bounded interval, macrocanonical distributions
are unique minimizers of the Lagrangian (\ref{lagrange}). 
When $d$ goes to $\infty$, the limit 
Gibbs measure is defined by normalizing this Lagrangian
so that it converges to a variational problem defined over
a stationary measure $\mu$.
Suppose that $\mu$ exists. Since $U x$ is equivariant to translations and
$\mu$ is stationary it results that
$\E_\mu (U x(u)) = \E_\mu(U x)$ does not depend upon
the grid point $u$. Suppose that $\mu$ has no long range correlation so that boundary
values have a negligible influence. Since $\Phi_\NN (x)$ is an 
average of $U x(u)$ in $\LaN$ it follows that
\[
\lim_{d \rightarrow \infty} \E_\mu (\Phi_\NN (x)) = \E_\mu (U x)~.
\]
The Lagrangian (\ref{lagrange}) includes a negative entropy term 
that diverges as $d \to \infty$ if $\mu$ has finite range correlations. 
The normalisation replaces the entropy by 
an entropy rate $\overline H (\mu)$, defined by considering the restriction
 $\mu_d$ of $\mu$ on the finite dimensional
configuration space $\domain$. 
Let $q_d$ be the density of $\mu_d$ relatively to the Lebesgue measure.
If $\mu$ has a finite range correlation we expect
that $H(q_d)$ grows linearly with $d$. The entropy rate is defined by
\begin{equation}
\label{entropyrate}
\overline H(\mu) = \lim_{d \rightarrow \infty} d^{-1}\, H(q_d) ~.     
\end{equation}

Normalizing the free energy Lagrangian (\ref{lagrange}) by 
$d$ and taking the limit when $d$ goes to $\infty$ defines
a new Lagrangian
\begin{equation}
\label{lagrangeinf}
{\cal L}_\infty (\mu,\beta) =  - \overline H (\mu) + \lb \beta , \E_{\mu} (U x) -y \rb~.
\end{equation}
Gibbs measures minimize this Lagrangian 
over the space of stationary measures for $\beta$ fixed.

If $U$ is a bounded, finite range and continuous potential, 
then one can prove \cite{stroock,Georgii} 
that the set of Gibbs measures which minimize this
Lagrangian is a non-empty, convex and compact set of measures.
In general the solution is not unique because contrarily to 
the finite Lagrangian (\ref{lagrange}) where
$-H(p)$  is strictly convex, the entropy
rate $\overline H (\mu)$ is affine \cite{stroock,Georgii}. This implies that
depending upon boundary
conditions in $\LaN$, macrocanonical densities may converge to different
Gibbs measures, which is a phase transition phenomena.

Periodic boundary conditions over the finite cube
$\LaN$ simplify computational algorithms,
but they are artificial. The limit Gibbs measure will not depend upon
these boundary conditions if it is unique, and hence
if there is no phase transition. This happens when there 
is no long range interactions, so that
boundary values do not condition the probability distributions of far
away values. 
In this paper, we concentrate on problems where there is no such phase transition.

\paragraph{Microcanonical convergence}
The main difficulty is to find conditions which guarantee that 
microcanonical measures converge to the same Gibbs measure, having
a maximum entropy rate conditioned by moment conditions.   
Suppose that $U$ is continuous, bounded and has a finite range. 
When $d$ goes to $\infty$ and $\epsilon$ goes to zero,
one can prove \cite{stroock,Georgii} that
microcanonical distributions converge for an appropriate topology, to a limit measure which minimizes the same Lagrangian (\ref{lagrangeinf}) as the one obtained  from macrocanonical densities. 
If there is no phase transition, 
so that the macrocanonical measure converges to a unique Gibbs measure $\mu$, then
this limit is the same for macrocanonical and microcanonical measures.
More specifically, if $f(x)$ is a bounded and continuous function defined for
any $x \in \margdomain^{\Z^\ell}$, then the expected value of $f$ computed over $\LaN$ with
microcanonical and macrocanonical measures converge to $\E_\mu (f(x))$
when $d$ goes to $\infty$. We thus have a convergence for all bounded moments.
However, it is not necessary to impose that $\margdomain$ is bounded
to verify the Boltzmann equivalence principle, 
as shown by the following Gaussian example. 

\paragraph{Gaussian processes}  Gaussian stationary
measures are important examples 
of Gibbs measures where $x$ takes its values
in $\margdomain = \R$. They are obtained with a quadratic potential $U x = \{ U_k x \}_{k \leq K}$ computed with convolutions so that it
is equivariant to translations over the grid $\Z^d$. Let us define
\[
U_k x(u) = |x \star h_k (u)|^2 = \Big| \sum_{m \in \Z^d} x(u-m)\, h_k (m) \Big|^2~,
\]
where each $h_k$ has a support in $[-\Delta,\Delta]$.

If $x \in \R^\LaN$ then $U x$ is computed by extending $x$ on $\Z^\ell$
with a periodic extension beyond boundaries.
Potentials can then be rewritten with circular convolutions of $x$ 
\begin{equation}
\label{posdf-sdf2}
U_k x (u) = | x \star h_{d,k}(u)|^2 = \Big| \sum_{m \in \LaN} x (m) \, h_{d,k} (n-m) \Big|^2~.
\end{equation}
with periodic filters
\begin{equation}
\label{posdf-sdf1}
h_{d,k} (n) = \sum_{m \in \Z^\ell} h_{k} (n - m d^{1/\ell} ) ~.
\end{equation}
The energy $\Phi_d (x)$ is thus a vector of normalized ${\bf l}^2$ norms:
\begin{equation}
\label{posdf-sdf}
\Phi_{d} (x) =  \Big\{ d^{-1}\, \sum_{u \in \LaN} | x \star  h_{d,k}(u)|^2  = d^{-1} 
\| x \star  h_{d,k}\|_2^2 \Big\}_{k \leq K}  .
\end{equation}

If $\hat h_k (\om)$ does not vanish for all $\om \in [0,2\pi]$ and
$k \leq K$ then
Varadhan and Donsker \cite{donsker1985large} proved that Boltzmann equivalence principle is satisfied when $d$ goes
to $\infty$. The microcanonical and macrocanonical models converge
to a Gaussian stationary process $\mu$ whose power-spectrum is
\begin{equation}
\label{posdf-wefsdf}
P_\mu(\om) = \Big(\sum_{k=1}^{K} \beta_k |\hat h_k (\om)|^2\Big)^{-1}~.
\end{equation}
The next section studies asymptotic properties of 
microcanonical models even though
the macrocanonical model may not exist.

%% file: sampling1.tex
\section{Microcanonical Models beyond Boltzmann Equivalence}
\label{sec_algo}

%
%
%
%
%

We can guarantee that a maximum entropy microcanonical measure exists by
making sure that microcanonical ensembles are compact. Even
if this valid, the macrocanonical measure may not exist if $x(u)$ is 
defined over an interval $\margdomain$ which is not bounded. In this case
the Boltzmann equivalence principle is violated. Section \ref{shiftpotent} gives an example with uniform measures over 
intersections of $\bf l^1$ and $\bf l^2$ balls, in the sparse regime.
Microcanonical models thus offer more
flexibility, particularly for signals having sparse behavior. 

In the rest of the paper, we  embed all processes over $\R$, 
including binary processes such as Ising and Bernoulli .
We thus consider that $x(u)$ takes its values in $\margdomain = \R$, in which case $\domain =\R^\NN$, where the grid topology is omitted for ease of notation. 
We study microcanonical properties independently from the 
corresponding macrocanonical measures which may not exist. For this
purpose, Section \ref{microcanonical} relates the
maximum entropy of a microcanonical measure to the Jacobian of the energy
potential. It gives sufficient conditions so that the entropy rate
converges when $d$ goes to $\infty$. However, sampling a maximum entropy 
microcanonical process is computationally very expensive. 
Section \ref{microgradsec} introduces a different class of microcanonical
processes obtained by transporting a maximum entropy measure with a 
gradient descent algorithm which converges towards the microcanonical set.
The transported measure does not have a maximum entropy but we 
prove that it has common symmetries with the maximum entropy measure.
Convergence to microcanonical sets
is studied in Section \ref{grad_descent_sect}.

\subsection{Microcanonical Entropy and Jacobian}
\label{microcanonical}

We study the convergence 
of maximum entropy
microcanonical models when $d$ goes to $\infty$ by studying the 
convergence of their entropy rate without supposing that there exists
a macrocanonical model. This is done by relating the maximum entropy rate
to the Jacobian of the energy $\Phi_d$. 

We consider a shift-equivariant and finite range
potential from Section \ref{shiftpotent},
and the corresponding microcanonical measure $\mumi_{d,\epsilon}$, 
defined as the uniform distribution on compact sets of the form
\[
\OOey = \{ x \in \domainr~:~\|\Phi_d (x) - y \| \leq \epsilon \}~.
\]
We saw in (\ref{eqnasdf198}) that the entropy of $\mumi_{d,\epsilon}$ is
\begin{equation}
\label{entropymicroeq}
H(\mumi_{d,\epsilon}) = - \int \pey(x)\, \log \pey(x)\, d x =
\log \Big(\int 1_{\Omega_{d,\epsilon}} (x) \, d x \Big)~.
\end{equation}

Since $\Phi_d (x) =  {d^{-1}} \sum_{u \in\LaN}  Ux(u)$ and $U x(u)$ only
depends on the values of $x(i)$ for $i \in [u-\Delta,u+\Delta]^\ell$, 
one can verify that 
the $i$-th column $J_i \Phi_d (x) = \partial_{x(i)} \Phi_{d} (x) \in \R^K$ 
only depends upon the restriction of $x$ in $[i - \Delta , i+ \Delta]^\dd$. Moreover, 
thanks to the equivariant structure of $U$, one can verify that
$$\forall~i \leq d~,~J_i \Phi_d (x) = d^{-1} \sum_{|m| \leq \Delta} \partial_{x(0)} U  (T_{-i}) x(m) ~,$$
so the global properties of the Jacobian $J \Phi_d(x)$ can be derived 
from the Jacobian of the potential, restricted on a window: \begin{eqnarray}
\label{columnjac}
{J} U ~ : \R^{(2\Delta+1)\ell} &\to& \R^K \\
\nonumber
{x} &\mapsto& \sum_{|m| \leq \Delta} \partial_{x(0)} U {x}(m) ~.
\end{eqnarray}

%
We denote by $\partial \mathcal{A}$ the frontier of a set $\mathcal{A}$ and by $\mathcal{A}^o = \mathcal{A} - \partial \mathcal{A}$
the interior of $\mathcal{A}$, and by $\overline{\mathcal{A}}$ the complement of $\mathcal{A}$.
We also denote by $|J \Phi_d(x)| = \sqrt{{\rm det} \left( J \Phi_d(x) J \Phi_d(x)^T \right)}$ the $K$-dimensional
determinant of $J \Phi_d$, and by $d {\cal H}(x)^L$ the $L$-dimensional Hausdorff measure.
We shall make the following assumptions on $U$:
\begin{enumerate}[label=(\Alph*)]
\item $U$ is  uniformly Lipschitz on compact sets, which implies that
for any compact $\mathcal{C} \subset \domainr$ there exists $\beta \geq 0$ such that
\begin{equation}
\label{condPhi}
\forall (x,x') \in \mathcal{C}^{2d}~~,~~\| \Phi_d (x) - \Phi_d(x') \|_2 \leq \beta\, \|x - x'\|_2~.
\end{equation}
It also implies that $| J \Phi_d (x)| \leq \beta^K$ for $x \in \mathcal{C}$. 
\item We shall also suppose that $\Phi_d^{-1}$
maps compact sets $\mathcal{C}$ to compact sets, with a controlled growth 
with respect to $d$. For each compact set $\mathcal{C}\subset \R^K$, there exists 
a constant $C$ independent of $d$ such that
\begin{equation}
\label{condPhi2}
\forall~d~,~\Phi_d^{-1}(\mathcal{C}) = \{x \in \domainr~:~\Phi_d(x) \in \mathcal{C}\} \subseteq B_{2,d}(C \sqrt{d})~, 
\end{equation}
where $B_{p,d}(R)$ denotes the $d$-dimensional $\bf{l}^p$ Euclidean Ball of radius $R$.
It follows that $\Phi_d^{-1}(y)$ is a compact and Lipschitz manifold whose dimension
is typically $d-K$, except for degenerated cases. For example, if $d^{-1} \| x \|_2^2$ is a component of the vector $\Phi_d$, this condition is satisfied. 
\end{enumerate}
Lastly, we need to control the integrability of $|J \Phi_\NN|^{-1}$ nearby microcanonical sets. More precisely,
 for each $y$ and any sufficiently small $\epsilon>0$, we require that $|J \Phi_\NN(x) |^{-1}$ is integrable in $\Omega^{y}_{\NN, \epsilon}$. The following gives a sufficient condition which depends only on the potential function.
\begin{enumerate}[label=(\Alph*),resume]
\item For some $R>0$, let $X$ be drawn from the uniform measure in the ball $B(2\Delta+1, R)$ and $Z = {J} U({X}) \in \R^K$ be
the random vector obtained by applying the mapping ${J} U$ defined in (\ref{columnjac}). We shall suppose that there exists $\eta > 0$ such that 
\begin{equation}
\label{condPhi3}
~\forall~\mathcal{S} \subset \R^K \text{  Lebesgue measurable} ~,~P( Z \in \mathcal{S}) \lesssim |\mathcal{S}|^{\eta}  ~. 
\end{equation}
%

\end{enumerate}
This condition assumes that the differential of $U$ does not concentrate too much on a low-dimensional subspace of $\R^K$, 
nor in a discrete subset, but it does not require that its distribution is absolutely continuous with respect to the Lebesgue measure. 
We shall see next that potentials of the form $U x = \{ | x \star h_k |^p \}_{k \leq K}$ with $p=1,2$ with complex filters $h_k$ define an integrable $|J \Phi_\NN|^{-1}$.

The following theorem computes the entropy of a microcanonical process 
from a change of variable metric, which depends upon
the Jacobian of the interaction energy $\Phi_d$. The theorem derives
a microcanonical entropy rate which converges when $d$ goes to $\infty$.

\begin{theorem}
\label{maxentth00}
Suppose $U$ verifies \textit{(A), (B) and (C)} above. 
Then the following properties are verified:
\begin{enumerate}[label=(\roman*)]
\item For sufficiently large $d$, 
\begin{equation}
\label{resundsf01}
H(\mumi_{d,\epsilon}) = \log {\int_{\|z - y\| \leq \epsilon} \h (z)  \, dz}~,
\end{equation}
where $\h$ is the change of variable metric which satisfies
\begin{equation}
\label{resundsf2}
\h(y) = \int_{\Phi_d^{-1}(y)} | J \Phi_{d}(x)|^{-1}\, d {\cal H}^{d-K}(x) < \infty~~a.e,
\end{equation}
where ${\cal H}^{d-K}$ is the $d-K$ dimensional Hausdorff measure. Moreover, $\h(y)$ has a finite integral on compact sets.
\item  The function $\h$ 
is strictly positive in the interior of 
$\Phi_d(\domainr)$, up to a thin shell on the boundary; ie, 
on sets $C_d \subset \Phi_d(\domainr)$ satisfying
$$\sup_{ y \in C_d} \mathrm{dist}(y, \overline{\Phi_d(\domainr)}) \leq c \cdot d^{-1/\ell}~,$$
for some constant $c$.
\item Suppose that either $\Delta=1$, or that the potential $U$ is H\"older continuous 
with parameter $\alpha < 2/\dd$: $|U(x) - U(x')| \leq C \| x - x' \|^\alpha$. 
Then, for each $\epsilon>0$, the entropy rate
$d^{-1} H(\mumi_{d,\epsilon})$
converges as $d\to \infty$ and satisfies
\begin{equation}
\label{theopartthree}
-\infty < \lim_{d\to \infty} d^{-1} H(\mumi_{d,\epsilon}) \leq C \log \|y \|^2~,
\end{equation}
where $C$ is a universal constant.
\end{enumerate}
\end{theorem}

The proof is in Appendix \ref{proof1}.
This theorem highlights the connection between the entropy and
the Jacobian through $\h(y)$, via the coarea formula. 
It defines the entropy rate  of a microcanonical ensemble
for general $\Phi_d$ in the thermodynamical limit $d \to \infty$, 
without relying on a macrocanonical model.
One can compare the conditions of Theorem \ref{maxentth00} 
with those that ensure the convergence of the microcanonical and macrocanonical 
measures. In \cite{stroockZeitouni, ZeitouniBook} this equivalence is 
established for bounded, finite-range potentials $U$. Our 
condition to prove that the entropy rate converges
is weaker ($U$ H\"older continuous), but we do not 
study convergence beyond the entropy rate. 
Studying the convergence of the microcanonical measure in more general conditions remains an open question.
Finally, notice that for positive integers $k$, the Hausdorff measure is 
equivalent to the $k$-dimensional Lebesgue measure up to a constant rescaling. 

The microcanonical thickness parameter $\epsilon$ 
is important to ensure appropriate convergence. The following 
corollary quantifies the effect of $\epsilon$ in the entropy rate, 
and proves that its contribution to the energy is small for sufficiently 
large $d$.

\begin{corollary}
\label{coroentropyepsilon}
Under the same conditions as Theorem \ref{maxentth00}, for $d$ fixed and when $\epsilon \to 0$, the entropy rate of the $\epsilon$-thick microcanonical model satisfies
\[
d^{-1} H(\mumi_{d,\epsilon}) \sim \frac{K}{d} \log \epsilon .
\]
\end{corollary}

As a consequence of this corollary, the entropy variation 
due to a change in the thickness from $\epsilon$ to $\epsilon'$ is 
of the order of $\frac{K}{d}\log\left(\frac{\epsilon}{\epsilon'}\right)$, 
which is negligible if $K \log\left(\frac{\epsilon}{\epsilon'}\right) 
\ll d$.



This paper concentrates on interaction energy vectors
$\Phi_d$ defined by ${\bf l}^2$ and ${\bf l}^1$ 
norms of convolutions of $x$ with multiple filters. 
The next proposition proves that such
interaction energies 
satisfy the assumptions of Theorem \ref{maxentth00}.
The proof is in Appendix \ref{proof2}.
\begin{proposition}
\label{maxentth01}
$\Phi_\NN$ satisfies assumptions (A), (B) and (C) in the following cases:
\begin{enumerate}[label=(\roman*)]
 \item $\Phi_d(x) = \{ d^{-1} \| x \star h_k \|_2^2 \}_{k \leq K}$ and the $\{ h_k \}_{k \leq K}$ are linearly independent. 
 \item
 $\Phi_d(x) = \{d^{-1} \|x\|^2, d^{-1} \|x\|_1 \}$. 
 \item $\Phi_\NN(x) = \{d^{-1} \|x\|^2, d^{-1} \| x \star h_k \|_1 \}_{k \leq K}$ and the $h_k$ are linearly independent 
with $|\hat{h}_k(-\om)| \neq |\hat{h}_k(\om) |$ for all $\om$. 
 \end{enumerate}
\end{proposition}

\subsection{Microcanonical Gradient Descent Model}
\label{microgradsec}
Computing samples of a maximum entropy
microcanonical model is typically done with 
MCMC algorithms or Langevin Dynamics 
\cite{creutz1983microcanonical}, which is computationally very expensive.
Computations can be considerably reduced by avoiding to enforce the
maximum entropy constraint over the microcanonical set. 
Microcanonical models computed with alternative projections
and gradient descents have been implemented to sample
texture synthesis models 
\cite{heeger1995pyramid, portilla2000parametric,gatys2015texture}. 
Another related sampling algorithm 
is the so-called Herding algorithm by Welling \cite{welling2009herding}, which 
produces `pseudo-samples' of a microcanonical model in a deterministic fashion by solving a sequence of primal-dual updates. 

We consider microcanonical
gradient descent models obtained by transporting
an initial measure towards a microcanonical set,
using gradient descent with respect to the distance to the microcanincal ensemble.
 We prove that the gradient descent preserves
many symmetries of the maximum entropy microcanonical measure. 


Let $\Phi_d$ be a shift-invariant function as defined in Section 
\ref{shiftpotent} and $y \in \Phi_d (\domainr) $. 
We transport an initial measure 
$\mu_0$ towards a measure
supported in a microcanonical set $\Omega_{d,\epsilon}$, 
by iteratively minimising
\begin{equation}
\label{potenener}
\En(x) = \frac{1}{2}\| \Phi_d (x) - y \|^2 
\end{equation}
with mappings of the form 
\begin{equation}
\label{yoy1}
\varphi_n(x) = x - \kappa_n \nabla \En(x) = x - \kappa_n J \Phi_d (x)^T 
( \Phi_d (x) - y)~,
\end{equation}
where $\kappa_n$ is the gradient step at each iteration $n$.

Given an initial measure $\mu_0$, 
the measure update is
\begin{equation}
\label{yoy2}
\mu_{n+1} := \varphi_{n,\#} \mu_n ,
\end{equation}
with the standard pushforward measure $f_{\#}(\mu) [\mathcal{A}] = \mu[ f^{-1}(\mathcal{A}) ]$ for any $\mu$-measurable set $\mathcal{A}$, 
where $f^{-1}(\mathcal{A})=\{ x; f(x) \in \mathcal{A}\}$.

Samples from $\mu_n$ are thus obtained by transforming samples $x_0$ from $\mu_0$ with the 
mapping $\bar{\varphi} = \varphi_n \circ \varphi_{n-1} \dots \circ \varphi_1$. It
corresponds to $n$ steps of a gradient descent initialized with $x_0 \sim \mu_0$:
$$x_{l+1} = x_l - \kappa_l J \Phi_d(x_l)^T ( \Phi_\NN(x_l) -y )~.$$

Next section studies the convergence of the gradient descent measures
$\mu_n$. Even if they converge to a measure supported in a microcanonical
set $\Omega_{d,\epsilon}$, in general they do not converge to a maximum entropy measure on this set. However, the next theorem proves that if
$\mu_0$ is a Gaussian measure of i.i.d Gaussian random variables then
they have a large class
of common symmetries with the maximum entropy measure.
Let us recall that a symmetry of a measure $\mu$
is a linear invertible operator
$L$ such that for any measurable set $\mathcal{A}$, $\mu [L^{-1}( \mathcal{A})] = \mu[\mathcal{A}]$. 
A linear invertible operator $L$ is a symmetry of $\Phi_d$ if 
for all $x \in \domainr$, $\Phi_d (L^{-1} x) = \Phi_d (x)$. 
It preserves volumes if its determinant satisfies $|{\rm det} L| = 1$.
It is orthogonal if $L^t L = L L^t = I$ and we say that it 
preserves a stationary mean if
$L {\bf 1} = {\bf 1}$ for $\mathbf{ 1} = (1,...,1) \in \R^\ell$.

\begin{theorem}
\label{invartheo}
(i) If $L$ is a symmetry of $\Phi_d$ which preserves volumes then it is
a symmetry of the maximum entropy microcanonical measure.\\
(ii) If $L$ is a symmetry of $\Phi_d$ and of $\mu_0$ then it is 
a symmetry of $\mu_n$ for any $n \geq 0$.\\
(iii) Suppose that $\mu_0$ is a Gaussian white noise measure
of $d$ i.i.d Gaussian random variables. Then, if $L$
is a symmetry of $\Phi_d$ which 
is orthogonal and preserves a stationary mean then it is a
symmetry of $\mu_n$ for any $n \geq 0$.
\end{theorem}

The theorem proof is in Appendix \ref{invartheoapp}. 
The initial measure $\mu_0$ is chosen so that it has many symmetries
in common with $\Phi_d$ and hence the gradient descent measures have
many symmetries in common with a maximum entropy measure. 
A Gaussian measure of i.i.d Gaussian variables 
of mean $m_0$ and $\sigma_0$ is a maximum entropy measure
conditioned by a stationary mean and variance. It is uniform over
spheres which guarantees that it has a large group of symmetries.
The stationary mean $m_0$ and variance 
$\sigma_0^2$ are adjusted so that
that microcanonical sets are nearly included over
the sphere of mean $m_0 \bf 1$ and radius $\sigma_0$, 
where $\mu_0$ concentrates and is uniform. We thus set $m_0$
and $\sigma_0^2$ to be the empirical stationary 
mean and variance calculated
from the realization $\bar x$ of $X$:
\begin{equation}
\label{initansdfmsd}
m_0 = d^{-1} \sum_{u \in \LaN} \bar x(u)~~\mbox{and}~~
\sigma_0^2 = d^{-1} \sum_{u \in \LaN} (\bar x(u) - m_0)^2~.
\end{equation}

Observe that periodic 
shifts are linear orthogonal operators and preserve a stationary
mean. The following corollary 
applies property (iii) of Theorem \ref{invartheo} to
prove that $\mu_n$ are circular-stationary.

\begin{corollary}
\label{cornsdf98we}
If $\Phi_d$ is invariant to periodic shift and $\mu_0$ is 
a Gaussian white noise 
then $\mu_n$ is circular-stationary for $n \geq 0$. 
\end{corollary}

\subsection{Convergence of Microcanonical Gradient Descent}
\label{grad_descent_sect}

This section studies conditions so that the 
gradient descent (\ref{yoy2}) converges to a
stationary measure  supported in a microcanonical ensemble, and we give
a lower bound of its entropy rate. 
To guarantee that the algorithm
is not trapped in local minima, we
use the characterization of stable solutions from \cite{lee2016gradient,panageas2016gradient} 
based on the second-order analysis of critical points of (\ref{potenener}).
Such analysis reveals that gradient descent methods do not get stuck at 
critical points which are \emph{strict saddles} --- in which 
at least one Hessian eigenvalue is strictly negative, since the 
set of initialization parameters corresponding to the non-negative spectrum 
has measure $0$ relative to $\mu_0$.

\begin{definition}
We say that $\Phi_d = (\phi_1, \dots, \phi_K)$ has the strict saddle condition if 
$\Phi_d$ is at least ${\bf C}^2$ and for each $v \in \text{Null}( J \Phi_d(x)^\top ) \subseteq \R^K$, $v \neq 0$, the matrix
\begin{equation}
\label{strictsaddlehere}
 \sum_{k \leq K} v_k {\nabla^2} \phi_k(x) + J \Phi_d(x)^\top J \Phi_d(x)
\end{equation}
has at least one strictly negative eigenvalue, where ${\nabla^2} \phi_k$ is the Hessian of $\phi_k$.
\end{definition}

The following theorem, proved in Appendix \ref{appendo2}, establishes 
basic properties of the distribution generated by gradient descent, 
including sufficient conditions for its convergence to the microcanonical 
ensemble.

\begin{theorem}
\label{convergencetheo}
Assume $\Phi_d$ is ${\bf C}^2$ and satisfies property (B) (\ref{condPhi2}). Suppose that $\Phi_d$ is Lipschitz with $\text{Lip}_{\Phi_d} = \beta$ and
that $\nabla \Phi_d$ is also Lipschitz, with $\text{Lip}_{\nabla \Phi_d }= \eta$.
Let $y \in \Phi_d(\R^d)^\circ$. 
Then:
\begin{enumerate}[label=(\roman*)]
\item If $\Phi_d$ satisfies the strict saddle condition, then (\ref{potenener}) 
has no poor local minima. Moreover, if $| J \Phi_\NN (x) | > 0$ for all $x\in \Phi_\NN^{-1}(y)$, then 
by choosing step-sizes $\kappa_n$ such that $\kappa_n < \eta^{-1}$ for all $n$, 
$\mu_n$ converges almost surely to a limit measure $\mu_{\infty}$ \footnote{Defined as ${\rm Prob}\Big[ \mu_n(A) \to \mu_{\infty}(A) \text{ for any } \mathcal{F}-\text{measurable set } A\Big] = 1$, where $\mathcal{F}$ is the Borel $\sigma$-algebra on $\R^d$.}
Moreover, $\mu_{\infty}$ is supported in the microcanonical ensemble  $\Phi_d^{-1}(y)$ 
with appropriate choice of learning rate $\kappa_n$; 
that is, $A \cap \Phi_d^{-1}(y) = \emptyset \Rightarrow \mu_{\infty}(A) = 0$. 

\item The entropy rate $d^{-1} H(\mu_n)$ satisfies 
\begin{equation}
\label{fla0}
d^{-1} H( \mu_n) \geq d^{-1} H( \mu_0) - \left(1 - \frac{K}{d}\right) \eta \sum_{n'\leq n}  \kappa_{n'} r_{n'}   - \frac{K}{d}  \beta^2 \sum_{n' \leq n} \kappa_{n'} ~,
\end{equation}
where $r_n =  \E_{\mu_{n}} \sqrt{\En(x)}$ is the average distance to the microcanonical ensemble at iteration $n$.
\end{enumerate}
\end{theorem}

Part ({\it i}) gives sufficient conditions for the gradient descent 
sampling to converge towards the microcanonical ensemble. 
Each gradient descent step can reduce the entropy rate. 
By computing an upper bound of this entropy reduction, 
part ({\it ii}) gives a lower bound of the entropy rate after $n$ 
iterations. 
Although the gradient descent converges to the microcanonical 
ensemble in general the resulting
measure will not have a maximum entropy. However, 
(\ref{fla0}) gives a lower bound of 
its entropy rate. By choosing a measure $\mu_0$ having
a maximum entropy, we maximize 
the entropy of the lower-bound (\ref{fla0}). 

Our current results rely on second-order stationarity assumptions, 
but first-order stationary condition $\nabla \En(x^*) = 0$ 
may be sufficient to characterize convergence as $d \to \infty$. 
Indeed, this condition implies that either we reached the microcanonical 
ensemble, $\Phi(x^*) =y$,  or that we have found a non-regular point, with $| J \Phi_\NN(x^*)|=0$. 
Such points occur with vanishing probability as $\NN \to \infty$, but 
the rigorous analysis of this phenomena is left for future work.

The sufficient condition for $\mu_n$ to converge to a limit measure $\mu_\infty$ requires $| J \Phi_\NN(x) | > 0$ for $x \in \Phi_\NN^{-1}(y)$, which for certain choices of $\Phi_d$ may be hard to check. 
The following corollary, proved in Appendix \ref{convergencecoroproof}, provides an alternative sufficient condition which is stronger but easier to evaluate. 

\begin{corollary}
\label{convergencecoro}
If $\Phi_\NN$ is ${\bf C}^\infty$ and Lipschitz and satisfies the strict saddle condition, then $\mu_n$ converges for any $y \in \Phi_d(\R^d)$ up to a set of zero measure, and $\mu_\infty$ is supported in the microcanonical ensemble. 
\end{corollary}

%

We now give examples of energies $\Phi_d$ which satisfy the 
assumptions of previous theorem. The next theorem, proved in Appendix \ref{globalconvtheoremi}, shows that
the ${\bf l}^2$ ellipsoid representation satisfies the strict saddle condition, 
and therefore that the microcanonical gradient descent measure is supported in the microcanonical ensemble.
\begin{theorem}
\label{convergencetheo2}
If $\Phi_d(x) = \{d^{-1} \|x \star h_k\|_2^2 \}_k$ with linearly independent and compactly supported $h_k$, then $\Phi_d$ satisfies the strict 
saddle condition and $|J \Phi_\NN(x)|>0$ 
for $x \in \Phi_\NN^{-1}(y)$ with $y \in \Phi_\NN(\R^d)^\circ$, and therefore $\mu_\infty$ is supported in the microcanonical ensemble. 
\end{theorem}

A current limitation of the convergence analysis is that it relies on smoothness properties of $\Phi_d$,
thus leaving out of scope the ${\bf l}^1$-based representations. This limitation is intrinsic 
to the convergence analysis of non-smooth, non-convex optimization methods, which provides
no guarantees using simple gradient descent. 
The analysis of other algorithms such as ADMM \cite{wang2015global} or gradient sampling \cite{burke2005robust} 
in such conditions is left for future work. 

\paragraph{Continuous-time limit dynamics:} The measure transport (\ref{yoy2})
defined by gradient descent can be seen as a discretization of 
an underlying partial differential equation 
in the space of measures, describing the behavior as the step-size $\kappa_n \to 0$. 
The resulting dynamics is described by the well-known \emph{continuity equation}, expressed in the  
distributional sense as
\begin{equation}
\label{conttime}
\partial_t \mu_t = \mathrm{div} ( \nabla E \cdot \mu_t)~, 
\end{equation}
or equivalently
$$\forall~\phi \in \mathbf{C}^1_c~,~\partial_t \left( \int \phi(x) \mu_t(dx) \right) = - \int \langle \nabla \phi(x), \nabla E(x) \rangle \mu_t(dx)~,$$
where $\mathbf{C}^1_c$ denotes the space of $\mathbf{C}^1$ compactly supported test functions. 
As opposed to MCMC algorithms, which are discretizations of diffusion Stochastic Differential Equations (SDEs), 
the dynamics in our case are deterministic, and the only source of randomness comes from the initial measure $\mu_0$.
Notice also that the symmetry preservation properties described in Theorem \ref{invartheo} directly apply to 
the Liouville equation above.
Equation (\ref{conttime}) can also be interpreted as a Wasserstein Gradient Flow over the functional energy 
$$\mathscr{E}[\mu] = \int E(x) \mu(dx)~.$$
Recent work \cite{chizat2018global,rotskoff2018neural} has established global convergence of such Wasserstein Gradient Flows in the cases where $E$ is positively homogeneous, for suitable initialization. Although in our case $E$ is not homogeneous, we leave for future 
work to exploit the homogeneity properties of $\Phi_\NN$ to derive similar convergence results that can generalize \ref{convergencetheo}.



%% file: scatt1.tex
\section{Multiscale Microcanonical Wavelet and Scattering Models}
\label{wavelet-sec}

We study multiscale microcanonical models obtained with 
energy vectors computed with a wavelet transform. 
Next section introduces energy vectors computed with $\bf l^2$ 
and $\bf l^1$ norms of wavelet coefficients.
Section \ref{scattsect} introduces scattering which provide complementary
$\bf l^1$ norm coefficient computed with a second wavelet transform. 

\subsection{Wavelet Transform $\bf l^2$ and $\bf l^1$ Norms}
\label{waveletl2}

A wavelet transform, 
computes signal variations at different scales through convolutions with dilated wavelets. 
Maximum entropy models conditioned by wavelet $\bf l^2$ norms define
Gaussian processes. 
Wavelet transforms define sparse representations of large classes 
of signals. This sparsity characterize non-Gaussian behavior which is
specified by wavelet $\bf l^1$ norms. We write $\widehat x$ the
Fourier transform of $x$.

\paragraph{Wavelet Transform}
Wavelet coefficients are convolutions $x \star \psi_{j,q} (u)$ 
for $u \in \R^\ell$, where
each wavelet $\psi_{j,q}$ is a dilated band-pass filter which covers different frequency domains:
\begin{equation}
\label{dilatonsdfsd}
\psi_{j,q} (u) = 2^{-\ell j} \psi_{q} (2^{-j} u)~~\Rightarrow~~
\widehat \psi_{j,q} (\om) = \widehat \psi_{q} (2^{j} \om). 
\end{equation}
We will focus our attention on  the compactly-supported case, where the $Q$ mother wavelets $\psi_q$ have a support in $[-C,C]^\ell$ 
so the support of $\psi_{j,q}$ is in $[-C 2^j , C2^j]^\ell$.
The Fourier transform $\hat \psi_q (\om)$ have an energy concentrated
in frequency intervals which barely overlap for different $q$. 

If $x$ is supported in a cube $\LaN \subset \Z^\ell$, 
then $u$ is discretized on this square grid.
Convolutions are defined by extending $x$ into a periodic 
signal over $\Z^\ell$.
We showed in (\ref{posdf-sdf2}) 
that it is equivalent to computing circular convolutions with periodic wavelet filters (\ref{posdf-sdf1}).
Discrete periodic wavelets $\psi_{j,q}$ are band-pass filters with
a zero average $\sum_{u \in \LaN} \psi_{j,q} (u) = 0$. 
The minimum scale $2^j$ is limited by the sampling 
interval normalized to $1$, whereas the maximum scale $2^J$ is limited by the width $d^{1/\ell}$ of $\LaN$.

Wavelet coefficients $x \star \psi_{j,q} (u)$ separate the
frequency components of $x$ in several frequency bands, at scales
$1 \leq 2^j \leq 2^J$.
The remaining low frequencies at scales larger than $2^J$ are carried by
a single 
low-pass filter which we write $\psi_{J,0}(u) = 2^{-Jd} \psi_0 (2^{-J} u)$,
whose support is also included in $[-C 2^J , C 2^J]^\ell$. 

The wavelet transform of $x$ is defined by
\begin{equation}
\label{nsdf8sydf0909s90}
Wx = \Big\{x \star \psi_{j,q} \Big\}_{1 \leq j \leq J , q \leq Q}~.
\end{equation} 
We impose that the frequency supports $\hat \psi_{j,q}$ cover uniformly the
whole frequency domain, which is captured by 
the following Littlewood-Paley condition.
 There exists $\gamma < 1$ such that 
\begin{equation}
\label{nsdf8sydf013320}
\forall \om~~,~~1 - \gamma
\leq |\hat \psi_{J,0} (\om)|^2 + \frac 1 2 \sum_{j ,q} (|\hat \psi_{j,q} (\om)|^2 +
|\hat \psi_{j,q} (-\om)|^2) \leq 1 + \gamma ~.
\end{equation}
The condition implies the following energy inequalities
for any $x \in I^{\LaN}$
\begin{equation}
\label{nsdf8sydf01332230}
(1-\gamma) \|x\|_2^2 \leq \|x \star \psi_{J,0}\|_2^2 + 
\sum_{j,q} \|x \star \psi_{j,q}\|_2^ 2 \leq (1 + \gamma)\, \|x\|_2^2 
~.
\end{equation}
This 
is proved by multiplying (\ref{nsdf8sydf013320}) with $|\hat x (\om)|^2$
and applying the Plancherel equality. This property implies that $W$ is 
a bounded and invertible operator, 
and its inverse has a norm smaller than $(1 - \gamma)^{-1/2}$. If $\gamma = 0$ then $W$ is an isometry.

For audio signals in dimension $\ell = 1$, each wavelet is a complex filter
whose Fourier transform $\hat \psi_{q} (\om)$ has
an energy concentrated in the interval $[2^{q/Q} , 2^{(q+1)/Q}]$. It follows 
that $\hat \psi_{j,q} (\om)$ covers the interval $[2^{-j+q/Q} , 2^{-j+(q+1)/Q}]$ and
satisfies the Littlewood-Paley condition (\ref{nsdf8sydf013320}).
The parameter $Q$ is the number of wavelets per octave, which adjusts their frequency resolution. Wavelet representations are usually computed
with about $Q = 12$ wavelets per octave, which are similar to half-tone musical notes. In numerical computations, we choose Gabor wavelets as in \cite{anden}. Although strictly speaking this wavelet family does not have spatially compact support, the decay is exponential and has no practical effect.

For images in $\ell = 2$ dimensions, each wavelet is computed
by rotating a single mother wavelet 
\begin{equation}
\label{dilasndf0wds}
\psi_{j,q} (u) = 2^{-\ell j}\, \psi(2^{-j} r_q^{-1} u)
~~\Rightarrow~~
\widehat \psi_{j,q} (\om) = \widehat \psi (2^{j} r_q \om), 
\end{equation}
where $r_q u$ is a rotation of $u \in \R^2$
by an angle $q \pi/Q$. We choose a complex mother wavelet $\psi(u)$ whose
Fourier transform $\hat \psi(\om)$ is centered at a frequency $\xi$
over a frequency domain of radius approximately $|\xi|/2$. The support
of each $\hat \psi_{j,q}$ is dilated and rotated according to (\ref{dilasndf0wds}). Wavelet coefficients $x \star \psi_{j,q}$
thus compute variations of $x$ at scales $2^j$ along different directions.
In numerical computations we use Morlet wavelets as in 
\cite{bruna2013invariant} with $Q = 8$ angles to satisfy the
Littlewood-Paley condition (\ref{nsdf8sydf013320}). As in the case of audio, these wavelets have exponentially decaying spatial envelop.

\paragraph{Wavelet $\bf l^2$ norms}
We saw in Section \ref{Gibbsconv} 
that microcanonical maximum entropy measures conditioned by
energy vectors (\ref{posdf-sdf}) of $\bf l^2$ norms 
converge to Gaussian processes. We can define such energy vectors with
wavelet $\bf l^2$ norms, with the quadratic potential
\begin{equation}
\label{posdf-sdf0}
U x = \{  |x \star \psi_{j,q}|^2 \}_{j \leq J, q \leq Q}~.
\end{equation}
Since each filter support is included in $[-C 2^J , C 2^J]^\ell$, this potential has a finite
range $\Delta  = C 2^J$. When $x$ is defined over a cube $\LaN$ then $U x$ is computed by periodizing $x$ which is equivalent to periodizing the wavelet filters and replacing convolutions with circular convolutions, as shown in 
(\ref{posdf-sdf2}). To simplify notations, the periodized filters are 
still written $\psi_{j,q}$.
According to (\ref{posdf-sdf}) the energy over a cube $\LaN$
is given by normalized ${\bf l}^2$ norms
\begin{equation}
\label{posdf-sdf9}
\Phi_{d} (x) =  \{ d^{-1}\, \| 
x \star \psi_{j,q}\|_2^2 \}_{j \leq J,q\leq Q} .
\end{equation}
It measures the energy of $x$ in the different frequency bands covered
by each $\hat \psi_{j,q}$.

\paragraph{Wavelet $\bf l^1$ norms for sparsity}
Non-Gaussian properties can be captured with statistics sensitive to sparsity, as observed in early works studying the statistics of natural images \cite{simoncelli2001natural}, and formalized on specific processes such as multifractals \cite{bruna2015intermittent}. 
Suppose that $X \star \psi_{j,q}(u)$ has few large amplitude coefficients
and a large proportion of negligible coefficients. 
For example, if $X(u)$
is piecewise regular then 
$X \star \psi_{j,q} (u)$ is negligible over
domains where $X(u)$ is regular and it has a large amplitude near
singularities and sharp variations.
The marginal probability density of $X \star \psi_{j,q}(u)$ is 
then highly concentrated near $0$. It
is thus better approximated by a Laplacian rather than a Gaussian
distribution. We saw in Section \ref{shiftpotent}
that Laplacian distributions are maximum entropy distributions
conditioned by first order moments. This suggests to 
estimate $\E_\mu (|x \star \psi_{j,q}(u)|)$ as opposed to
$\E_\mu (|x \star \psi_{j,q}(u)|^2)$, with a normalized $\bf l^1$ norm
\[
d^{-1} \|x \star \psi_{j,q}(u)\|_1 = d^{-1} \sum_{u \in \La_d}
|x \star \psi_{j,q}(u)|~.
\]
A wavelet $\bf l^1$ norm energy is defined by replacing the 
quadratic potential (\ref{posdf-sdf0}) by a modulus potential
\begin{equation}
\label{posdf-sdf12}
U x = \{ |x \star \psi_{j,q}| \}_{j \leq J, q \leq Q}~,
\end{equation}
which also has a finite range 
$\Delta  = C 2^J$. The resulting energy over a cube $\LaN$
is 
\begin{equation}
\label{posdf-sdf912}
\Phi_{d} (x) =  \{ \, d^{-1}\, \| 
x \star \psi_{j,q}\|_1 \}_{j \leq J,q\leq Q} .
\end{equation}
It captures the sparsity of wavelet coefficients for each scale and
orientation.

\subsection{Scattering Transform}
\label{scattsect}

Wavelet $\bf l^1$ norm measure the sparsity of wavelet coefficients but
do not specify the spatial distribution of 
large amplitude wavelet coefficients.
Scattering transforms provide information about this geometry by 
computing interaction terms across scales, with an iterated wavelet
transform. Their mathematical properties are described 
in \cite{mallat2012group, bruna2015intermittent}, and applications to image and audio classification 
are studied in \cite{bruna2013invariant,anden}. We review important properties needed to define microcanonical models, including the energy conservation
allowing to recover wavelet $\bf l^2$ norms.


The mean of $x$ is estimated
over a cube $u \in \LaN$ by $d^{-1} \sum_{u \in \LaN} x(u)$.
The modulus of a wavelet coefficient $|x \star \psi_{j,q} (u)|$ measures
the variation of $x$ around its mean, 
in a neighborhood of $u$ of size proportional to 
$2^j$. A normalized ${\bf l}^1$ norm is the average of 
$|x \star \psi_{j,q} (u)|$ 
\[
d^{-1} \|x \star \psi_{j,q} \|_1 = 
d^{-1} \sum_{u \in \LaN} |x \star \psi_{j,q} (u)|~.
\]

Similarly, we can capture the variability of $|x \star \psi_{j,q} (u)|$ 
around this mean by convolving $|x \star \psi_{j,q} (u)|$ 
with a new set of wavelets: 
\[
||x \star \psi_{j,q}|\star \psi_{j',q'} (u)| .
\]
It measures the variations of $|x \star \psi_{j,q} (u)|$ in a
neighborhood of size $2^{j'}$. 
We shall consider the second wavelet $\psi_{j',q'}$ 
is calculated from the same mother wavelet than $\psi_{j,q}$
but for different $j',q'$, 
although the second mother wavelet may be changed as in \cite{anden}.

The maximum scales $2^j$ and $2^{j'}$ remain below a cut-off scale $2^J$
which specifies the maximum interaction range of the model. 
Incorporating first and second order coefficients defines a new potential
which captures the multiscale variations of $x$ as well as interaction
terms across scales:
\begin{equation}
\label{posdf-sdf1sc}
U x = \{ x \,,\, |x \star \psi_{j,q}| \,,\,||x \star \psi_{j,q}| \star 
\psi_{j',q'}| \}_{j,j' \leq J , q,q' \leq Q} .
\end{equation}
The corresponding energy vector is
\begin{equation}
\label{ScatRepnsdfm}
\Phi_\NN (x) = \Big\{ d^{-1}\sum_{u \in \LaN} x(u)\,,\,d^{-1}\, 
\|  x \star  \psi_{j,q} \|_1 \,,\,
d^{-1}\, \| |x \star \psi_{j,q}| \star \psi_{j',q'} \|_1 \Big\}_{1 \leq j,j' \leq J , q,q' \leq Q}~.
\end{equation}
It includes $K = 1+J Q + J^2 Q^2$ coefficients.

The following proposition,
shows that wavelet 
$\bf l^2$ norms can be closely
approximated from $\bf l^1$ norm scattering coefficients. 
As a result, we will be able to approximate Gaussian process as well as
non-Gaussian processes with a scattering energy vector. 
It is proved in Appendix \ref{enenpropproof},

\begin{proposition}
\label{enenprop}
Suppose that the wavelets satisfy (\ref{nsdf8sydf013320}) 
with $\gamma = 0$ then for $J = \log_2 d$
\begin{eqnarray}
\label{nsdf8sydf0133223}
\| x \star \psi_{j,q}\|_2^2 &=& \| x \star \psi_{j,q}\|_1^2 + 
\sum_{j'=1}^{\log_2 d} \sum_{q'=1}^Q 
\| |x \star \psi_{j,q}| \star \psi_{j',q'} \|_1 ^2\\
\nonumber
& & + \sum_{j'=1}^{\log_2 d} \sum_{q'=1}^Q \sum_{j''=1}^{\log_2 d} \sum_{q''=1}^Q 
\| |x \star \psi_{j,q}| \star \psi_{j',q'}| \star \psi_{j'',q''} \|_2 ^2 .
\end{eqnarray}
\end{proposition}

This proposition proves 
that $\bf l^2$ of wavelet coefficients are 
approximated by sums of first and second
order scattering coefficients plus a third order
term $ \sum_{j',q',j'',q''} \| |x \star \psi_{j,q}| 
\star \psi_{j',q'}| \star \psi_{j'',q''} \|_2 ^2$. 
For most stationary process this third order term is much smaller than
the first two and can be neglected \cite{bruna2013invariant}.
The theorem hypothesis supposes that 
wavelets satisfy the Littlewood inequality (\ref{nsdf8sydf013320})
with $\gamma=0$. If $\gamma$ is non-zero, it creates corrective terms
proportional to $(1 - \gamma)^2$. 
Observe also that we set $J = \log_2 d$. In microcanonical models,
$2^J$ is a fixed scale so that the
number of scattering coefficients does not increase with $d$.

%% file: experiments.tex
\section{Approximations of Stationary Processes}
\label{isingsec}

We study approximation of probability measures 
associated with stationary processes $X(u)$, $u \in \Z^\ell$, 
taking its values in $\R$,  with gradient descent 
microcanonical models calculated with shift-invariant energy vectors.
We first concentrate on Gaussian, Ising and point processes whose properties are well understood mathematically. We then consider the synthesis of image and audio textures from a single example.

\subsection{Microcanonical Approximation Errors}

This section analyzes the approximation
errors of a
stationary process $X$ of probability measure $\mu$
by a gradient descent microcanonical model of measure $\mu_n$.
The gradient descent is initialized with a Gaussian white measure
$\mu_0$ whose mean and variance are defined in (\ref{initansdfmsd}).
Since the energy $\Phi_d$ is shift-invariant,
Corollary \ref{cornsdf98we} proves that the gradient descent measures
$\mu_n$ are stationary.

\paragraph{Concentration}
Section \ref{microsec} explains that a microcanonical model 
is based on a concentration hypothesis, which needs to be verified.
For almost all realization $x$ of $X$, $\Phi_d (x)$ should remain in a
ball of radius $\epsilon_d$ which converges to zero when $d$ goes to 
$\infty$. We can verify this convergence in probability from a mean-square
convergence, by calculating the variance
\[
\overline \sigma^2_\mu = \E_\mu \Big(\| \Phi_d(x) - \E_\mu (\Phi_d(x)) \|^2\Big).
\]
The Markov inequality implies that if  
$\lim_{d \rightarrow \infty} \overline \sigma_\mu (\Phi_d (x))/\epsilon_d  = 0$  then 
\[
\lim_{d \rightarrow \infty} {\rm Prob} \Big(
\| \Phi_d (X) - \E_\mu ( \Phi_d (x))\| \leq \epsilon_d \Big) = 0~.
\]
This means that when $d$ increases there is a probability converging to $1$ 
that a realization of $X$ 
belongs to a microcanonical set computed
from a single realization $\bar x$ with $y = \Phi_d (\bar x)$:
\[
\Omega_{d,\epsilon_d} = 
\Big \{x \in \R^{\La_d}~:~ \|\Phi_d (x) - \Phi_d (\bar x) \| \leq \epsilon_d \Big\}~.
\]

In numerical calculations, we 
stop the gradient descents after a fixed number $n$ 
of iterations so that the resulting gradient descent measure is supported
in a microcanonical set $\Omega_{d,\epsilon}$ for $\epsilon$ small enough.  
If $\epsilon/\overline \sigma^2_\mu(\Phi_d(x)) \gg 1$ then 
nearly all realizations
of $X$ are included in $\Omega_{d,\epsilon}$. However, the microcanonical
set may become too large and hence include points which are not 
typical realizations of $X$. We thus typically wait to reach
a smaller $\epsilon$ width

Since $\Phi(x)$ is in a space of dimension $K$,
Corollary \ref{coroentropyepsilon}
proves that reducing $\epsilon$ by a factor $\gamma$
reduces the maximum 
entropy of the microcanonical model by a factor of the order
of $K \log \gamma$. In the extensive case,
this maximum entropy is proportional to $d$ so
the entropy reduction is negligible if 
$K\, |\log(\epsilon / \overline \sigma_\mu (\Phi_d(x))| \ll d$.
In all numerical calculations of this paper 
$\epsilon / \overline \sigma_\mu(\Phi_d(x))$ is of the order of $10^{-3}$. 
We evaluate the concentration of $\Phi_d(X)$
by computing the normalized variance
\begin{equation}
\label{sigmanormeq}
\sigma_\mu^2(\Phi_d) =  \frac{\E_\mu \Big(\| \Phi_d(x) - \E_\mu (\Phi_d(x)) \|^2\Big)} {\E_\mu \Big( \| \Phi_d(x) \|^2 \Big)}.
\end{equation}

\paragraph{Microcanonical gradient descent entropy}
Since $I = \R$, the gradient descent is initialized with a 
Gaussian white noise measure $\mu_0$ of variance 
$\sigma_0^2 = d^{-1} \|\bar x \|_2^2$.
The convergence of the gradient descent algorithm to the
microcanonical set is checked by verifying that for almost all 
Gaussian white realization $x_0$,
after a sufficient large number $n$ of gradient steps
\[ 
\|\Phi_d (x_n) - \Phi_d (\bar x) \| \leq \epsilon, 
\]
and hence $x_n \in \Omega_{d,\epsilon}$.
Convergence issues may be due to existence of local minima or because
the Hessian of $\Phi_d (x)$ is too ill-conditioned.
Let $\mu_{n}$ be the resulting 
microcanonical gradient descent measure.
If $\mu_n$ 
is supported in $\Omega_{d,\epsilon}$ then it has a smaller entropy than
the maximum entropy microcanonical measure, which is uniform in
$\Omega_{d,\epsilon}$. Theorem \ref{convergencetheo}
gives an upper bound on the reduction of entropy.

\paragraph{Model error}
Suppose that the restriction of $X$ to $\Lambda_a$ has a maximum entropy
measure $\mu$ associated to a 
known energy $\Phi^\mu_{d}(x)$.
This will be the case for Gaussian or Ising processes.
The typical sets where the realizations of $X$ are almost all 
concentrated are 
sets where $\|\Phi^\mu_{d}(x) - \E_\mu (\Phi^\mu_d(x))\|$
is sufficiently small. 
In this case we can
verify that the gradient descent microcanonical
measure $\mu_{n}$ computed with a model energy
$\Phi_d$ is also included in such a typical set with high probability.
This concentration property is satisfied if
the mean-square variation of the process energy
$\E_{\mu_{n}} (\| \Phi^\mu_d(x) - \E_{\mu} ( \Phi^\mu_d (x))\|^2)$
converges to $0$ when $d$ increases. 
This convergence
is evaluated by computing the concentration of $\Phi^\mu_d(x)$ around
$\E_{\mu} ( \Phi^\mu_d (x))$ for $\mu_n$:
\begin{equation}
\label{modelerreq}
e^2_{\mu_n}(\Phi_d) = \frac{\E_{\mu_{n}} (\| \Phi^\mu_d(x) - \E_{\mu} ( \Phi_d^\mu (x)) \|^2)}
{\E_{\mu_{n}} ( \|\Phi^\mu_d (x)\|^2) } .
\end{equation}

If $\mu_{n} = \mu$ then
$e^2_{\mu_n}(\Phi_d) =  \sigma^2_\mu(\Phi^\mu_d)$  but the reverse is not
true. It would be true only if the
microcanonical gradient descent measure had a maximum entropy, which
is not valid in general. On the other hand, if 
$e^2_{\mu_n}(\Phi_d) \gg  \sigma^2_\mu(\Phi^\mu_d)$ then it indicates 
that there is a model error.

\subsection{Approximation of Gaussian Processes}
\label{GaussApproSec}

We study approximations of stationary 
Gaussian random processes with gradient descent 
microcanonical models, defined with 
wavelet and scattering energy vectors. 

We consider a scalar quadratic potential $U x = |x \star h(u)|^2$
for $u \in \Z^2$. As in (\ref{posdf-sdf1}), we define a periodic 
filter $h_d (n) = \sum_{m \in \Z^2} h(n - m d^{1/2})$ 
over square images of $d$ pixels and an energy
\begin{equation}
\label{posdf-sdf}
\Phi^\mu_{d} (x) = d^{-1} 
\| x \star  h_{d}\|_2^2 = d^{-1} 
\sum_\om |\hat x(\om)|^2\, |\hat h_d (\om)|^2  .
\end{equation}
If $\inf_{\om} |\hat h (\om)| > 0$ then
we saw in (\ref{posdf-wefsdf}) that microcanonical and macrocanonical models converge to a Gaussian stationary process $\mu$ 
over $\Z^2$ whose power spectrum is
\begin{equation}
\label{posdf-wefsdf2}
P_\mu(\om) =  \beta^{-1}\, |\hat h (\om)|^{-2}~.
\end{equation}
In numerical experiments, we choose a discrete filter 
$h(n) = c\, e^{-|n|/\xi}$ with $\xi = 0.5$, 
whose Fourier transform satisfies for $\om \in [-\pi,\pi]^2$
\begin{equation}
\label{posdf-wefsdf289sdf}
|\hat h(\om)|^2 = c^2 \sum_{m \in \Z^2} {(\xi^2 + |\omega + 2 m \pi|^2)^{-2}} .
\end{equation}

\begin{figure}
\centering
\includegraphics[width=0.19\textwidth,trim=1.7in 0.7in 1.5in 0.5in, clip]{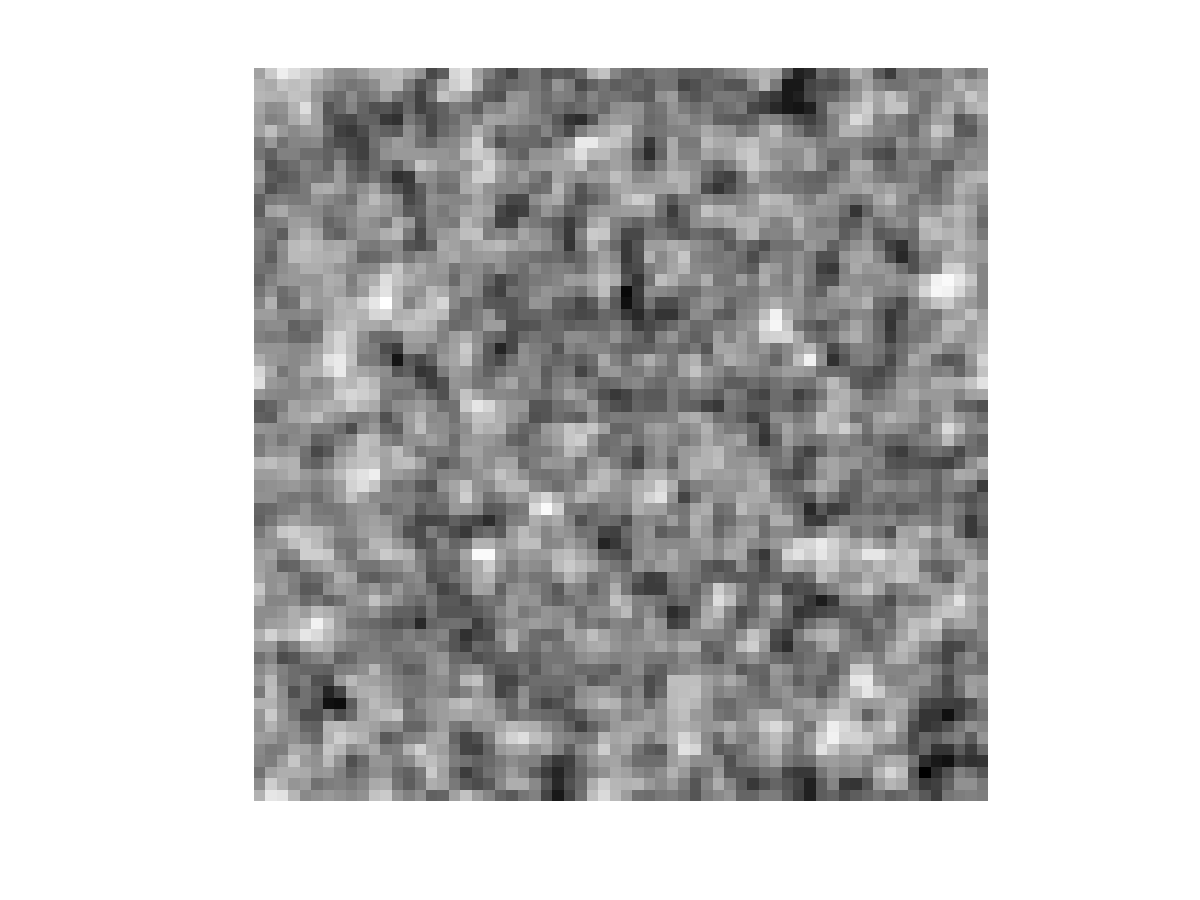}
\includegraphics[width=0.19\textwidth,trim=1.7in 0.7in 1.5in 0.5in, clip]{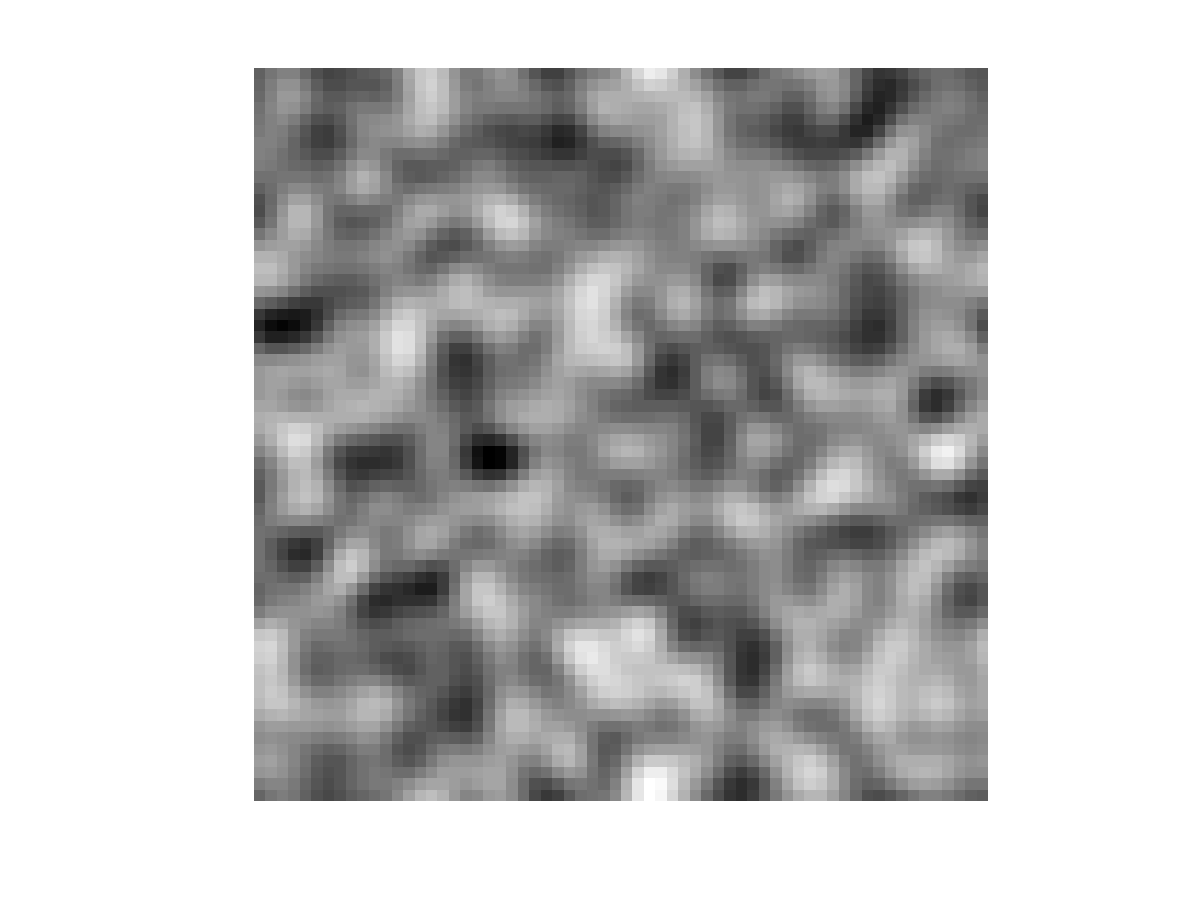}
\includegraphics[width=0.19\textwidth,trim=1.7in 0.7in 1.5in 0.5in, clip]{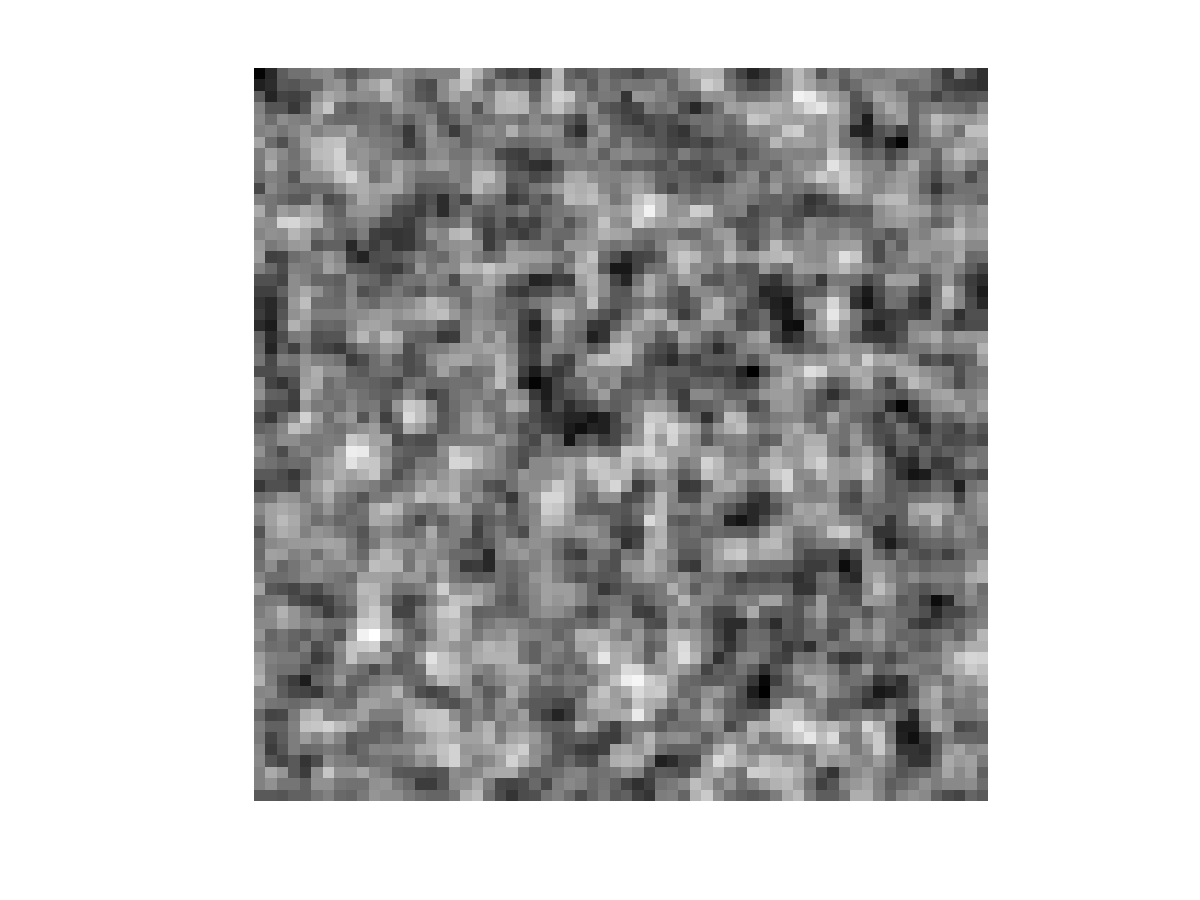}
\includegraphics[width=0.19\textwidth,trim=1.7in 0.7in 1.5in 0.5in, clip]{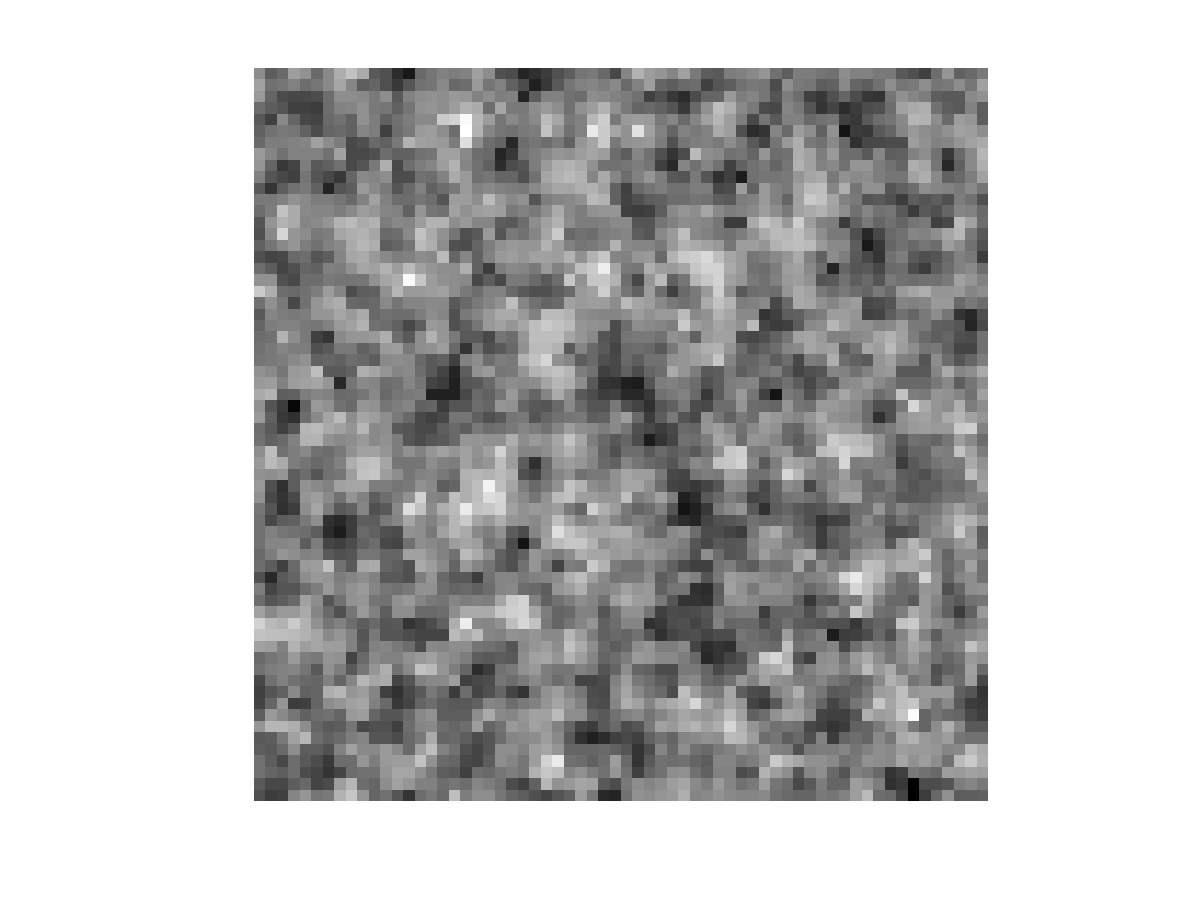}
\includegraphics[width=0.19\textwidth,trim=1.7in 0.7in 1.5in 0.5in, clip]{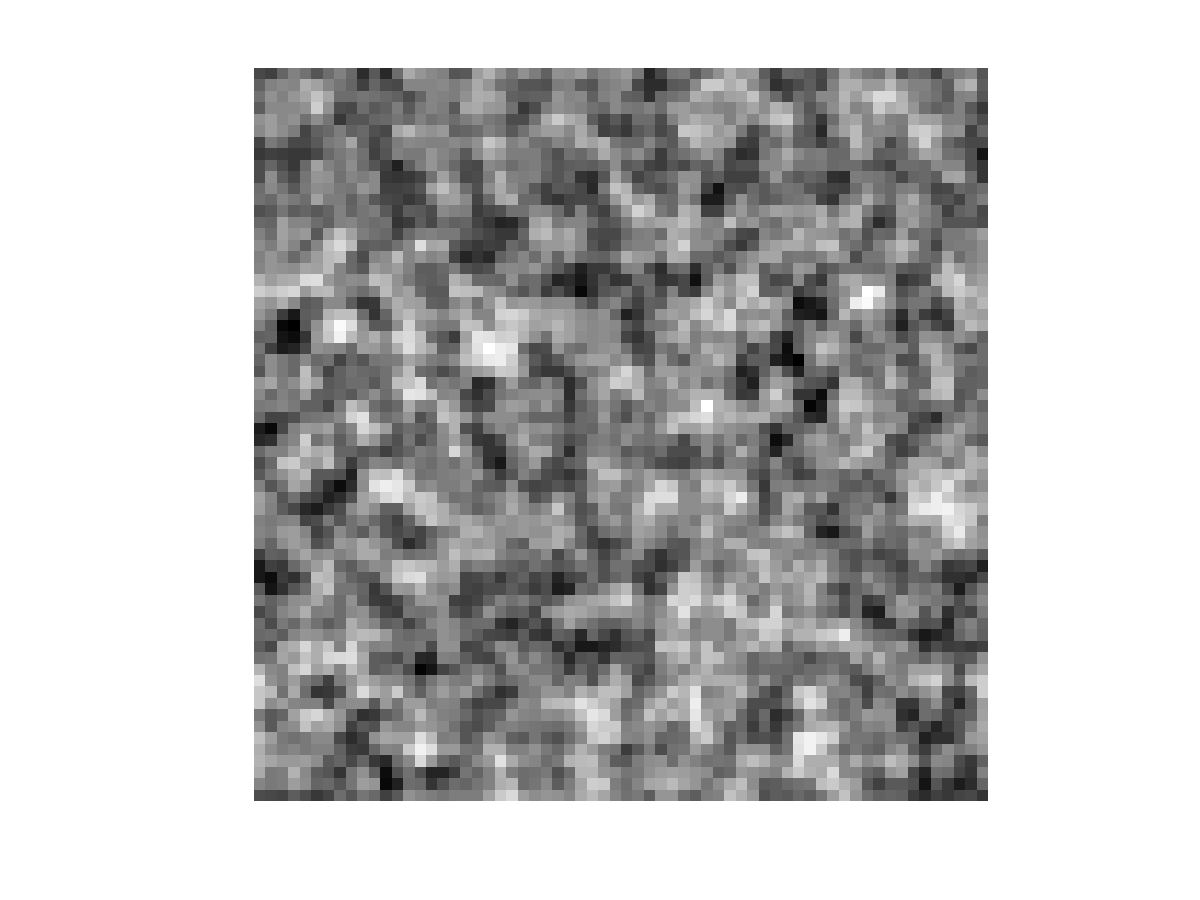}\\
(a)~~~~~~~~~~~~~~~~~~(b)~~~~~~~~~~~~~~~~~~(c)~~~~~~~~~~~~~~~~~~(d)~~~~~~~~~~~~~~~~~~~(e)
\caption{(a):
Realization of the  Gaussian process
(\ref{posdf-wefsdf2}). (b):
Realization of the microcanonical gradient descent computed
with $\Phi_d(x) = \Phi^\mu_d (x) = \|x \star h \|_2^2$.
(c): Realization computed with a vector $\Phi_d(x)$ 
of $\bf l^2$ wavelet norms. (d): $\Phi_d(x)$ is composed of
$\bf l^1$ wavelet norms. (e): $\Phi_d(x)$ is a scattering transform.}
\label{gaussian_fig}
\end{figure}

\begin{figure}
\centering
\includegraphics[width=0.19\textwidth,trim=1.7in 0.7in 1.5in 0.5in, clip]{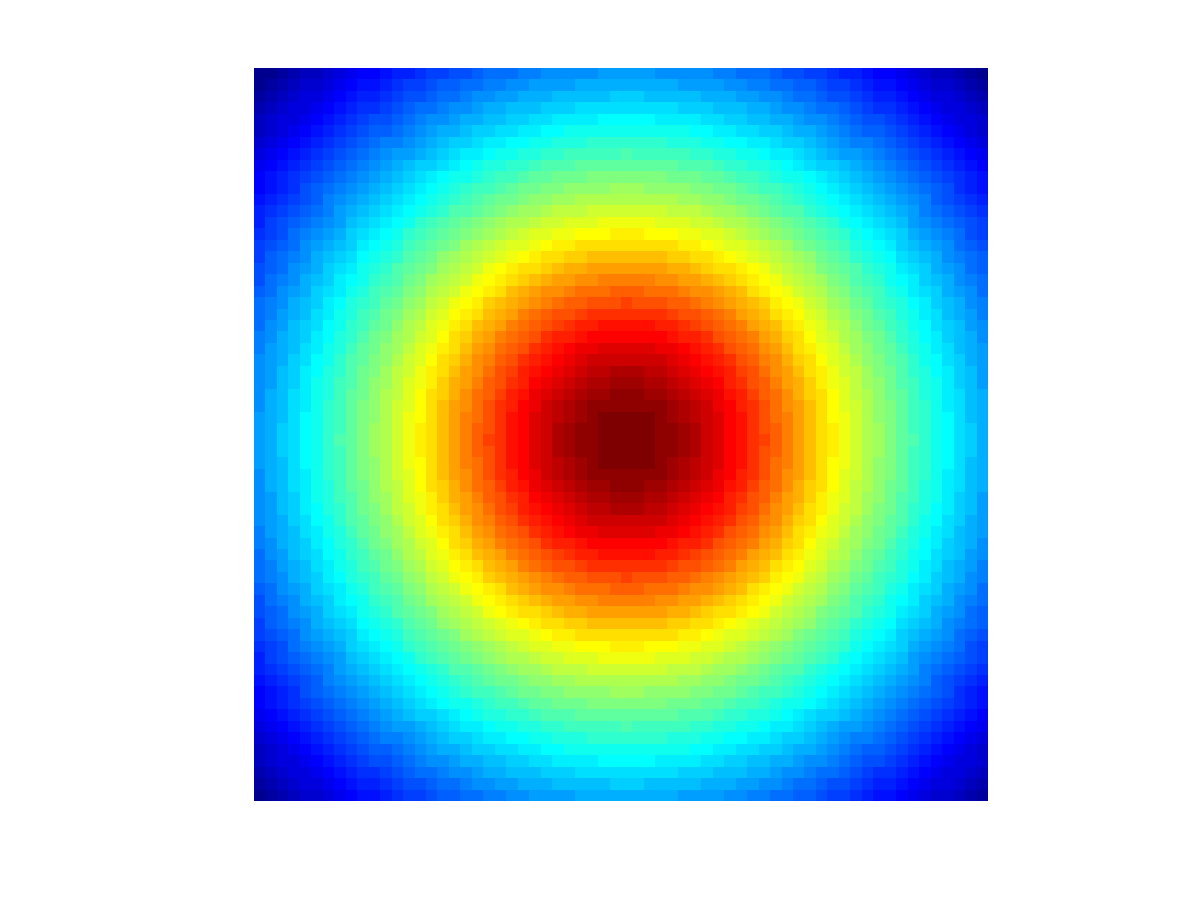}
\includegraphics[width=0.19\textwidth,trim=1.7in 0.7in 1.5in 0.5in, clip]{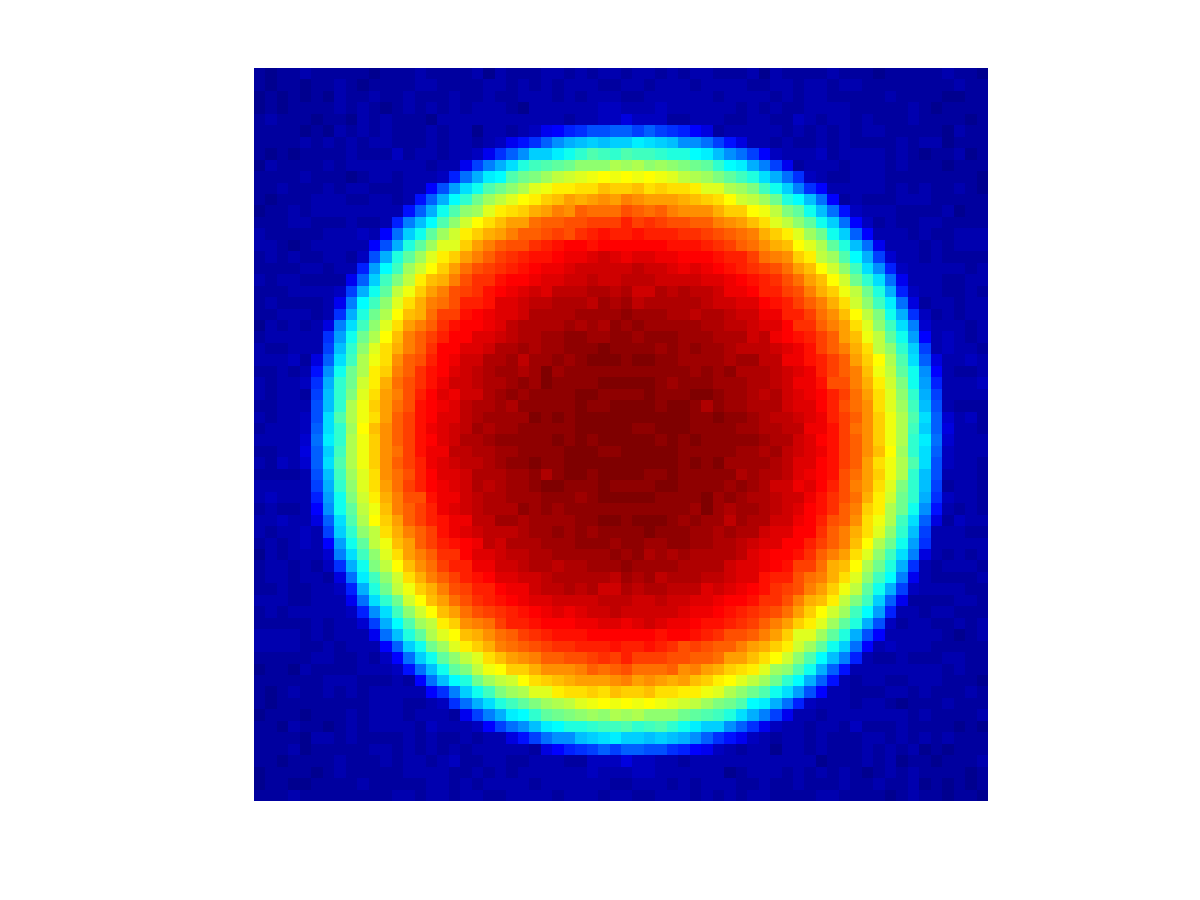}
\includegraphics[width=0.19\textwidth,trim=1.7in 0.7in 1.5in 0.5in, clip]{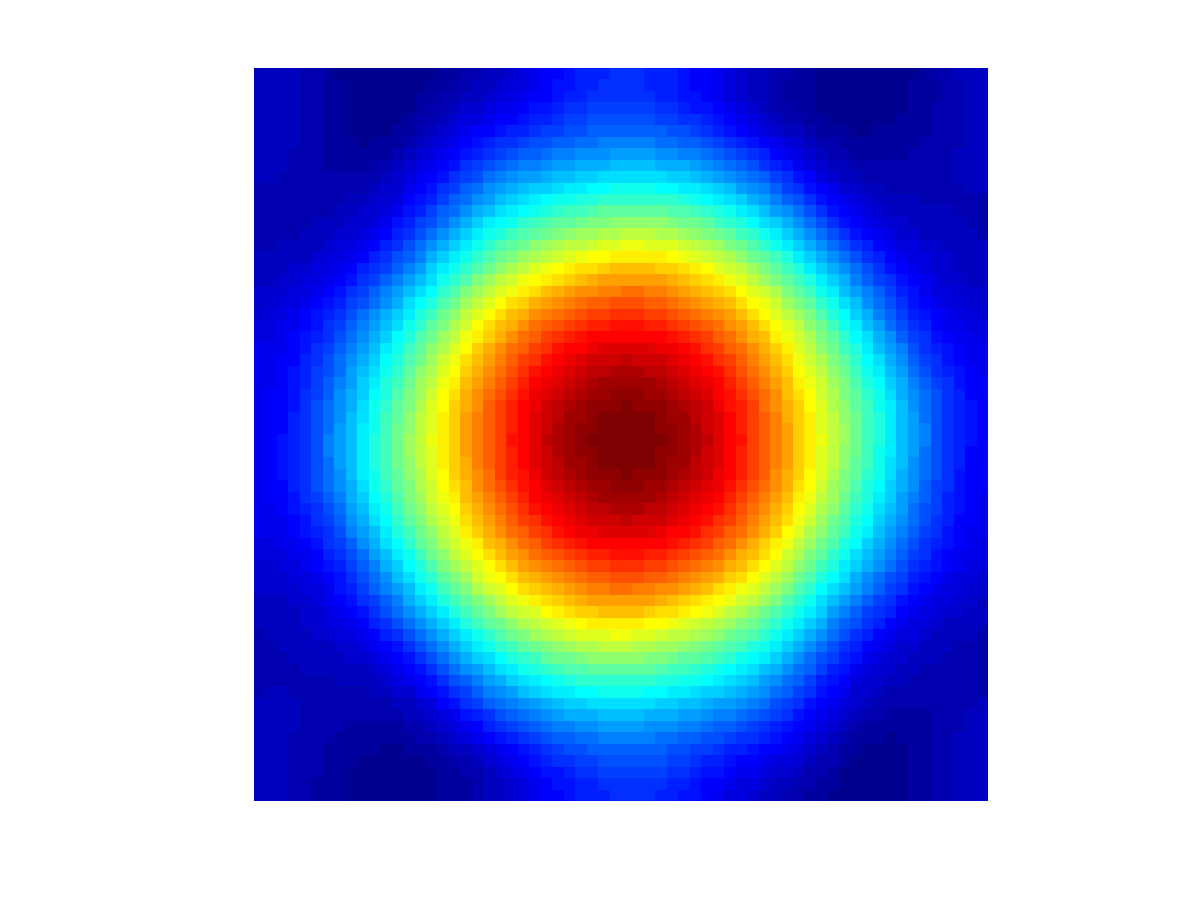}
\includegraphics[width=0.19\textwidth,trim=1.7in 0.7in 1.5in 0.5in, clip]{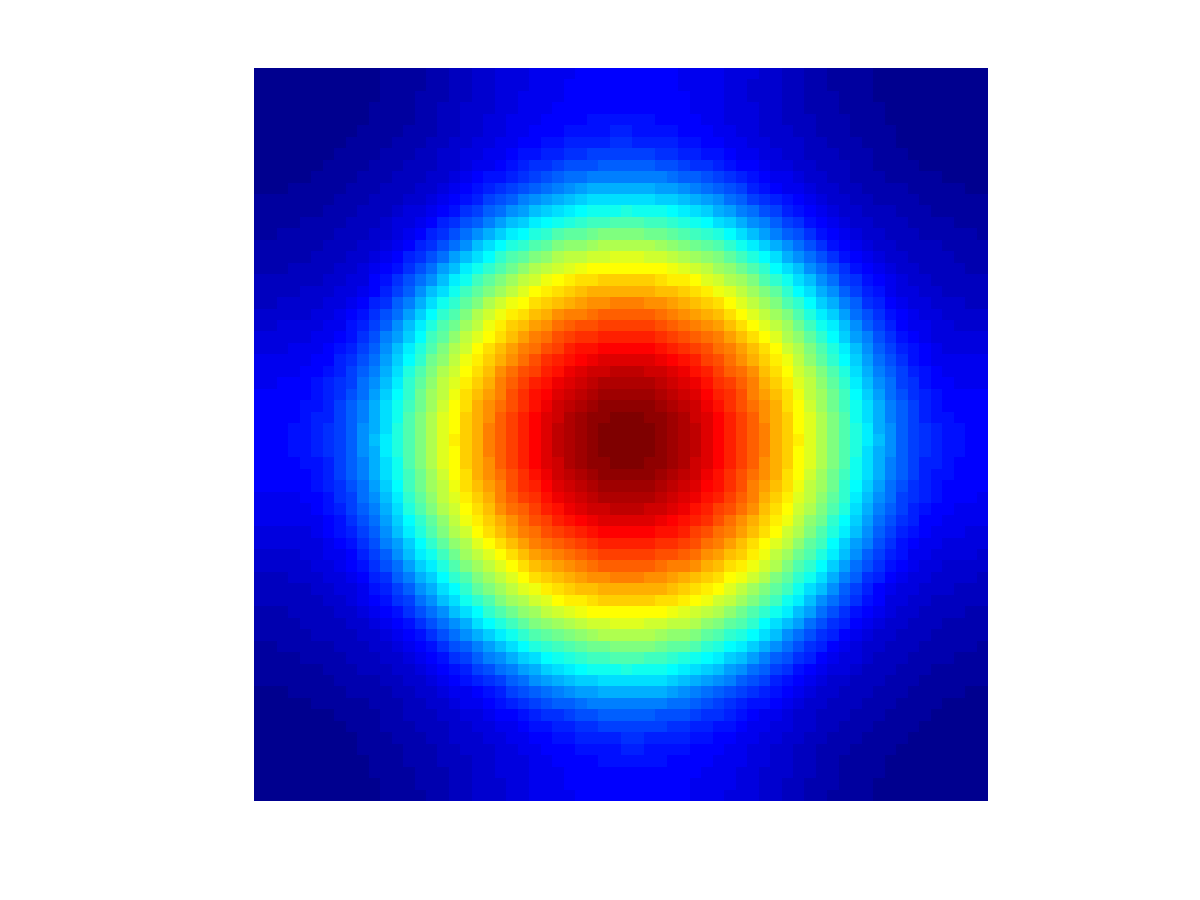}
\includegraphics[width=0.19\textwidth,trim=1.7in 0.7in 1.5in 0.5in, clip]{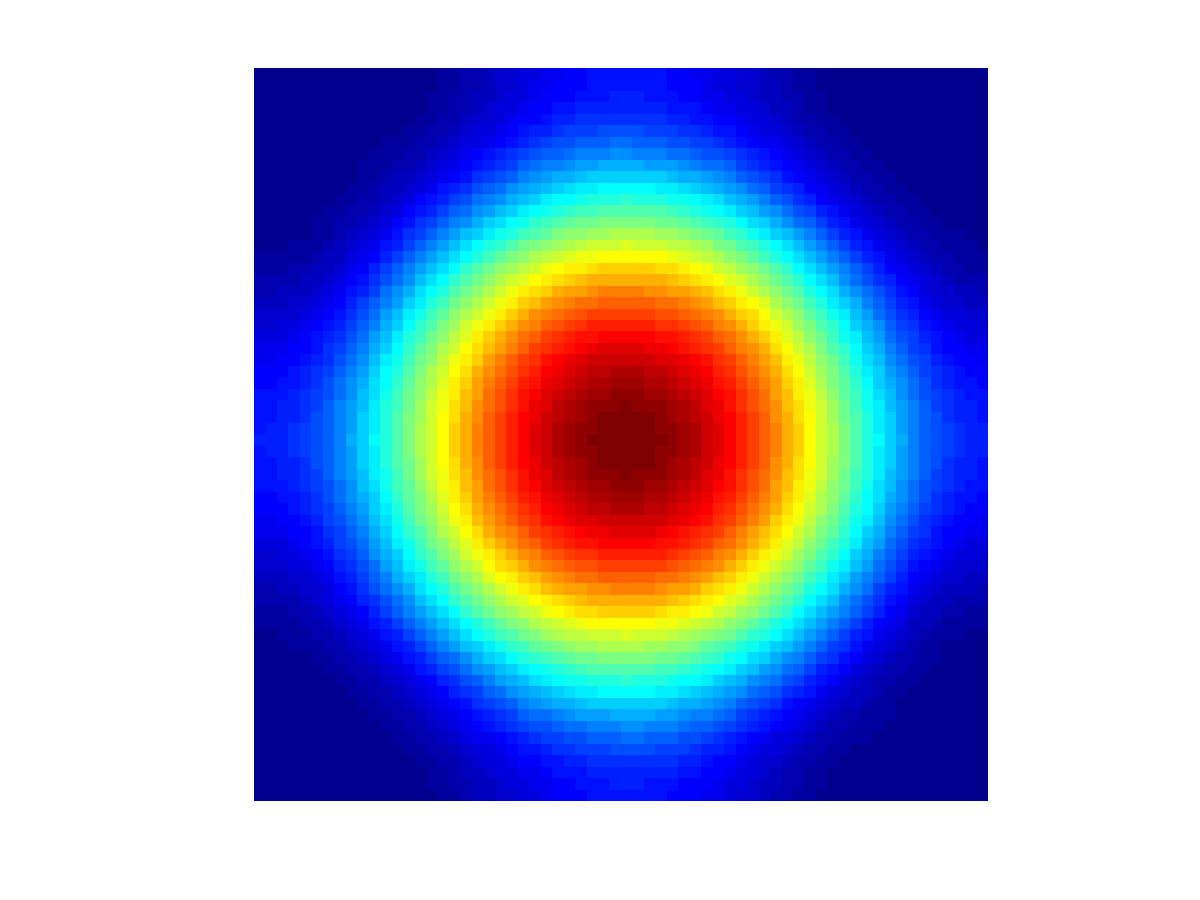}\\

(a)~~~~~~~~~~~~~~~~~~(b)~~~~~~~~~~~~~~~~~~(c)~~~~~~~~~~~~~~~~~~(d)~~~~~~~~~~~~~~~~~~~(e)
\caption{
(a): Power spectrum of the original Gaussian process. 
(b): Estimation of the
spectrum of a microcanonical gradient descent computed
with the energy vector $\Phi_d (x) = \phi_d(x) = \|x \star h \|_2^2$.
(c): The energy vector $\Phi_d(x)$ consists of
 $\bf l^2$ wavelet norms. (d): $\Phi_d(x)$ includes
$\bf l^1$ wavelet norms. (e): $\Phi_d(x)$ includes $\bf l^1$ scattering norms.}
\label{gauss-moment}
\end{figure}

Figure \ref{gaussian_fig}(a) shows realizations
of the Gaussian process of power spectrum $P_\mu(\om)$, 
which is nearly the same as the
maximum entropy microcanonical process computed
with the scalar energy $\Phi^\mu_d$.
Since $\Phi^\mu_d$ is an $\bf l^2$ energy,
Theorem \ref{convergencetheo2} proves that the gradient descent 
is not trapped
in a local minima and thus converges to a microcanonical set
of $\Phi^\mu_d$. This is verified by 
Table \ref{Gauss-Concentr} where
$e^2_{\mu_n} (\Phi^\mu_d) = \sigma^2_\mu(\Phi^\mu_d)$. 
However Figure \ref{gaussian_fig}(b)shows that realizations of the 
microcanonical gradient descent process are different from 
realizations of the original Gaussian process and hence of the
maximum entropy microcanonical process. 
Figure \ref{gauss-moment}(a,b) show that
the maximum entropy microcanonical process has a power spectrum 
which is different from the spectrum of 
the microcanonical gradient descent process.

Observe that the power spectrum in 
Figure \ref{gauss-moment}(a,b) 
are invariant by rotations in the Fourier plane. 
These rotations are orthogonal operators and they preserve the
stationary mean which corresponds to the Fourier transform value
at $\om = 0$. If $\widehat h_d (\om)$ is invariant 
by a rotation of $\om$ then (\ref{posdf-sdf}) implies that
$\Phi^\mu_d(x)$ is invariant to these rotations, and
Theorem \ref{invartheo} proves that
$\mu^{\rm min}_{d,\epsilon}$ and $\mu_n$ are invariant to these rotations.
This rotation invariance is not strictly valid at the highest frequencies 
because of the square grid sampling.

\begin{table}
\centering
\begin{tabular}{|c|c|c|c|c|}
\hline
 & $\Phi_d = \Phi^\mu_d$ & $\Phi_d = $Wavelet $\bf l^2$ & $\Phi_d =$ Wavelet $\bf l^1$  & $\Phi_d =$ Scattering \\
\hline
${\rm dim}(\Phi_d)$ & 1  & 40 & 40 & 114 \\ 
\hline
$ \sigma_\mu ^2 (\Phi_d )$ &5e-4&4e-3 & 4e-3&5e-3\\
$e_{\mu_n}^2(\Phi_d)$ &5e-4& 2e-2& 0.15 & 2e-2\\
\hline
\end{tabular}
\caption{
The first line gives the
dimension of each energy vectors $\Phi_d (x)$.
The next lines give the normalized variance
$ \sigma_\mu^2(\Phi_d)$ 
and the process energy concentration $e_{\mu_n}^2(\Phi_d)$, 
depending upon the microcanonical energy vector $\Phi_d$,
for the Gaussian process (\ref{posdf-wefsdf2}).}
\label{Gauss-Concentr}
\end{table}

\paragraph{Wavelet $\bf l^2$ norms}
Let us now compute
the gradient descent microcanonical measure $\mu_n$ 
with a wavelet $\bf l^2$
norm energy vector $\Phi_d$ in (\ref{posdf-sdf9}). We shall see that
it can provide good approximations of Gaussian processes.
The normalized variance 
$ \sigma^2(\Phi_d)$ in Table \ref{Gauss-Concentr} remains small which
indicates that this energy vector remains concentrated around its mean.
Figure \ref{gaussian_fig}(c) shows a realization of the resulting
microcanonical gradient descent model and  
Figure \ref{gauss-moment}(c) gives an estimation of 
the power spectrum of this stationary process. This power spectrum is now much closer to the original power spectrum.

To understand this, observe that
wavelet $\bf l^2$ norms specify the signal energy in the
different frequency bands covered by 
each band-pass wavelet filter $\hat \psi_{j,q}(\om)$:
\begin{equation}
\label{nsdfoi8ysdf}
\|x \star \psi_{j,q} \|^2 = \sum_\om |\hat x(\om)|^2\, |\hat \psi_{j,q} (\om)|^2  .
\end{equation}
The fact that the power spectrum 
remains nearly constant over the support
of each $\hat \psi_{j,q}$ is a consequence of Theorem \ref{invartheo}(iii). 
Indeed, suppose that $L x$ is a linear operator which
performs a permutation of the values of 
$\hat x(\om_1)$ and
$\hat x(\om_2)$, for two non-zero 
frequencies $\om_1$ and $\omega_2$ such that
$\hat \psi_{j,q} (\om_1) = \hat \psi_{j,q}(\om_2)$ for all $j,q$.
It is an orthogonal operator which preserves the mean (zero frequency) and 
it is a symmetry of $\Phi_d$.
Theorem \ref{invartheo}(iii) 
implies that the gradient
descent measure $\mu_n$ is also invariant to the action of $L$ and is thus
a stationary process whose power spectrum is the same at
$\om_1$ and $\om_2$. This property is approximately valid for
any frequencies $\om_1$ and $\om_2$ located near the center of the support
of each $\hat \psi_{j,q}$, where it remains nearly constant
and where all other $\hat \psi_{j',q'}$ nearly vanish. It implies that 
the spectrum of $\mu_n$ remains nearly constant in these frequency domain.

The energy concentration $e^2_{\mu_n}$ in Table \ref{Gauss-Concentr} 
is small although not as small as $\sigma^2_\mu (\Phi^\mu_d)$ which indicates the
presence of a bias. To reduce this bias we must reduce the support size
of each wavelet $\hat \psi_{j,q}$ where the spectrum must remain nearly
constant. Appropriate wavelet design can yield arbitrarily small errors when
$d$ increases. 

Besides having an appropriate power spectrum, these microcanonical gradient
descent models are also nearly Gaussian processes. 
This can be shown with a phase symmetry argument, which is explained without
a formal proof. 
The wavelet norms in (\ref{nsdfoi8ysdf}) and hence $\Phi_d (x)$
are invariant if we preserve $|\hat x (\om)|$ but change the
complex phase of $\hat x(\om)$ for $\om \neq 0$. 
Arbitrary rotations of the Fourier complex phases which transform
real signals into real signals are linear orthogonal operators which preserve the stationary mean. As a result, Theorem \ref{invartheo} proves 
that the gradient descent process is invariant to any such
Fourier phase rotation.
This means that Fourier transforms
of realizations of these microcanonical gradient descent processes 
have phases which are
independent and uniformly distributed. Given a fixed power spectrum,
a standard result based on 
the central limit theorem proves that stationary random processes 
with independent and uniformly distributed Fourier phases
converge to a Gaussian processes when
the dimension $d$ goes to $\infty$ \cite{PhaseMorel}. 
Under appropriate hypotheses, 
microcanonical gradient descent processes conditioned
by $\bf l^2$ wavelet norms will thus converge to Gaussian processes.

\paragraph{Wavelet $\bf l^1$ norms}
Maximum entropy models conditioned by
wavelet $\bf l^1$ norms capture sparsity with Laplacian distributions
but do not approximate Gaussian processes accurately. 
Figure \ref{gaussian_fig}(d) shows samples 
of the microcanonical gradient processes computed with a wavelet
$\bf l^1$ norm energy (\ref{posdf-sdf912}). 
The $\bf l^1$ norm constraints produce wavelet
coefficients which are more sparse than a true Gaussian process.
It creates images which are more piece-wise regular than in
Figure \ref{gaussian_fig}(c).
Errors are also visible 
in the resulting power spectrum shown in 
Figure \ref{gauss-moment}(d). Table \ref{Gauss-Concentr} 
shows that the resulting model error $e^2_{\mu_n}$ 
for the $\bf l^1$ norm wavelet vector is about $10$ times
larger than with the $\bf l^2$ wavelet energy vector.

\paragraph{Scattering energy}
The scattering energy vector (\ref{ScatRepnsdfm}) includes high order
multiscale terms which can nearly reproduce the $\bf l^2$ norms of wavelet
coefficients, as proved by Proposition \ref{enenprop}. 
Table \ref{Gauss-Concentr} gives
the normalized
variance $\sigma^2_\mu(\Phi_d (x))$ which shows that it concentrates 
nearly as well as wavelet $\bf l^2$ norm energy vectors, despite the fact
that it is much larger. 
Figure \ref{gaussian_fig}(e) shows a realization of the scattering
microcanonical gradient descent model and 
Figure \ref{gauss-moment}(e)
gives its power spectrum. It is nearly
as precise as the $\bf l^2$ norm microcanonical model and 
the model error $e^2_{\mu_n}$ in Table \ref{Gauss-Concentr} has about the
same amplitude.

\subsection{Ising Processes}
\label{Ising}

We consider a two-dimensional 
Ising process with no outside magnetization, over a
two-dimensional square lattice with periodic boundary conditions. 
We denote by $x(u)$ the spin values in $\{-1,1\}$. 
The Ising probability of a configuration $x$ is
\begin{equation}
\label{IsiHam}
p(x) = {\cal Z}^{-1} \exp\left(-\beta \,\phi_d (x) \right)~~\mbox{with}~~\phi_d (x) = d^{-1} \sum_{u \in \LaN} 
\sum_{u' \in {\cal N}_u} x(u) x(u') ,
\end{equation}
where ${\cal N}_u$ is the $4$ point neighborhood of $x(u)$ in the two-dimensional grid. The constant
$\beta= (k_B T)^{-1}$ is the inverse temperature 
scaled by the Boltzmann constant $k_B$.
In two dimension, the free energy can be exactly
computed with the method of Onsager \cite{onsager1944crystal}. 
It has a phase transition when $T$ 
reaches a critical value $T_c \approx 2.27$. 
We study the approximation of Ising for several values of the temperature. 

The complex behavior of Ising arises from the 
conjunction of the quadratic Hamiltonian with the binary constraint. This
binary condition may be replaced by a condition on a fourth order moment
to obtain the same critical behavior but we shall impose it here
through first and second order moments. 
For all $x \in \R^d$, 
one has $\|x \|_2 \leq \|x \|_1 \leq \sqrt{d} \|x \|_2$, 
and $\| x \|_1 = \sqrt{d}\, \|x \|_2$ if and only if $|x(u)|$ 
is constant. 
It follows that  
$$
 \forall u~,~ x(u) = \pm 1 
\Leftrightarrow \|x \|_1 = \| x\|_2^2 = d~.$$
We can thus impose that $x$ is binary by
adding $d^{-1} \|x \|_2^2$ and $d^{-1} \|x \|_1$ 
into the energy vector.
The resulting 
microcanonical interaction energy 
for $x \in  \R^\LaN$ is
\begin{equation}
\label{isdfnsdf-}
\Phi^\mu_d(x)=\{d^{-1} \| x\|_2^2\,,\, d^{-1} \| x\|_1\,,\, \phi_d (x)\} .
\end{equation}
If we remove the ${\bf l}^1$ term, this energy is quadratic and the
maximum entropy model is therefore a stationary Gaussian process.

The Ising model has a phase transition at the critical temperature $T_c \approx 2.27$, 
from an `ordered' to a `disordered' state. The spin spatial correlation 
exhibits a characteristic scale $\xi(T)$ for $T > T_c$ and
$\E{\{X(u)X(u+r)\}} \simeq e^{-|r|/\xi(T)} $ \cite{kopietz2010mean},
with $\xi(T_c) = 0$.  
The correlation is self-similar at $T=T_c$
and $\E{\{X(u)X(u+r)\}} \simeq |r|^{-1/2}$.

Figure \ref{Ising-Concentr}(a) gives two realizations
of Ising for a large temperature (bottom) and a temperature just
above the critical temperature (top). 
Figure \ref{Ising-Concentr}(b) shows realizations of
the microcanonical gradient descent process computed with the
Ising energy vector $\Phi^\mu_d$. 
The first column of Table \ref{Ising-Concentr} shows that
$e^2_{\mu_n}(\Phi^\mu_d) \gg  \sigma^2(\mu (\Phi^\mu_d)$ which means
that the microcanonical gradient descent does not converge 
to a microcanonical set for $\epsilon$ small. 
Near the critical temperature, 
the gradient descent microcanonical model is unable to 
recover low-frequency long-range structures which appear in Ising.
This is due to a well-known instability near criticality. 

\begin{table}
\centering
\begin{tabular}{|c|c|c|c|c|}
\hline
 & $\Phi_d = \Phi^\mu_d$ & $\Phi_d = $Wavelet $\bf l^2$ & $\Phi^d =$ Wavelet $\bf l^1$  & $\Phi_d =$ Scattering \\
\hline
${\rm dim}(\Phi_d(x))$ & 3  &42 &42 &116\\
\hline
$ \sigma_\mu ^2 (\Phi_d)$ $T = 2.2$ &6e-6&3e-4&  4e-4&6e-4\\
$e_{\mu_n}^2(\Phi_d) $ $T = 2.2$ &2e-2 &7e-2  & 5e-2&9e-3\\
\hline
$ \sigma_\mu ^2 (\Phi_d )$ $T = 3$ &3e-6&2e-5&4e-5&4e-5\\
$ e_{\mu_n}^2(\Phi_d)$ $T = 3$ &7e-3&4e-2&5e-2&5e-3\\
\hline
\end{tabular}
\caption{
The first line gives the
dimension of each energy vectors $\Phi_d (x)$.
We consider two Ising processes (\ref{IsiHam}),
computed near the critical temperature $T = 2.2$ and at a larger
temperature $T = 3$. The table 
gives the normalized variance
$ \sigma_\mu^2(\Phi_d)$ 
and the Ising energy concentration $e_{\mu_n}^2(\Phi_d)$, 
for different $\Phi_d (x)$.}
\label{Ising-Concentr}
\end{table}


\begin{figure}
\centering
\includegraphics[width=0.19\textwidth]{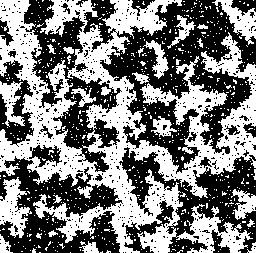}
\includegraphics[width=0.19\textwidth,trim=1.7in 0.7in 1.5in 0.5in, clip]{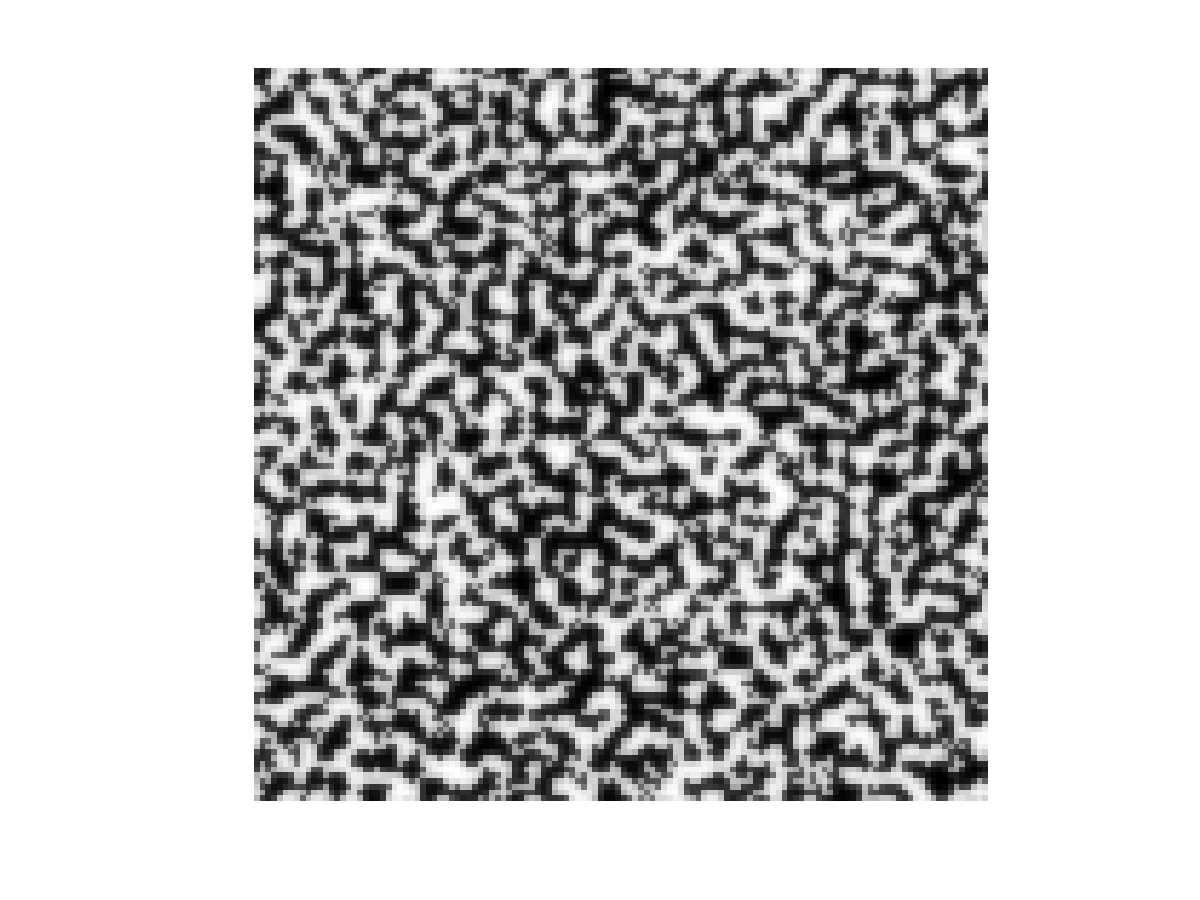}
\includegraphics[width=0.19\textwidth]{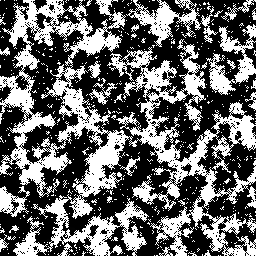}
\includegraphics[width=0.19\textwidth]{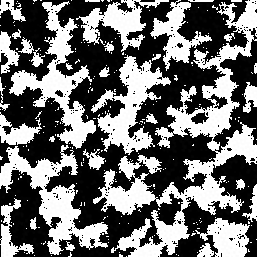}
\includegraphics[width=0.19\textwidth]{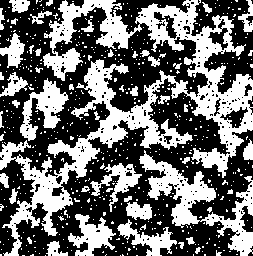}\\
\includegraphics[width=0.19\textwidth]{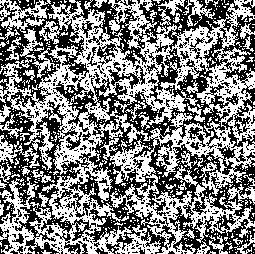}
\includegraphics[width=0.19\textwidth,trim=1.7in 0.7in 1.5in 0.5in, clip]{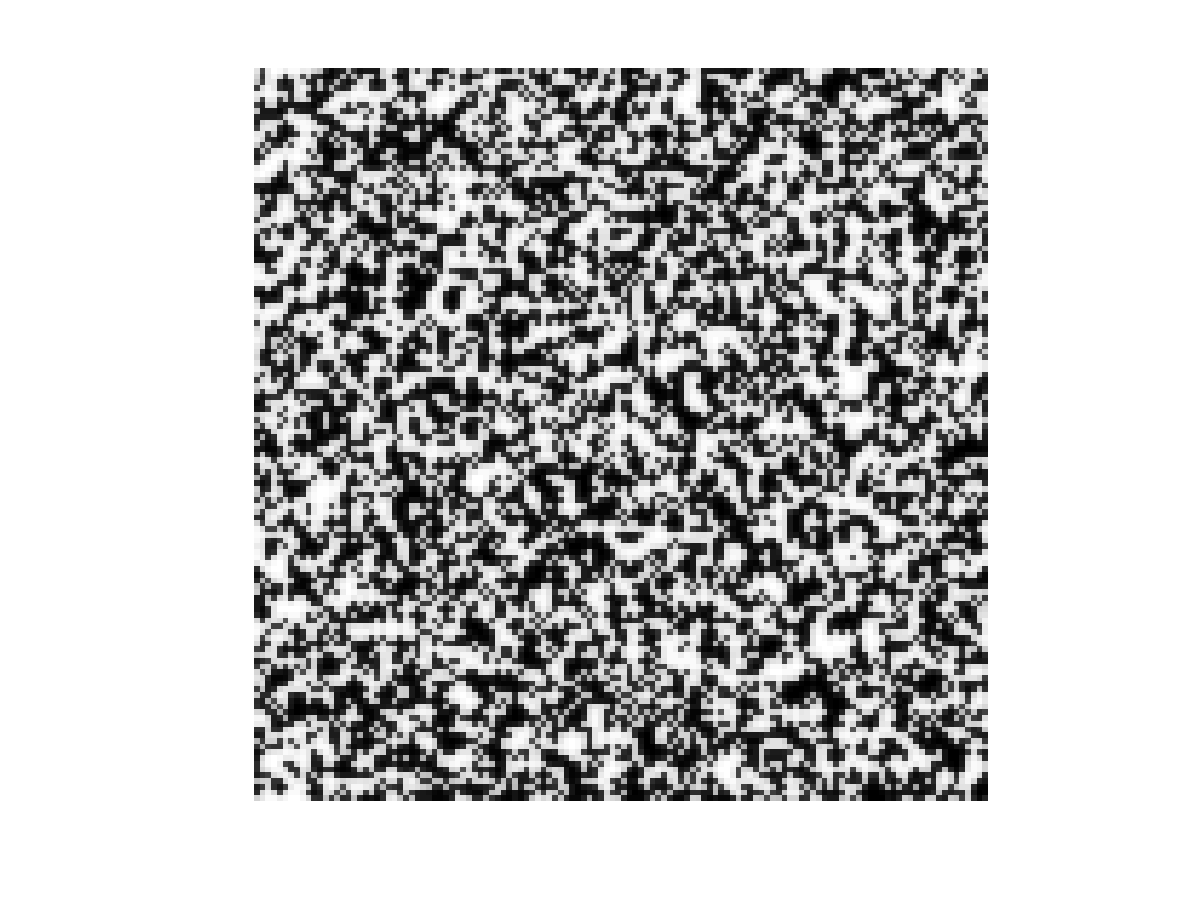}
\includegraphics[width=0.19\textwidth]{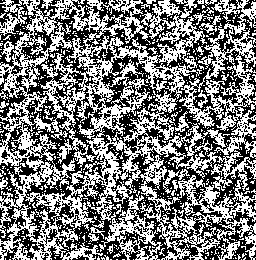}
\includegraphics[width=0.19\textwidth]{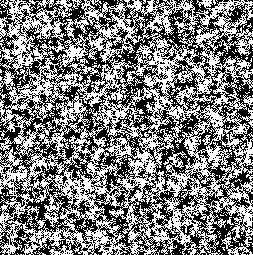}
\includegraphics[width=0.19\textwidth]{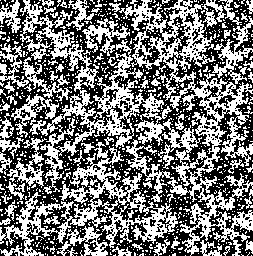}\\
(a)~~~~~~~~~~~~~~~~~~(b)~~~~~~~~~~~~~~~~~~(c)~~~~~~~~~~~~~~~~~~(d)~~~~~~~~~~~~~~~~~~~(e)
\caption{(a): Realizations of an Ising process 
near the critical temperature $T=2.2$ (top), and for $T=3$ (bottom). (b):
Realizations computed with the microcanonical
gradient descent with $\Phi_d = \Phi^\mu_d$.
(c): $\Phi_d (x)$ includes $\bf l^2$ wavelet norms. (d): $\Phi_d (x)$ includes $\bf l^1$ wavelet norms. (e): $\Phi_d (x)$ includes $\bf l^1$ scattering norms.}
\label{ising_simple_fig}
\end{figure}


\paragraph{Renormalization and wavelets}
As in Wilson renormalization group,
wavelets separate the frequency components of $x$ into dyadic 
frequency annulus. 
Relations between wavelets and renormalization group decompositions 
were studied by Battle \cite{Battle}. 
In the following, we give a qualitative argument to explain how
to approximate the Ising potential with wavelet norms.

Since $x(u) \in \{-1,1\}$, for an integer $p$
\[ 
x(u) x(u') = 1 - 2^{-1}\,{|x(u) - x(u')|^p} 
\] 
so we can rewrite the Ising energy $\phi_d (x) = d^{-1} \sum_{u \in \LaN} 
\sum_{u' \in {\cal N}_u} x(u) x(u')$ satisfies
\begin{equation}
\label{IsiHam21}
d-  \phi_d(x) = 
2^{-1} \sum_{u \in \LaN} \sum_{u' \in {\cal N}_u} |x(u) - x(u')|^p 
= \|\Delta_1 x\|_p^p +\|\Delta_2 x\|_p^p ,
\end{equation}
with $\Delta_1 x (u_1,u_2) = x(u_1,u_2) - x(u_1,u_2-1)$ and
$\Delta_2 x (u_1,u_2) = x(u_1,u_2) - x(u_1-1,u_2)$.

The equivalence of $\bf l^p$ norms of increments and $\bf l^p$ norms of
wavelet coefficients is established in \cite{Meyer}. For any $p > 1$
there exists $A_p > 0$ and $B_p > 0$ so that for any 
$x \in {\bf l^2} (\Z^2)$
\begin{equation}
\label{IsiHam212}
A_p \sum_{j,q} 2^{- j p} \|x \star \psi_{j,q} \|_p^p \leq
\| \Delta_1 x \|_p^p +\|\Delta_2 x\|_p^p \leq
B_p \sum_{j,q} 2^{-j p} \|x \star \psi_{j,q} \|_p^p .
\end{equation}
For $p = 1$ the upper-bound remains valid but to get a lower-bound
we must replace the sum over $j,q$ by a sup operator. However, we conjecture
that there exists $A_1$ which verifies the lower bound for $p = 1$ 
when the values of $x(u)$ are restricted to $\{-1,1\}$.
With equations (\ref{IsiHam21}) and (\ref{IsiHam212}) one can approximate
the Ising energy $\phi_d (x)$
with discrete wavelet ${\bf l}^p$ norms computed at all scales
$2^j \leq 2^J = d$. We limit the maximum scale $2^J$
independently of $d$, which is set to be the largest correlation length
of the process.

As in Section \ref{Ising}, 
we capture the fact that $x(u) \in \{-1, 1\}$ by including a condition on
$d^{-1} \| x\|_1$ and $d^{-1}\| x\|_2^2$.
The resulting energy vector
for $p = 1$ and $p = 2$ is
\begin{equation}
\label{posdf-sdf1nsddf}
\Phi_{d} (x) =  \Big\{ d^{-1} \|x\|_2^2\,,d^{-1} \|x\|_1\,,\,
d^{-1}\, \| x \star \psi_{j,q}\|_p^p \Big\}_{j \leq J,q\leq Q} .
\end{equation}
Table \ref{Ising-Concentr} shows the normalized variance
$ \sigma^2 (\Phi_d)$ is smaller at high temperature than
near critical temperature but the separation of scale still provides
a high concentration of $\Phi_d(x)$ for an Ising process,
close to the critical temperature.
Figure \ref{ising_simple_fig}(c,d) show
realizations of a microcanonical gradient descent Ising model computed
with the wavelet energy (\ref{posdf-sdf1nsddf}) for $p = 1$ and $p = 2$.
Near critical temperature, the microcanonical gradient descent still converges where as it was not the case when the energy was calculated directly
with the Ising Hamiltonian energy $\phi_d (x)$ in 
Figure \ref{ising_simple_fig}(b). 
The scale
separation avoids having an ill-conditioned gradient descent. 
The Ising approximation with an $\bf l^2$ energy
vector for $p=2$ amounts to compute a Gaussian approximation of Ising, which
is not precise, when we are 
close to the critical temperature \cite{kopietz2010mean}.
One can indeed visualize important
differences with the statistical distribution of original Ising 
in Figure \ref{ising_simple_fig}(a). 
Table \ref{Ising-Concentr} shows that the model error $e^2_{\mu_n}$ is 
smaller at higher temperature. 

The Ising approximation with an $\bf l^1$ energy vector has about the same
error as the model computed with 
an $\bf l^2$ energy vector. 
Near the critical temperature, the microcanonical models obtained with
$\bf l^1$ wavelets norms shown in 
Figure \ref{ising_simple_fig}(d) are more piecewise regular than the
ones in Figure \ref{ising_simple_fig}(c) obtained with
wavelet $\bf l^2$ norms. This is due to the wavelet coefficient sparsity
imposed by these $\bf l^1$ norms. 

\paragraph{Scattering energy}
A scattering energy vector is defined for Ising process, 
by complementing 
the scattering energy vector (\ref{ScatRepnsdfm}) with $\bf l^1$ and
$\bf l^2$ norms of $x$ in order to impose that $x(u)$ takes binary values: 
\begin{equation}
\label{ScatRepnsdfmIsing}
\Phi_\NN (x) = \Big\{ d^{-1} \|x\|_2^2\,,\,d^{-1} \|x\|_1\,,\,d^{-1}\sum_{u \in \LaN} x(u)\,,\,d^{-1}\, 
\|  x \star  \psi_{j,q} \|_1 \,,\,
d^{-1}\, \| |x \star \psi_{j,q}| \star \psi_{j',q'} \|_1 \Big\}_{j,j' \leq J , q,q' \leq Q}~.
\end{equation}
Table \ref{Ising-Concentr}
shows that the normalized variance of the scattering energy
is about twice larger than for $\bf l^2$ wavelet energy vectors. 
Figure \ref{ising_simple_fig}(e) shows 
realizations of microcanonical 
gradient descent models computed with this scattering energy vector. 
They are visually difficult to 
distinguish from realization of the original
Ising process above the critical temperature and close to the critical
temperature. 
Table \ref{Ising-Concentr} shows 
that the model error $e^2_{\mu_n}$ is about $10$ times smaller 
than with $\bf l^2$ or $\bf l^1$ wavelet energies. 

These numerical experiment seem to indicate that 
scattering microcanonical gradient descents can provide accurate model
of Ising even close to critical temperature. However, this needs to be
sustained by a better mathematical of these approximations, by
analyzing the preservation of symmetries.

\subsection{Point Processes}

Point processes provide powerful models of stochastic geometry, with 
applications in many areas of astrophysics, neuroscience, finance and computer vision. Realizations of point processes have a support reduced to
isolated points. We first show that this sparsity
can be captured by wavelet $\bf l^1$ norms. We then study
approximations of point processes and shot noises with microcanonical models
defined by scattering coefficients.

\paragraph{Support from wavelet $\bf l^1$ norms}
We prove that wavelet $\bf l^1$ norms capture important geometric properties
of the support of point processes.
Young's inequality implies that
\[
\|x \star \psi_{j,q} \|_1 \leq \|x\|_1\, \| \psi_{j,q} \|_1 .
\]
If $x$ is a Dirac in $\LaN$ then 
this inequality is an equality. Conversely, the following theorem, proved in Appendix \ref{sparsewavetheo}
proves that if this inequality is an equality then $x$ is a sum of
Diracs, with conditions on their distances. 
The inner product and norm of $v$ and $v'$ in $\R^\ell$ is written
$v.v'$ and $\|v\|$.

We suppose that wavelets are defined from 
a mother wavelet $\psi(u)$ which is continuous with
$\psi(0) \neq 0$. We suppose that 
$\psi(u) = |\psi(u)|\, e^{i\, \varphi(\xi.u)}$ where
$\xi \in \R^\ell$ and the complex phase $\varphi$ is a bi-Lipschitz function.
We may choose linear phase $\varphi (\xi.u) = \xi.u$. 
This wavelet is rotated and dilated
$\psi_{j,q} (u) = 2^{-j \ell} \psi (2^{-j} r_q^{-1} u)$,
where the $r_q$ are $Q \geq \ell$ different rotations in $\R^\ell$.  
The following theorem applies to these wavelets.

\begin{theorem}
\label{theorexmsparse} 
(i) If $\|x \star \psi_{j,q} \|_1 = \|x \|_1 \,\|\psi_{j,q} \|_1$ then
$x$ is non-zero at $u$ and $u'$ only if
$\xi_q .(u-u') = 0$ with $\xi_q = r_q \xi$ or if $|\xi_q . (u - u')| \geq  C\,2^{j}~$,
where $C > 0 $ does not depend on $x$. \\
(ii) Suppose that $\psi$ has a compact support, and that $x$ has a support which
is a union of isolated points with distances larger than $\Delta$. If
$x'$ satisfies
\begin{equation}
\label{corcond}
\forall q \leq Q\,,\, \forall j \leq \log_2 \Delta~,~\|x' \star \psi_{j,q} \|_1 = \|x \star \psi_{j,q} \|_1 ~~\mbox{and}~~\|x'\|_1 = \|x\|_1~
\end{equation}
then the support of 
$x'$ is a set of isolated points of distances larger than $C\,\Delta$,
where $C > 0$ does not depend on $x$.
\end{theorem}

In dimension $\ell = 2$,
property (i) of Theorem \ref{theorexmsparse}
proves that the support of $x$ is included in
straight lines perpendicular to $\xi_q$, 
whose distances are larger than $C\,2^j$. If this is valid for several $q$
then the support is included over intersections of non-parallel lines
and hence reduced to isolated points, as proved by property
(ii).  

If $x$ is a realization of a point process, its support is a union of
isolated points whose minimum distance depends the point process
distribution. If we construct an $\epsilon = 0$ microcanonical model 
with wavelet $\bf l^1$ norms then  property (ii) proves that
all realizations of this microcanonical model will also be a
point process with a similar separation between points.

\paragraph{Microcanonical models of point processes}
We study microcanonical models of point processes with
wavelet $\bf l^1$ norms and scattering coefficients.
A point process $N$ on $\R^\dd$ is a measure whose support is 
composed of isolated 
points. Second-order point processes \cite{bremaud2003power} are those satisfying $\E[ N(C)^2 ] < \infty$ for 
all bounded Borel sets $C \subset \R^\dd$. If $N$ is a stationary, second-order point process then 
one can define its associated Bartlett spectral measure \cite{bremaud2003power} $P_N$, which generalizes
the power spectrum of second-order stationary processes.

Given a non-negative stationary process $\lambda(t)$, $t \in \R^\dd$, a Cox process $N$ is defined as a Poisson process conditional on $\lambda$ with intensity $\lambda(t)$. Important geometric information of $N$ is 
captured by its Bartlett power spectrum, which satisfies $P_N(d\om) = P_\lambda(d\om) + \E(\lambda)\, \delta(d\om)$ \cite{bremaud2003power}. 
Shot noises are classes of random processes defined by convolutions
of point processes with a filter $h(t)$
$$X(t) = N \star h(t)~.$$
The filter $h(t)$ can be interpreted as a pattern which is randomly
translated at point locations and added. It may also be the transfer
function of a detector measuring the point-process. 
In this case, the power spectrum of $X$ is $P(d\om) = P_N(d\om) \,| \hat{h}(\om) |^2$, which
mixes the geometric information of $N$ with the profile of the filter $h$. 
We will show that they
can be disentangled by a wavelet scattering transform. 

The loss of information in the power spectrum is due to the fact that 
it does not measure scale interactions. 
When there is a scale separation between $N$ and $h$, ie  
\begin{equation}
\label{ppeq1}
\E({\lambda})^2 \gg \int u^2 |h(u)|^2 du
\end{equation}
then for sufficiently small scales $2^j$, one can verify \cite{bruna2015intermittent} that
\begin{equation}
\label{ppeq2}
| X \star \psi_{j,q} | = | N \star (\psi_{j,q} \star h) | \approx N \star | \psi_{j,q} \star h|~ 
\end{equation}
with high probability, due to the fact that the events in $N$ rarely interact at spatial scales
$j$ such that $2^j \ll \E({\lambda})$. 
From this approximation, 
it follows that for sufficiently large scale gap $j' \gg j$, we have 
\begin{equation}
\label{ppeq3}
|| X \star \psi_{j,q} | \star \psi_{j',q'}| \approx C_{j,q} | N \star \psi_{j',q'} |~,
\end{equation}
since $| \psi_{j,q} \star h| \star \psi_{j',q'} \approx C_{j,q} \delta \star  \psi_{j',q'}$. 
Second order scattering coefficients, indexed with pairs $(j,q,j',q')$, thus provide measurements that convey spectral information about the point process $N$ as $(j',q')$ varies, disentangled from the spectral information of $h$.

\begin{figure}
\centering
\includegraphics[width=0.95\textwidth]{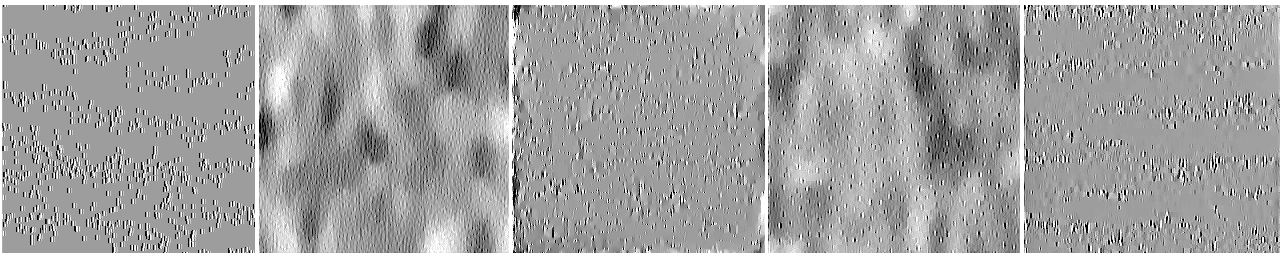} \\
(a)~~~~~~~~~~~~~~~~~~(b)~~~~~~~~~~~~~~~~~~(c)~~~~~~~~~~~~~~~~~~(d)~~~~~~~~~~~~~~~~~~~(e)
\caption{(a): Realization of a shot noise computed with a Cox process. (b,c): Realizations of a gradient descent process, 
computed with an energy $\Phi_\NN$ including wavelet $\bf l^1$ norms of maximum scale respectively $2^J=8$ and $2^J=64$. 
(d,e): Same computed with an energy $\Phi_\NN$ including scattering $\bf l^1$ norms of maximum scale respectively $2^J=8$ and $2^J=64$.
}
\label{pp_fig}
\end{figure}

We illustrate this phenomena by considering a two-dimensional Cox point process $N(u)$, whose rate $\lambda(u)$ is a stationary Gaussian process 
whose power spectrum is concentrated in the low-frequencies, and with an integral scale of $100$ pixels. 
 This Cox process is convolved with a pattern $h(u)$ with zero mean and small spatial support of $5$ pixels. 
We build microcanonical models with 
energy vectors $\Phi_d (x)$ defined by
wavelet $\bf l^1$ norms or scattering coefficients, computed up to a maximum
scale $2^J$.
For the shot noise measure $\mu$ shown in Figure \ref{pp_fig}(a), 
Table \ref{pointprocess_variance} gives the normalized variances 
$ \sigma^2_\mu = 
{{\E_\mu}( \| \Phi_d(x) -  \E (\Phi_d(x)) \|^2 ) } /
{\|  {\E}_\mu( \Phi_d(x)) \|^2 }$
as a function of the maximum scale $2^J$.
Although the size of scattering vectors for large $J$ becomes relatively
large, the normalized variance remains small which proves that these 
energy vectors remain concentrated around their mean, for images
of size $d = 256^2$. We can thus define
microcanonical models from an energy vector $\Phi_d (\bar x)$ calculated
from the realization $\bar x$ shown in  Figure \ref{pp_fig}(a).

\begin{table}
\centering
\begin{tabular}{|c|cc|cc|cc|}
\hline
 & ~~~$J=2$ & & ~~~$J=4$ & & ~~~$J=6$ &  \\
\hline
 & $\sigma_\mu^2(\Phi_d)$ & $\dim(\Phi_d)$ 
& $\sigma_\mu^2(\Phi_d) $ & $\dim(\Phi_d)$  & $\sigma_\mu^2(\Phi_d)$ & $\dim(\Phi_d)$   \\
$\Phi_d$: Wavelet $\bf l^1 $ & $4 10^{-6}$ & $21$ & $3 \,10^{-6}$ & $38$ & $3 \,10^{-6}$ & $52$ \\ 
$\Phi_d$: Scattering & $8 \,10^{-6}$ & $88$ & $10^{-5}$ & $422$ & $10^{-5}$ & $580$  \\ 
\hline
\end{tabular}
\caption{Estimated normalized variance 
for wavelet $\bf l^1$ norm and scattering energy vectors $\Phi_\NN$, at different maximum scales $2^J$. They are computed 
for a shot noise of size $d=256^2$ defined from a Cox point process. Figure \ref{pp_fig}(a) shows
a realization.}
\label{pointprocess_variance}
\end{table}

Figure \ref{pp_fig} gives realizations of microcanonical gradient descent
models computed from wavelet $\bf l^1$ norms and scattering energies,
at different maximum scales $2^J$.
Figure \ref{pp_fig}(b,d) are computed with $2^J = 8$. 
These microcanonical models can only capture
sparsity properties up to this maximum scale. At larger scale, the entropy maximisation creates Gaussian random process like variations having a uniform low-frequency spectrum. 
Figure \ref{pp_fig}(c,e) are microcanonical realizations computed at a larger maximum scale  $2^J = 64$. In this case, wavelet $\bf l^1$ norm and scattering microcanonical models capture the point process sparsity. 
The geometry of the shot noise is 
defined by the stationary rate $\lambda(u)$ which has relatively high frequency oscillations vertically but low frequency variations horizontally.
The scattering model Figure \ref{pp_fig}(e) captures this distribution thanks to second order coefficients. This is not the case for the $\bf l^1$ norm model in Figure \ref{pp_fig}(c) which can not reproduce the low-frequency
horizontal alignments. 

\subsection{Image and Audio Texture Synthesis}
\label{sec_numerics}

An image or an audio texture is usually
modeled as the realization of a stationary process.
Modeling textures amounts to compute an approximation of this
stationary process given a single realization. A texture synthesis
then consists in calculating new realizations from this stochastic model,
which are hopefully
perceptually identical to the original texture sample, although
different if considered as deterministic signals.
As opposed to the Gaussian, Ising or point process examples, since
we do not know the original stochastic process, 
perceptual comparisons are the only criteria
used to evaluate a texture synthesis algorithm.
Microcanonical models can be considered as
texture models computed from an energy function $\Phi_d (x)$ 
which concentrate close to its mean. We review previous work 
and give results obtained with a scattering microcanonical 
gradient descent model.

Geman and Geman \cite{geman1984stochastic} have introduced macrocanonical
models based on Markov random 
fields. They provide good texture models as long as these textures
are realizations of random processes 
having no long range correlations. Several
approaches have then been introduced 
to incorporate long range correlations.
Heeger and Bergen \cite{heeger1995pyramid} 
capture texture statistics through the marginal distributions
obtained by filtering images with
oriented wavelets. This approach has been generalized by
the macrocanonical Frame model of Mumford and Zhu \cite{zhu1998filters},
based on marginal distributions of filtered images. The filters are
optimized by trying to minimize the maximum entropy conditioned
by the marginal distributions. Although the Cramer-Wold theorem proves that 
enough marginal probability distributions characterize any random vector
defined over $\R^d$ the number of such marginals is typically intractable,
which limits this approach. 

Portilla and Simoncelli \cite{portilla2000parametric} 
made important improvements to these texture models, with
wavelet transforms. They capture
the correlation of the modulus of wavelet coefficients with a covariance
matrix which defines an energy vector $\Phi_d(x)$. Although they use a
macrocanonical maximum entropy formalism,
their algorithm computes 
a microcanonical estimation from a single realization, with 
alternate projections as opposed to a gradient descent. 
This approach was extended to audio textures by
McDermott and Simoncelli \cite{mcdermott2011sound}. 
A scattering representation is related to Portilla and Simoncelli
model but covariance coefficients are replaced by a much smaller number of
scattering ${\bf l}^1$ norms.

Excellent texture synthesis have recently been obtained with 
deep convolutional neural networks. In \cite{gatys2015texture}, 
the authors consider a deep VGG convolutional network, trained 
on a large-scale image classification task. The energy vector
$\Phi_d(x)$ is defined as the spatial cross-correlation 
values of feature maps at every layer of the VGG networks.
This energy vector is calculated on a particular texture image.
Texture syntheses of very good perceptual quality are calculated 
with a gradient descent microcanonical algorithm initialized on random 
noise. However, the dimension of this energy
vector $\Phi_d(x)$  is larger than the dimension $d$ of $x$.
These estimators are therefore
not statistically consistent and have no asymptotic limit. 

In the following, we give results obtained with different
wavelet microcanonical models computed on a collection of natural 
image and auditory textures. The Brodatz image texture dataset \footnote{Available at \url{http://sipi.usc.edu/database/database.php?volume=textures}} 
consists of 155 texture classes, with a single 512 $\times$ 512 sample per class. Auditory textures are taken from McDermott and Simoncelli \cite{mcdermott2011sound},  which contains 1 second samples of different sounds. 

\begin{figure}
\centering
\includegraphics[width=0.8\textwidth]{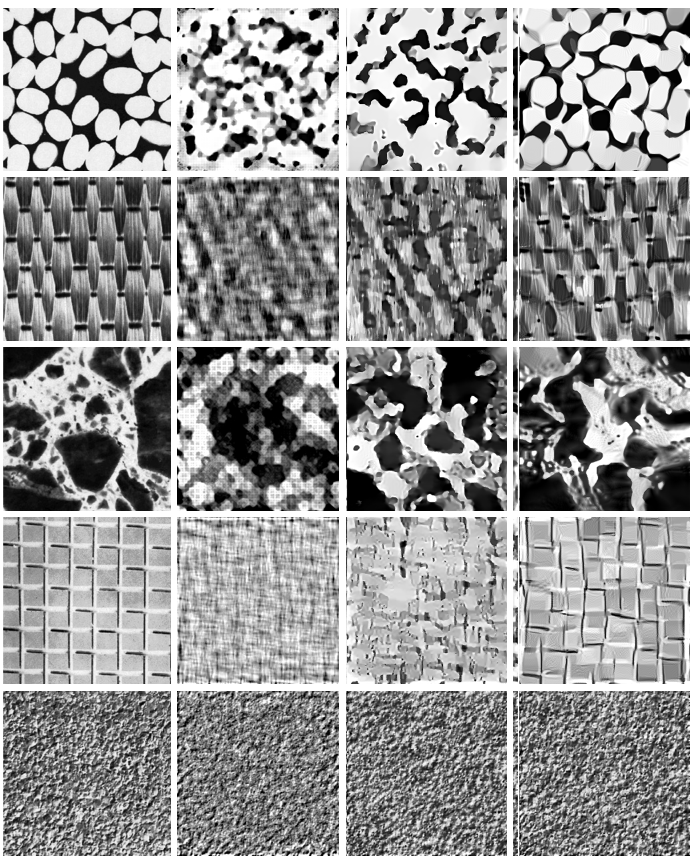}
(a)~~~~~~~~~~~~~~~~~~~~(b)~~~~~~~~~~~~~~~~~~~(c)~~~~~~~~~~~~~~~~~~~(d)
\caption{(a): Original texture. (b): texture synthesized with a microcanonical gradient descent model with a vector $\Phi_d(x)$ of wavelet
$\bf l^2$ norms. (c): $\Phi_d(x)$ has wavelet $\bf l^1$ norms.
(d): $\Phi_d(x)$ has wavelet scattering coefficients.}
\label{brodatz_fig}
\end{figure}


\begin{figure}
\centering
\includegraphics[width=0.8\textwidth]{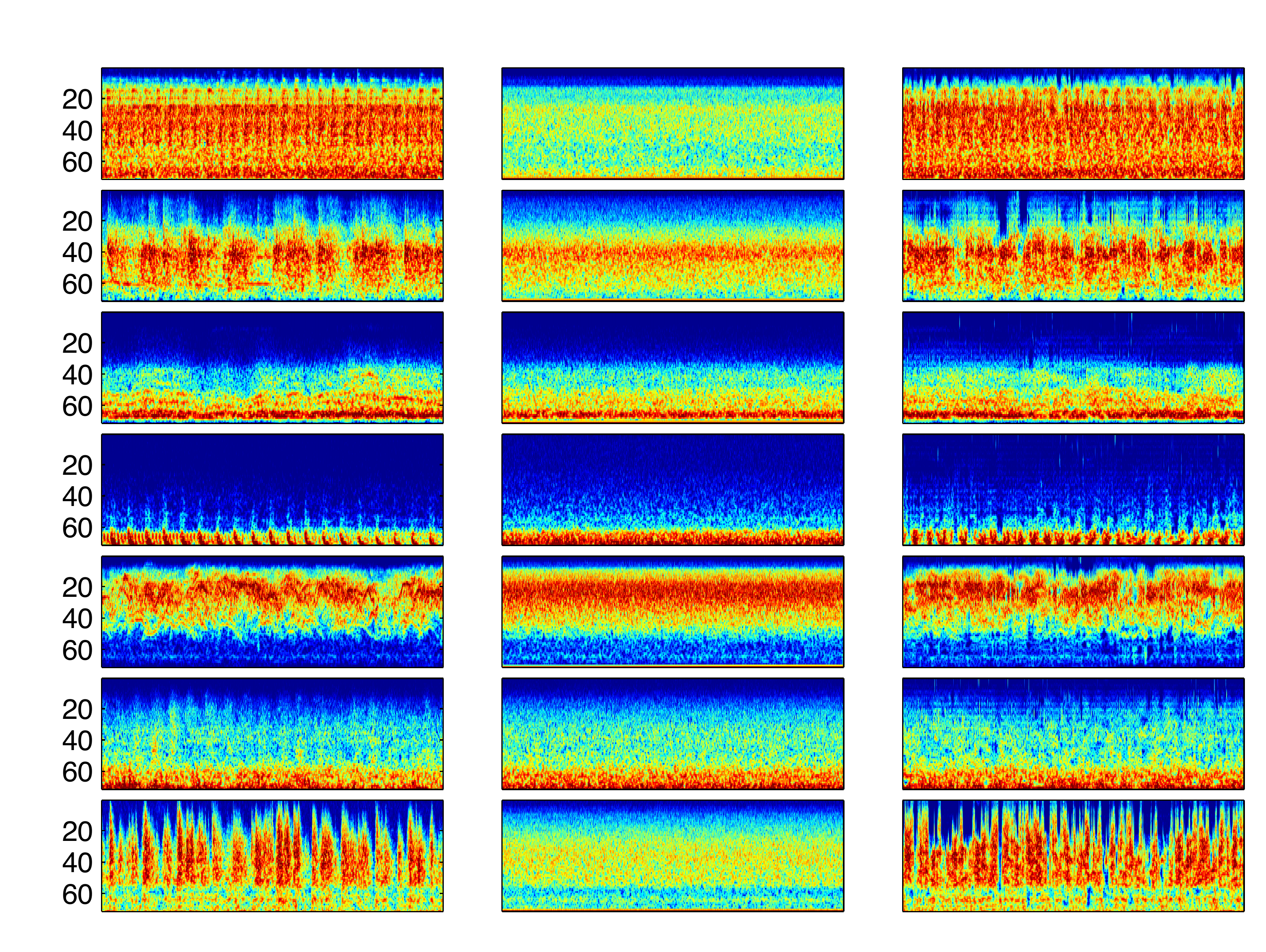}
(a)~~~~~~~~~~~~~~~~~~~~(b)~~~~~~~~~~~~~~~~~~~(c)
\caption{(a): Spectrograms of original audio textures
produced (from to to bottom) by jackhammer, applause, wind, helicopter, sparrows, train, rusting paper. (b): Spectrograms of an audio
texture synthesized with a microcanonical gradient descent model with a vector $\Phi_d(x)$ of wavelet $\bf l^2$ norms. (c): Spectrogram
produced with a vector $\Phi_d(x)$ of wavelet scattering coefficients.}
\label{audiotext_fig}
\end{figure}

Since we have a single realization of each texture,
we can not compute the concentration 
properties of energy vectors over these textures. 
Figure \ref{brodatz_fig}(a) gives input examples $\bar{x}$ 
corresponding to realizations of different stationary processes $X(u)$.
Figure \ref{brodatz_fig}(b) shows texture
samples obtained with a microcanonical gradient descent computed
with an energy vector $\Phi_d (x)$ of wavelet $\bf l^2$ norm. 
It provides a good model for the bottom texture which is nearly Gaussian but it otherwise destroys the texture geometry.
Figure \ref{brodatz_fig}(c) displays textures obtained
with a vector $\Phi_d (x)$ of wavelet $\bf l^1$ norms. Their
wavelet coefficients are more sparse than in 
Figure \ref{brodatz_fig}(b) which produces more 
``piecewise regular'' images, 
but it does improve the texture geometry. On the contrary,
scattering microcanonical textures
in Figure \ref{brodatz_fig}(d) have a geometry which is much closer to
original textures. Scattering coefficients can be interpreted as convolutional deep neural networks computed with predefined wavelet filters 
\cite{bruna2013invariant} as opposed to filters learned on a supervised image classification problem as in VGG.

The reconstruction of auditory textures is computed with 
a one-dimensional Gabor wavelet transform \cite{bruna2013audio} with $Q = 12$
scales per octave. Auditory textures have
a rich mixture of homogeneous and impulsive, transient components, 
as well as amplitude and frequency modulation phenomena. 
Figure \ref{audiotext_fig}(a) displays the spectrograms of original 
auditory textures $\bar{x}$.
Figure \ref{brodatz_fig}(b) shows the spectrogram of Gaussian texture
models calculated with a microcanonical gradient descent computed
with an energy vector $\Phi_d (x)$ of wavelet $\bf l^2$ norm. The global
spectral energy is preserved but the time variations which destroys
ability to recognize these audio textures. On the contrary,
Figure \ref{brodatz_fig}(c) shows that 
audio textures synthesized with a scattering energy vector 
have spectrograms with the same type of time intermittency as the original
textures. The resulting audio textures are perceptually difficult
to distinguish from the original ones. 

Synthesis from scattering energy vectors can also destroy some certain 
structures which affect their perceptual quality. This is the case for speech or music backgrounds which have harmonic alignments which are not reproduced by scattering coefficients. 
Deep convolutional network reproduce image and audio textures of better
perceptual quality than scattering coefficients, 
but use over 100 times more parameters.
Much smaller models providing similar perceptual quality
can be constructed with
wavelet phase harmonics for audio
signals \cite{mallat2018phase} or images \cite{zhangharmonic}, which
capture alignment of phases across scales. However, understanding
how to construct low-dimensional multiscale energy vectors to approximate
random processes remains mostly an open problem.


%% file: conclusion.tex
\section{Conclusion}

This paper shows that gradient descent
microcanonical models computed with  multiscale energy vectors can provide powerful models to approximate
large classes of stationary processes. Realizations of such models are 
calculated with a gradient descent algorithm which is much faster than
MCMC algorithms, used to sample from macrocanonical models. 

We introduced 
a mathematical framework to analyze the statistical and
algorithmic properties of these microcanonical gradient
descent models. Our analysis reveals that, whereas micrcocanonical
gradient descent measures do not generally agree with the microcanonical
maximum entropy measure, they have rich regularities through shared 
symmetries, and, under appropriate conditions, are shown to converge to the microcanonical 
ensemble. In the high-dimensional setting, gradient descent microcanonical 
models are therefore valid alternatives to classic macrocanonical and microcanonical 
maximum entropy measures, thanks to their computational tractability. 

However, many mathematical questions remain open. For instance, on the
convergence properties of this gradient descent algorithm,
on the choice of the energy vector to obtain accurate approximations of random processes, and 
on the extension to locally stationary processes.

%% file: proofs.tex

\section{Proof of Theorem \ref{maxentth00}}
\label{proof1}

\subsection{Proof of part (i)}
The main technical challenge to prove (\ref{resundsf01}) 
is to show that assumption (C) is sufficient to guarantee that 
$|J \Phi_d x|^{-1}$ is integrable. 
Since $\Phi_\NN$ is Lipschitz from assumption (A), the coarea formula proves that
for any integrable function $g(x)$ 
\begin{equation}
\label{coarea}
\int_{\B} g(x)\, 
|J \Phi_\NN x|\, dx = \int_{\R^K}\, \int_{\Phi_\NN^{-1}(y)} g(x)\,d {\cal H}^{\NN-K}(x)\, dy~.
\end{equation}

In order to apply (\ref{coarea}) to $|J\Phi_\NN (x)|^{-1}$ 
and obtain the expression of $H(\mumi_{d,\epsilon})$, 
we need to show that $|J\Phi_\NN (x)|^{-1}$ is integrable in $\Phi^{-1}_{\NN, \epsilon}(y)$. 
Using the notation for each Jacobian column (\ref{columnjac}), we verify 
that $|J\Phi_\NN (x)|$ satisfies
\begin{equation}
\label{bla0}
|J\Phi_\NN (x)| \geq d^{-\dd}\max \left\{ |\text{det} [ {J}U(\bar{X}_1), \dots,  {J}U(\bar{X}_K) ]|, \dots, |\text{det} [ {J}U(\bar{X}_{\tilde{\NN}+1}), \dots,  {J}U(\bar{X}_{\tilde{\NN}+K}) ]| \right \} ~,
\end{equation}
where $\bar{X}_i$ is a projection of $x$ onto disjoint subsets of $2\Delta + 1$ coordinates, and $\tilde{\NN} \geq \NN (2\Delta + 1)^{-1}=\Theta(\NN)$.

We will show that for $\NN$ large enough and arbitrary $R>0$, 
\begin{equation}
\label{bla1}
\int_{|x|_\infty < R} |J\Phi_\NN (x)|^{-1} dx < \infty~,
\end{equation}
by interpreting (\ref{bla1}) as proportional to the expected value of $\E_{ X \sim \mathrm{Unif}( \NN, R)} |J\Phi_\NN(X)|^{-1}$. 
Since $\Phi^{-1}_{d,\epsilon}(y)$ is a compact set thanks to assumption (B), it is bounded, so $\Phi^{-1}_{\NN, \epsilon}(y) \subseteq \{ x\,;\, |x|_\infty < R\}$ for some $R$, 
which proves that  $|J\Phi_\NN (x)|^{-1}$ is integrable in $\Phi^{-1}_{\NN, \epsilon}(y)$. 

For that purpose, let us prove that assumption (C) from (\ref{condPhi3}) is sufficient to guarantee (\ref{bla1}). 
If $F_V(y)$ denotes the cumulative distribution function of a random variable $V$, and 
$Y$ denotes the r.v. $Y =  |\text{det} [ {J}U(\bar{X}_1), \dots,  {J}U(\bar{X}_K) ]|$, 
we first observe that thanks to (\ref{bla0}) it is sufficient to show that 
\begin{equation}
\label{bla2}
F_Y(y) \lesssim y^\eta~, ~\text{for some } \eta > 0~, ~(y \to 0)~.
\end{equation}
Indeed, since $V = |J\Phi_\NN(X)| \geq \max( Y_1, \dots, Y_{\tilde{\NN}} )$ with $Y_i$ independent and identically distributed, we have 
that 
$$F_V(y) \leq F_Y(y)^{\tilde{\NN}} \simeq y^{\eta \tilde{\NN}}~.$$
It follows that 
$$\E_{X \sim \mathrm{Unif}(\NN, R)} |J \Phi_\NN(X)|^{-1} \leq \int v^{-1} f_V(v) dv = C + \int_0^{R'} v^{-2} F_V(v) dv < \infty$$
as soon as $\tilde{\NN} \eta > 1$, which will happen for large enough $\NN$.

Let us thus prove (\ref{bla2}) by induction on $K$. When $K=1$, 
$V = | \text{det} {J}U(\bar{X}_1)| = |{J}U(\bar{X}_1)| $ and assumption (C) directly 
implies that
$$F_V(y) = P ( V \leq y) \lesssim y^\eta~. $$
Now, suppose (\ref{bla2}) is true for $K-1$ and let us prove it for $K$. 
We use the following lemma:
\begin{lemma}
\label{lembla}
We say that a bounded random vector $Z$ in $B(K, R) \subset \R^K$ has property (*) if there exists $\eta>0$ such that 
$$~\forall~\mathcal{S} \subset \R^K \text{  Lebesgue measurable} ~,~P( Z \in \mathcal{S}) \lesssim |\mathcal{S}|^{\eta}  ~.$$
If $Z$ has property (*) and $K>1$, then $Z_H$, the orthogonal projection of $Z$ onto any hyperplane, also has property (*), and
\begin{equation}
\label{bla44}
\E ( \| Z \|^{-\eta} ) < C_{R, \eta}~.
\end{equation}
\end{lemma}
Before proving the lemma, let us conclude with (\ref{bla2}). 
By denoting $Z_i = {J}U(\bar{X}_i)$, $i=1\dots K$, and assuming $\| Z_1 \| >0$, one Gram-Schmidt iteration yields
$$|\text{det} \left[ Z_1, \dots, Z_K \right]| = \| Z _1\| |\text{det} \left[ \tilde{Z}_2, \dots, \tilde{Z}_K \right]|~,$$
where $\tilde{Z}_i$ is the projection of $Z_i$ onto the orthogonal complement of $Z_1$. 
Using assumption (C), we use lemma \ref{lembla} to observe that $\tilde{Z}_i$, $i=2,\dots K$ also satisfies assumption (C), since we compute it with an orthogonal projection 
that depends only on $Z_1$, which is independent from all the $Z_i$, $i \geq 2$. Thus by induction hypothesis and using (\ref{bla44}) we obtain
\begin{eqnarray*}
F_Y(y) = P \left( | \text{det} \left[ Z_1, \dots, Z_K \right] | \leq y \right) &=& P \left( \| Z _1\| |\text{det} \left[ \tilde{Z}_2, \dots, \tilde{Z}_K \right]| \leq y \right) \\
&=& \E_{Z_1} P \left(  |\text{det} [ \tilde{Z}_2, \dots, \tilde{Z}_K ] | \leq y \| Z_1 \|^{-1} ~ | ~ Z_1 \right) \\
& \leq & \E_{Z_1} y^\eta \| Z_1 \|^{-\eta} \lesssim y^{\eta}~,
\end{eqnarray*}
which proves (\ref{bla2}). 

Let us finally prove lemma \ref{lembla}. Let $\mathcal{S}_H$ be a measurable set in a given hyperplane $H$ of dimension $K-1$, and let 
$\tilde{\mathcal{S}} = \mathcal{S}_H \times (-R, R)$ be the corresponding cylinder in $B(K, R)$. By definition, we have 
$$P ( Z_H \in \mathcal{S}_H ) = P( Z \in \tilde{\mathcal{S}} ) \leq | \tilde{\mathcal{S}} |^\eta = |\mathcal{S}_H|^\eta (2R)^\eta $$
which proves that $Z_H$ also has the property (*). 

Finally, let us show that $\E ( \| Z \|^{-\eta} ) < C_{R,\eta}$. For positive random variables we have
\begin{eqnarray*}
\E ( \| Z \|^{-\eta} ) &=& \int_0^R r^{-\eta} f_{\| Z \|}(r) dr \\
&=& R^\eta - \lim_{r \to 0} r^{-\eta} P( \| Z \| \leq r) + \eta \int_0^R r^{-\eta - 1} P( \| Z \| \leq r ) dr \\
&\leq & R^\eta + C \eta \int_0^R r^{-\eta-1 +\eta K} dr \leq C_{R, \eta}~,
\end{eqnarray*}
since $K>1$ and $\eta>0$. This proves lemma \ref{lembla} and thus (\ref{resundsf01}). $\square$


%
%
%
%
%

To prove that $\h(y)$ is integrable on any bounded set, we 
apply the coarea formula to (\ref{coarea}) to
$g(x) = |J_{K} \Phi x|^{-1} \,1_{\mathcal{A}} (\Phi x)$ where $\mathcal{A}$ is bounded:
\[
\int_{\domainr} 1_{\mathcal{A}} (\Phi x)\, dx = \int_{\mathcal{A}}\, \int_{\Phi^{-1}(z)} 
|J_{K} \Phi x|^{-1}\, d {\cal H}^{\NN-K}\, dz = \int_{\mathcal{A}} \h(z)\, dz~.
\]
If $\mathcal{A}$ is a compact set then
by assumption (B) it follows immediately that 
\begin{equation}
\label{internasd8fs}
\int_{\domainr} 1_{\mathcal{A}} (\Phi x)\, dx = \int_{\Phi^{-1}(\mathcal{A})} dx \leq | B_{2,d}(C \sqrt(d))| < \infty,  
\end{equation}
which proves that $\h$ is integrable on a compact.

\subsection{Proof of part (ii) }

%
%

Let us now prove that for each $\NN$, $\gamma_\NN(y)$ can only vanish 
when $\mathrm{dist}(y, \overline{\Phi_d(\domainr)}) \leq c/d$ for some fixed constant $c$.
We will exploit the relationship between the sets
$\Phi_\NN(\R^d)$ and $\Phi_{\NN/2}(\R^{d/2})$ 
thanks to the fact that $\Phi_\NN$ is an average potential over the domain.


The inequality (\ref{resundsf2}) proves that $\h(y) = 0$ only
if $\int_{\Phi^{-1}(y)} d {\cal H}^{\NN-K} = 0$. 
Since in finite integer dimensions the Hausdorff measure ${\cal H}^{\dd}$ is
a multiple of the Lebesgue measure in $\R^{\dd}$, it is sufficient to show 
that whenever $y \in (\Phi_\NN(\domainr))^\circ$ , 
the set $\Phi^{-1}(y)$ has positive Lebesgue measure of dimension $\NN-K$. 

Without loss of generality, assume that $\Phi=(\phi_1,\dots,\phi_K)$ are linearly independent functions.
Otherwise, if there were a linear dependency of the form
$$\sum_{k \leq K} \alpha_k \phi_k(x) \equiv 0~,$$
then $\Phi_\NN(\domainr) = \partial \Phi_\NN(\domainr)$, 
thus $\Phi_\NN(\domainr)^o$ is empty and there is nothing to prove. 

Let us write $\NN = r^\dd$, with $r$ denoting the length of the cube $\LaN$. Suppose first that $r$ is even. 
Given $y \in (\Phi_{2^{-\dd} \NN}(\R^{2^{-\dd} \NN}))^\circ$  
we will see that there exists $x \in \Phi^{-1}(y)$ whose 
Jacobian $J\Phi (x)$ has rank $K$. Then, by the Implicit Function 
Theorem, one can find a local reparametrization of $\Phi^{-1}(y)$ in  
a small neighborhood $V$ of the form $x = (v, \varphi(v))$ such that
$$\{ (v, \varphi(v) ); v \in V \subset \R^{\NN-K}, \varphi: V \to \R^K \} = \{ (v,v') \in V \times \varphi(V) ;\, \Phi(v,v') = y \}~,$$
which has positive Lebesgue measure of dimension $\NN-K$.

Suppose first that $\Delta =1$. 
Then the sets 
$\mathcal{S}_\NN = \Phi_\NN(\domainr) \subset \R^K$ satisfy 
$\mathcal{S}_{\NN} \subseteq \mathcal{S}_{q^\dd \NN}$ for $q =1,2,\dots$. 
Indeed, given $y \in \mathcal{S}_\NN$, by definition there exists 
$x \in \domainr$ with $\Phi_\NN(x) = y$. 
Consider $\tilde{x}=(x, \dots, x)^{\otimes \dd} \in \R^{q^\dd \NN}$, a tiling of $x$, $q$ times along each dimension. 
By construction, 
$\tilde{x}$ satisfies $\Phi_{q^\dd \NN}(\tilde{x}) = y$ 
and therefore $y \in \mathcal{S}_{q^\dd \NN}$. 

Now, consider $y \in \mathcal{S}_\NN^\circ \subseteq \mathcal{S}_{2^\dd \NN}^\circ$. 
If $\Phi_\NN$ was a smooth $C^s$ map, with $s > d - K$, then
by Sard's theorem, the 
image of critical points $\{ x \in \domainr; | J \Phi_\NN(x) | < K \}$ has 
zero Lebesgue measure in $\mathcal{S}_\NN$. Although one can extend Sard's 
theorem to weaker regularity assumptions \cite{barbet2016sard}, 
for our purposes we will use a weaker and simpler property that does not require the smoothness
assumption, as described in the following lemma:
\begin{lemma}
\label{lemmasard}
Under the assumptions of the theorem, the set $\mathcal{A}=\{ y \in \R^K; ~ 0 < \h(y) < \infty \}$ is dense in $\Phi_\NN( \domainr)$, 
and for each $y \in \mathcal{A}$ there exists $x \in \Phi^{-1}_\NN(y)$ with $|J \Phi_\NN(x)|>0$.
\end{lemma}
It follows that for a sufficiently small $\delta>0$, a neighborhood 
$B(y, \delta) \subset \mathcal{S}_\NN$ of $y$ necessarily contains two points 
$y_1 =y+\eta$, $y_2= y - \eta$ such that $\Phi_\NN^{-1}(y_1)$ or 
$\Phi_\NN^{-1}(y_2)$ contain a regular  point. Let 
$x_1 \in \Phi_\NN^{-1}(y_1)$ and $x_2 \in \Phi_\NN^{-1}(y_2)$ be two points such that at least one is regular.
The point $\tilde{x} = (x_1^{\otimes \dd}, x_2^{\otimes \dd}) \in \R^{2^\dd \NN}$, obtained by concatenating $x_1$ and $x_2$ along the first coordinate, and tiling them along the rest, satisfies 
$$\Phi_{2^\dd \NN}(\tilde{x}) = \frac{1}{2}\left( \Phi_{\NN}(x_1) + \Phi_\NN(x_2)\right) = y~,\text{ and} $$ 
$$| J \Phi_{2^\dd \NN}(\tilde{x})| \geq \max( | J \Phi_{\NN}(x_1)|, | J \Phi_{\NN}(x_2| ) > 0~,$$
which shows that we have just found an element $\tilde{x}$ of $\Phi_{2^\dd \NN}^{-1}(y)$ with 
$\text{rank}( J \Phi_{2^\dd \NN}(\tilde{x}) ) = K$.

Suppose finally that $\Delta > 1$. The proof follows the same strategy, but we need
to handle the border effect introduced by the support $\Delta$.
 In that case, given $y \in \mathcal{S}_\NN$, we consider $\tilde{x}=(x, u, x)^{\otimes \dd}$, where $u$ has $2(\Delta-1)$ zero coordinates
and $x \in \Phi_\NN^{-1}(y)$. That is, we consider $2^{\dd}$ copies of $x$ separated by $2(\Delta - 1)$ zeroes along each dimension 
so that their potential functions do not interact.

Let $\tilde{\NN} = (2r + 2(\Delta-1))^\dd$.
It follows that 
$$\Phi_{\tilde{\NN}}(\tilde{x}) = \frac{2^\dd \NN \Phi_\NN(x)}{\tilde{\NN}} = \left(1+\frac{\Delta-1}{r}\right)^{-\dd} \Phi_\NN(x) = \left(1+\frac{\Delta-1}{\NN^{1/\dd}}\right)^{-\dd}  y~,$$
which shows that $\text{dist}(y; \mathcal{S}_{\tilde{\NN}}) \lesssim C \dd \|y \| / {\NN^{1/\dd}}$ for any $y \in \mathcal{S}_\NN$. 

Now consider $y$ in the open set $C_\NN = \mathcal{S}_\NN \cap \mathcal{S}_{\tilde{\NN}}$, such that $\text{dist}(y, \partial \mathcal{S}_\NN) \geq \|y \| \dd \Delta \NN^{-1/\dd}$. 
 It follows from the previous argument that there exists small $\delta>0$ and 
$x_1 \in \Phi_\NN^{-1}(y_1)$ with $| J \Phi_\NN(x_1) | > 0$ and $y_1 \in B(y, \delta) \cap \mathcal{S}_\NN \cap \mathcal{S}_{\tilde{\NN}}$. 
We verify from the assumption that 
$$y_2 = 2 \left(1 + \frac{\Delta - 1}{r} \right)^{\dd} y - y_1 \in \mathcal{S}_\NN~,$$
and therefore for any $x_2 \in \Phi_\NN^{-1}(y_2)$ 
the point $\tilde{x}=(x_1; u, x_2)^{\otimes \dd}$ that contains $2^{\dd-1}$ copies of $x_1$
and $2^{\dd-1}$ copies of $x_2$ satisfies by construction
$$\Phi(\tilde{x}) = \frac{\NN 2^{\dd-1} y_1 + \NN 2^{\dd-1} y_2}{\tilde{\NN}} = y~\text{ and }$$
$\text{rank}( J \Phi_{2\NN}(\tilde{x}) ) = K$. 
Finally, the case where $r$ is odd is treated analogously, but splitting the coordinates into $\lfloor \frac{r}{2}\rfloor$ and $\lceil \frac{r}{2} \rceil$ parts.

It remains to prove Lemma \ref{lemmasard}.
We know from part (i) that thanks to the coarea formula,  
$$\forall~\epsilon~\forall y \in (\Phi_\NN(\domainr))^\circ,~0 <  \int_{\| z - y \| \leq \epsilon } \gamma_\NN(z) dz =   \int_{ \| \Phi(x) - y \| \leq \epsilon} dx < \infty~.$$

It follows that $\mathcal{A}=\{ z; ~ 0 < \gamma_\NN(z) < \infty \}$ is dense in $\Phi_\NN( \domainr)$. 
But if $y \in \mathcal{A}$, by definition this implies that $\Phi_\NN^{-1}(y)$ has positive $(\NN-K)$-Hausdorff measure, and that there is necessarily $x \in \Phi_\NN^{-1}(y)$ 
with $|J \Phi_\NN(x)|^{-1} < \infty$, therefore with a full-rank Jacobian. $\square$

\subsection{Proof of part (iii)}

In order to prove (\ref{theopartthree}), we will 
again exploit the relationships between the sets $\mathcal{S}_\NN = \Phi_\NN ( \domainr)$ 
as $\NN$ grows.
We also first establish the result for $\Delta=1$, and then 
generalize it to $\Delta>1$. 
Denote $F_{\NN,\epsilon} = \NN^{-1} H(\mumi_{d,\epsilon})$ the entropy rate 
associated with $y$ and $\epsilon$ and $\Omega_{\NN,\epsilon}(y) = \{ x; \| \Phi_\NN(x) - y \| \leq \epsilon\} $.

In the last section we proved that when $\Delta=1$, $\mathcal{S}_\NN \subseteq S_{q^\dd \NN}$ for $q=1,2,\dots$. 
For any $\epsilon > 0$ and $y \in \mathcal{S}_\NN$, observe that 
\begin{equation}
\label{pokpok1}
\Omega_{\NN,\epsilon}(y) \underbrace{\otimes \dots \otimes}_{\text{$2^{\dd}$ times}}  \Omega_{\NN,\epsilon}(y) \subseteq \Omega_{2^\dd \NN,\epsilon}(y) ~.
\end{equation}
Indeed, if $x \in \Omega_{\NN,\epsilon}(y) \underbrace{\otimes \dots \otimes}_{\text{$2^{\dd}$ times}}  \Omega_{\NN,\epsilon}(y)$, then by definition 
$x=(x_1, \dots, x_{2^{\dd}})$ with 
$$\| \Phi_\NN(x_i) - y \| \leq \epsilon~.$$
But 
$$\Phi_{2^\dd \NN}(x) = 2^{-\dd} \sum_{i=1}^{2^{\dd}} \Phi_\NN(x_i)$$
and $\| \Phi_{2^\dd \NN}(x) - y \| \leq \epsilon$
by the convexity of the ${\bf l}^2$ norm, thus $x \in \Omega_{2^\dd \NN,\epsilon}(y)$. 
It follows that 
\begin{equation}
F_{2^\dd \NN,\epsilon} = \NN^{-1} 2^{-\dd} H(\mumi_{2^\dd \NN,\epsilon})\geq \NN^{-1} 2^{-\dd}\log \left( \left[\int_{\| \Phi_\NN(x)-y\| \leq \epsilon} dx \right]^{2^{\dd}}\right) = F_{\NN, \epsilon}~.
\end{equation}
Thus, for any fixed $\NN_0$, $y \in \mathcal{S}_{\NN_0}$ and $\epsilon>0$, 
the sequence $F_k = F_{2^{k\dd} \NN_0, \epsilon}$ is increasing. 
Also, thanks to assumption (B), we have that 
$$\forall~\NN~,~x \in \Omega_{\NN,\epsilon}(y) \implies \|x \| \leq C \sqrt{d} (\|y \| + \epsilon)~,$$
which implies that $| \Omega_{\NN,\epsilon}(y) | \leq | B_\NN( \sqrt{d} R_0)| $. 
Therefore 
$$\forall~\NN~, F_{\NN, \epsilon}^y \leq \NN^{-1} \log | B_\NN ( \sqrt{d} R_0) | ~,$$
and we verify from $|B_\NN ( R)| = \frac{\pi^{d/2}}{\Gamma(d/2+1)} R^d$ that 
$|B_\NN ( \sqrt{d} R_0)| \simeq \tilde{K}^d$ with $\tilde{K} = 2\pi R_0^2 e$, 
which shows that $\lim_{d\to \infty} d^{-1} \log |B_\NN ( \sqrt{d} R_0)| = \log \tilde{K}$ 
and thus that the entropy rate 
$F_k$ is also upper bounded, and 
therefore its limit exists $\lim_{k \to \infty} F_k = \bar{F}~.$
We shall see later that the limit does not depend upon the choice of $\NN_0$.

Let us now prove the case when $\Delta > 1$. The idea is to show that
(\ref{pokpok1}) is now valid up to an error that becomes small as $\NN$ increases, 
provided that the potential $U$ is Holder continuous. 

Consider $y \in \mathcal{S}_\NN$. Given $\epsilon>0$, we form 
$$\Psi_{2^\dd \NN,\epsilon}(y) = \left(\Omega_{\NN,\epsilon}(y)\right)^{\otimes 2^{\dd}} $$
as the Cartesian product of $2^{\dd}$ copies of $\Omega_{\NN,\epsilon}(y)$. 
When $\Delta=1$, we just saw that 
\begin{equation}
\label{pokpok3}
\Psi_{2^\dd \NN, \epsilon}(y) \subseteq \Omega_{2\NN,\tilde{\epsilon}}(y) 
\end{equation}
with $\epsilon = \tilde{\epsilon}$, but when $\Delta>1$, let us see how to increase $\tilde{\epsilon}$ 
so that (\ref{pokpok3}) is verified. 
Given $x \in \Psi_{2^\dd \NN, \epsilon}(y)$, we write $x=(x_1, \dots, x_{2^{\dd}})$ to denote its projections 
into each of the $2^{\dd}$ subdomains $C_{1, \NN}, \dots, C_{2^{\dd}, \NN}$ of size $\NN$. We have
\begin{eqnarray}
\label{fr1}
\Phi_{2^\dd \NN}(x) &=& \frac{\sum_n U x(n)}{2^{\dd} \NN} \nonumber \\
&=& 2^{-\dd} \sum_{k=1}^{2^{\dd}}  \NN^{-1} \left( \sum_{n \in C_{k,\NN}^\circ} Ux(n) +  \sum_{n \in \partial C_{k,\NN}} Ux(n) \right)~,
\end{eqnarray}
where each $C_{k,\NN}^\circ$ contains the interior of the domain that does not interact with the other domains, and 
$\partial C_{k,\NN} = C_{k,\NN} \setminus C_{k,\NN}^\circ$. 
We have 
$|\partial C_{k,\NN}| = \NN - ( \NN^{1/\dd} - 2\Delta)^{\dd}$, thus 
\begin{equation}
\label{fr2}
\NN^{-1} | \partial C_{k,\NN}| = 1 - (1- 2\Delta \NN^{-1/\dd})^{\dd} \lesssim \frac{\ell \Delta}{\NN^{1/\dd}}~.
\end{equation}
Since $|U x(n) | \leq B \| x\|^\alpha$ with $\alpha < 2/\dd$ by the Holder assumption, and
$\| x\| \leq C \sqrt{d}$ by assumption (B), we have $|U x(n) | \leq B' d^{\alpha/2}$. 
It follows from (\ref{fr1}) and (\ref{fr2}) that
\begin{eqnarray*}
\| \Phi_{2^\dd \NN}(x) - y \| &=& \left\| 2^{-\ell} \sum_{k=1}^{2^{\dd}} \left[ \NN^{-1} \left( \sum_{n \in C_{k,\NN}^\circ} Ux(n) +  \sum_{n \in \partial C_{k,\NN}} Ux(n) \right) - y \right] \right\| \\
&\leq & 2^{-\dd} \sum_{k=1}^{2^{\dd}} \left( \| \Phi_\NN(x_i) - y \| + 2 B' d^{\alpha/2} (1 - (1- 2\Delta \NN^{-1/\dd})^{\dd} )  \right) \\
 & \leq & \epsilon + o\left(\NN^{\frac{\alpha}{2}-\frac{1}{\dd}} \dd \Delta \right)~,
\end{eqnarray*}
Thus by taking $\tilde{\epsilon} = \epsilon +  o\left( \NN^{\frac{\alpha}{2}-\frac{1}{\dd}} \dd \Delta \right)$ (\ref{pokpok3}) is verified. 
By denoting $\nu = \frac{\alpha}{2}-\frac{1}{\dd}$, 
it follows that the entropy rate $F_{\NN, \epsilon}$ satisfies
$$F_{\NN, \epsilon} \leq  F_{2^\dd \NN, \epsilon+ \tilde{\ell}\NN^{\nu}}~,$$
with $\tilde{\ell} = C \Delta \dd$, and $\nu < 0$ since $\alpha < 2/\dd$.
By repeating the inequality for sufficiently large $\NN$ and $k=1,2,\dots $ and $\epsilon>0$ we have
\begin{equation}
F_{\NN, \epsilon} \leq F_{\NN2^{k\dd}\,,~ \epsilon+\tilde{\dd}\NN^{\nu}\sum_{k'=0}^k 2^{k'\dd \nu} } \leq F_{\NN2^{k\dd}, 2\epsilon} \leq \tilde{C}~,
\end{equation}
and thus by defining 
\begin{equation}
\label{yuy}
F_{\infty ,\epsilon} := \lim_{k \to \infty} F_{\NN_0 2^{k\dd}, \epsilon_k}~,\text{ with } \epsilon_k = \epsilon+\tilde{\dd}\NN_0^{\nu}\sum_{k'=0}^k 2^{\dd \nu k'}
\end{equation}
we have shown that its entropy rate is well-defined for each $\epsilon>0$ and $\NN_0$ sufficiently large.


It remains to be shown that this limit does not depend upon $\NN_0$. Suppose 
$F_{\infty, \epsilon,0} \neq F_{\infty, \epsilon,1}$ where $F_0$ is associated with 
$\NN_0$ and $F_1$ is associated with $\NN_1$, and suppose $\NN_1 > \NN_0$ without
loss of generality.
Let $r_i = \NN_i^{1/\dd}$ for $i=0,1$. 

Observe that an analogous argument to (\ref{fr1}) shows that if $r = r_a + r_b$, then
\begin{equation}
\label{yui1}
F_{r^\dd, \tilde{\epsilon}} \geq \frac{r_a}{r} F_{r_a^\dd, \epsilon} + \frac{r_b}{r} F_{r_b^\dd, \epsilon} ~,
\end{equation}
and 
\begin{equation}
\label{yui2}
F_{l^\dd \NN, \tilde{\epsilon}} \geq F_{\NN, \epsilon}~\text{for } l =1,2,\dots~,
\end{equation}
with $\tilde{\epsilon} = \epsilon + o\left(\NN^\nu \dd \Delta \right)$.
Consider now large integers $k$ and $\tilde{k} \simeq \sqrt{k}$, and let $q, \tilde{q}$ 
denote respectively the quotient and residual such that
$$r_1 2^k = r_0 2^{\tilde{k}} q + \tilde{q}$$
with $0\leq \tilde{q} < r_0 2^{\tilde{k}}$. 
Then, for any $\delta >0$, by choosing $k$ large enough we obtain from (\ref{yui1}) and (\ref{yui2}) that 
\begin{eqnarray}
\label{yui4}
|F_{\NN_1 2^{k\dd}, \tilde{\epsilon}} - F_1  | &\leq& \delta/4~, \nonumber \\
|F_{\NN_0 2^{\dd\tilde{k}} q^\dd , \tilde{\epsilon}} - F_0  | &\leq& \delta/4~, \text{ and }\nonumber \\
|F_{\NN_1 2^{k\dd}, \bar{\epsilon}} - F_{ \NN_0 2^{\tilde{k}\dd} q^\dd , \tilde{\epsilon}}| &\leq& \delta/4~,
\end{eqnarray}
with $\bar{\epsilon} = \tilde{\epsilon} + o\left(\NN^\nu \dd \Delta \right)$. 

Finally, let us show that $F_{\NN, \epsilon}$ is continuous with respect to $\epsilon$ for $\epsilon>0$.
Let us denote $\gamma_{\NN,\epsilon} = \int_{\|z - y \| \leq \epsilon} \h(z) dz $. 
Since $F_{\NN, \epsilon} = \NN^{-1} \log( \gamma_{\NN, \epsilon}^y ) $ and 
$\gamma_\NN(y) > 0$ for all $y \in \mathcal{S}_\NN^\circ$ from the previous section,
it is sufficient to show that $\gamma_{\NN,\epsilon}$ is continuous with respect to $\epsilon$. 
Let $\tilde{\epsilon}=\epsilon + \delta$ with $\epsilon>0$, and suppose $\delta>0$ without loss of generality. By denoting $Q(\delta, \epsilon, y) = \{z\,;\,\epsilon < \|z - y \| \leq  \epsilon + \delta  \}$, 
we have
\begin{eqnarray*}
| \gamma_{\NN,\tilde{\epsilon}} - \gamma_{\NN,\epsilon} | &=& \int_{\epsilon < \|z - y \| \leq  \epsilon + \delta} \h(z) dz = \int \h(z) {\bf 1}_{Q(\delta, \epsilon, y)}(z) dz \\
&:=& \int \h_\delta(z) dz
\end{eqnarray*}
For each $z$, $\h_\delta(z) = \h(z) {\bf 1}_{Q(\delta, \epsilon, y)}(z)$ converges pointwise to $0$ as $\delta \to 0$, except for a set of measure zero, $\{z ; \| z - y\| = \epsilon\}$. 
Also, $|\h_\delta | \leq \h $, which is integrable in $\Phi(\OO)$ by part (i). 
We can thus apply the dominated convergence theorem, and conclude that 
$$\lim_{\delta \to 0} \int \h_\delta(z) dz = \int \left(\lim_{\delta \to 0} \h_\delta(z) \right) dz = 0~,$$
which shows that $\gamma_{\NN,\epsilon}$ is continuous with respect to $\epsilon$.




It follows from (\ref{yui4}) that 
$$| F_{\NN_1 2^{k\dd}, \tilde{\epsilon}} - F_{ \NN_0 2^{\tilde{k}\dd} q^\dd , \tilde{\epsilon}}| \to 0~\text{as } k \to \infty~,$$
but $F_{\NN_1 2^{k\dd}, \tilde{\epsilon}} \to F_1$ and $F_{\NN_0 2^{\tilde{k}\dd} q^\dd , \tilde{\epsilon}} \to F_0$ as $k \to \infty$, which 
is a contradiction with the fact that $F_0 \neq F_1$. $\square$

\section{Proof of Corollary \ref{coroentropyepsilon}}

We saw in Theorem \ref{convergencetheo} that the entropy rate of 
the microcanonical measure can be measured with the co-area formula 
as $\NN^{-1} H(\mumi_{d,\epsilon}) = \NN^{-1} \log \int_{\| z - y \| \leq \epsilon} \h(z) dz~$
and that $\h(z) >0 $ in the interior of $\Phi_\NN(\domainr)$.
As $\epsilon \to 0$, we can interpret the previous formula in terms of an 
$L^1(\R^K)$ approximate identity $h_\epsilon(z) = C_K \epsilon^{-K} {\bf 1}_{\| z \| \leq \epsilon}(z)$: 
$$C_K \epsilon^{-K} \int_{\| z - y \| \leq \epsilon} \h(z) dz = \h \star h_\epsilon (y) \to \h(y) ~~\text{as }\epsilon \to 0~$$
in $L^1(\R^K)$. One can verify that, by possibly reparametrising $\epsilon$, 
this implies pointwise convergence for almost every $y$, so 
\begin{equation}
    \left| \log\left(C_K \epsilon^{-K} \gamma_{d,\epsilon}^y \right) - \log  \gamma_d(y) \right| \underset{\epsilon \to 0}{\to} 0~, ~a.e.~,
\end{equation}
which shows that $d^{-1} H(p_{d,\epsilon}^y) \simeq \frac{-K}{d} \log \epsilon$ as $\epsilon \to 0$ $\square$. 

\section{Proof of Proposition \ref{maxentth01}}
\label{proof2}

Properties (A) and (B) are verified for (i-ii) 
because the potentials $U$ are continuous and the resulting 
features $\Phi$ always include $\NN^{-1} \| x\|^2$ respectively. 
We thus focus on proving property (C). 

Part (i) is easily obtained, since
the ${\bf l}^2$ wavelet model has a Jacobian $J \Phi(x)$ that is linear 
with respect to $x$, and therefore it has absolutely continuous
density relative to the Lebesgue measure.

Part (ii) is proved by directly controlling $| J \Phi_\NN (x) |^{-1}$. A direct computation 
shows that $| J \Phi_\NN (x) | =  \NN^{-1} \sqrt{ \NN \| x \|^2 -  \| x \|_1^2}$, which 
only vanishes when $|x|$ is a constant vector. Therefore, for $y\neq ( \alpha, \LaN \alpha)$, 
$\Phi^{-1}_{\NN, \epsilon}(y)$ does not contain those points for sufficiently small $\epsilon$.

Let us now show part (iii). 
The Jacobian matrix in that case is given by
$$J \Phi_\NN(x)_j = \NN^{-1} \text{Re}\left\{\left( \frac{x \star h_j}{|x \star h_j|} \right) \star h_j^*\right\}~,$$
with $j \leq K$.
We proceed by induction over the scale $K$. 
Suppose first $K=1$. 
Since $h_j$ has compact spatial support, its Fourier transform only
contains a discrete number of zeros. Denote by $\Delta_j$ the spatial support of $h_j$.
 We can thus generate all but a zero-measure set 
of unitary signals $z$ with $z_s =e^{i \theta_s}$, $s=1\dots \Delta_j$ 
from the uniform measure over $x$ using
$z =  \frac{x \star h_j}{|x \star h_j|}$. 
 In the uniform phase space defined by $\theta_1, \dots, \theta_{\Delta_j}$, 
the event $| \text{det} \bar{J}U(\bar{X}_1)| \leq y$ has a probability 
proportional to $y$, since it is equivalent to 
$$\left| \sum_s \cos (\theta_s) \mathrm{Re}(h_j^*(s)) - \sum_s \sin(\theta_s) \mathrm{Im}(h_j^*(s)  \right| \leq y~.$$
Suppose now the result holds for the $K-1$ filters in the family with smallest spatial support, 
and let us show how to extend it to an extra filter $h_K$ with strictly larger spatial support. 
Among the variables $\bar{X} \in \R^{2\Delta+1}$, a subset of them, say $R_K$, only affect the $K$-th output 
corresponding to filter $h_K$. It follows that a set $S \subset \R^K$ with shrinking measure necessarily 
introduces constraints on the variables in $R_K$, and therefore $P( Z \in S) \leq |S|^{1/K}$  $~\square~. $

\section{Proof of Theorem \ref{invartheo}}
\label{invartheoapp} 

(i) Let us first prove that volume preserving
symmetries of $\Phi_d (x)$ are
symmetries of the microcanonical maximum entropy measure. 
If for all $x \in \R^d$, $\Phi_d (L^{-1} x) = \Phi_d (x)$ then a microcanonical set
$\Omega_{d,\epsilon}$ is invariant to the action of $L$ and $L^{-1}$. 
Since $L$ preserves volume and hence the Lebesgue measure
of a set, for any
measurable set $\mathcal{A}$, since $\mu^{\rm mi}_{d,\epsilon}$ is supported over
$\Omega_{d,\epsilon}$ and uniform relatively to the Lebesgue measure, 
we have
\[
\mu^{\rm mi}_{d,\epsilon} [L^{-1} \mathcal{A}] = 
\mu^{\rm mi}_{d,\epsilon} [L^{-1} \mathcal{A} \cap \Omega_{d,\epsilon}] =
\mu^{\rm mi}_{d,\epsilon} [L^{-1} (\mathcal{A} \cap \Omega_{d,\epsilon})] =
\mu^{\rm mi}_{d,\epsilon} [\mathcal{A} \cap \Omega_{d,\epsilon}] =
\mu^{\rm mi}_{d,\epsilon} [\mathcal{A}] ,
\]
so $L$ is a symmetry of $\mu^{\rm mi}_{d,\epsilon}$.

(ii) We prove that symmetries of $\Phi_d (x)$ and $\mu_0$ 
are symmetries of $\mu_n$, by induction on $n$. 
It is trivially valid for $n = 0$.
Suppose now by induction that $\mu_n$ is
invariant to the action of $L$ which is a symmetry of $\Phi_d$. 
From (\ref{yoy2}), 
$\mu_{n+1} = \varphi_{n,\#} \mu_n$, 
with 
$$\varphi_n(x) = x - \kappa_n J \Phi_d(x)^\top 
( \Phi_d (x) - y)~.$$
Let us verify that $\varphi_n$ is equivariant to the action of $L$: $\varphi_n L^{-1} x = L^{-1} \varphi_n x$ for all $x$.
 Since $\Phi_\NN (L^{-1} x) = \Phi_\NN (x)$, and since $L$ is linear
\begin{equation}
\label{rufu}
J \Phi_\NN( L^{-1} x)^\top = L^{-1} (J \Phi_\NN(L^{-1} x))^\top  =  L^{-1} (J \Phi_\NN(x))^\top    
\end{equation}
so
\begin{eqnarray*}
\varphi_n L^{-1} x &=& L^{-1} x - \kappa_n J \Phi_\NN(L^{-1} x)^\top (\Phi_\NN(L^{-1} x) - y) \\
&=& L^{-1} x - L^{-1} \kappa_n J \Phi_\NN(x)^\top (\Phi_\NN(x) - y) \\ 
&=& L^{-1} \varphi_n x~,
\end{eqnarray*}
which proves that $\varphi_n$ is equivariant to the action of $L$. 
Moreover, if $\varphi_n$ is equivariant to the action of $L$ then 
we verify that it
is equivariant to the action of $L^{-1}$. 
Also, observe that 
\begin{eqnarray*}
\varphi_n^{-1}( L^{-1}(\mathcal{A}) ) &=& \{ x; \varphi_n(x) \in L^{-1} \mathcal{A} \} \\
&=& \{ x; L \varphi_n(x) \in \mathcal{A} \}\\
&=& \{ x; \varphi_n(L x) \in \mathcal{A} \}\\
&=& L^{-1} \varphi_n^{-1}(\mathcal{A})~. 
\end{eqnarray*}

Finally, using the definition of pushforward measure, 
$\mu_{n+1} = \varphi_{n,\#} \mu_n$, for any measurable $\mathcal{A}$, the induction hypothesis yields
\begin{eqnarray*}
\mu_{n+1}[L^{-1} \mathcal{A}] &=& \mu_n [\varphi_n^{-1}(L^{-1} \mathcal{A})] \\
&=& \mu_n [ L^{-1} \varphi_n^{-1}(\mathcal{A})] = \mu_n [ \varphi_n^{-1}(\mathcal{A}) ] \\
&=& \mu_{n+1}[\mathcal{A}]~,
\end{eqnarray*}
which proves that $\mu_{n+1}$ is also invariant to the action of $L$.

(iii) We prove that an orthogonal operator which preserves a stationary mean is a symmetry of 
a Gaussian measure $\mu_0$ of $d$ i.i.d Gaussian random variables.
Applying the statement (ii) then implies the statement (iii).
Let $m_0$ be the mean of each of the $d$ Gaussian random variables. 
The Gaussian measure
$\mu_0$ is uniform over all spheres of $\R^d$ centered over the stationary
mean $m_0\, {\bf 1}$. An orthogonal operator $L$ which preserves 
the stationary mean leaves invariant all spheres centered in $m_0\, {\bf 1} \in \R^d$. Indeed $L (m_0\, {\bf 1}) = m_0\, {\bf 1}$ and $\|L x \|^2 = \|x\|^2$ so 
\[
\|L x - m_0 {\bf 1}\|^2 = \|L (x - m_0 {\bf 1})\|^2 = 
\| x - m_0 {\bf 1}\|^2 .
\]

If $S(m {\bf 1},r)$ is a sphere centered in $m {\bf 1}$ of radius $r$ then
$\R^d = \cup_{(m,r) \in \R \times \R^+} S(m {\bf 1},r)$. 
So for any measurable set $\mathcal{A}$
\begin{eqnarray*}
\mu_{0} [L^{-1} \mathcal{A}] &=& 
\mu_{0} [L^{-1} \mathcal{A} \cap \cup_{(m,r) \in \R \times \R^+} S(m {\bf 1},r)] =
\mu_{0} [\cup_{(m,r) \in \R \times \R^+}  L^{-1} (\mathcal{A} \cap S(m {\bf 1},r))]\\
& =&
\mu_{0} [\cup_{(m,r) \in \R \times \R^+}  \mathcal{A} \cap S(m {\bf 1},r)] = 
\mu_{0} [\mathcal{A}] ,
\end{eqnarray*}
so $L$ is a symmetry of $\mu_0$.


\section{Proof of Theorem \ref{convergencetheo}}
\label{appendo2}

\subsection{Proof of part (i)}

Let us first show how the strict saddle condition (\ref{strictsaddlehere}) implies that 
the minimisation $\mathcal{E}(x)$ has no poor local minima. The statement follows directly from 
\cite{lee2016gradient}, which shows that when the saddle points are strict, gradient descent does not converge to those saddle points, up to a set of initialization values with Lebesgue measure $0$. 
Observe first that $\kappa_n < \eta^{-1}$ ensures that $\varphi_n(x) = x - \eta \nabla E(x)$ 
is a diffeomorphism for each $n$.
Observe also that a critical point $x$ such that 
$\nabla \En(x) = J \Phi_\NN(x)^T ( \Phi_\NN(x) - y) = 0 $ necessarily falls into two categories. Either
$\Phi_\NN(x) = y $, which implies that $x$ is a global optimum, or $x$ is such that
$J \Phi_\NN(x)^T v = 0$ with $v = \Phi_\NN(x) - y \neq 0$. We verify that assumption (\ref{strictsaddlehere}) 
implies that in that case $x$ is a strict saddle point by observing that the Hessian of $\En$ satisfies
$$\nabla^2 \En(x) = \sum_{k=1}^K \nabla^2 \Phi_k(x) v_k + J \Phi(x)^T J \Phi(x)~.$$


Since $\mu_0$ is absolutely continuous with respect to the Lebesgue measure, 
we can apply Theorem 2.1 from \cite{panageas2016gradient}, and establish that gradient descent does not converge to any saddle point with probability 1. 

Let us now prove that the hypothesis that 
$|J \Phi_\NN(x) | > 0$ for $x \in \Phi_\NN^{-1}(y)$ with $y \in \Phi_\NN(\domainr)^\circ$, together with the strict saddle condition, implies that 
the gradient descent sequence $x_n$ has a limit $\lim_{n\to \infty} x_n$ (that may depend upon $x_0$). 
For that, we will apply the following result from \cite{absilslides}:
\begin{theorem}
\label{theogradconv}
If $\En(x)$ is twice differentiable, has compact sub-level sets, and the Hessian $\nabla^2 \En(x)$ is non-degenerate on the normal space to the level set of local minimisers, then $x_n$ has a limit, denoted 
$x_\infty := \lim_{n \to \infty} x_n $. 
\end{theorem}

Indeed, since $\Phi_\NN$ satisfies assumption (B), it follows that the sub-level
sets of $\En$, $\{ x; \En(x) \leq t\}$
are compact for each $t$. 
We need to show that the Hessian of $\En$ is non-degenerate on the normal space of $\Phi_\NN^{-1}(y)$. Since $\h > 0$ for $y \in \Phi_\NN(\domainr)^\circ$ for sufficiently large $\NN$ from Theorem \ref{maxentth00}, $\Phi_\NN^{-1}(y)$ has positive $\NN-K$-dimensional Hausdorff measure, hence it is sufficient to show that $\nabla^2 \En(x)$ has $K$ strictly positive eigenvalues when $x \in \Phi^{-1}(y)$. But by definition, 
$$\nabla^2 \En(x) =  \sum_{k\leq K} \nabla^2 \phi_k(x) (\phi_k(x) - y_k) + J \Phi_\NN(x)^T J \Phi_\NN(x)~,$$
thus
\begin{equation}
    \nabla^2 \En(x) = J \Phi_\NN(x)^T J \Phi_\NN(x) ~~\text{for }~x \in \Phi_\NN^{-1}(y)~.
\end{equation}
Therefore, if $|J \Phi_\NN(x) | > 0$ for $x \in \Phi_\NN^{-1}(y)$, we 
can apply Theorem \ref{theogradconv}, and conclude that the iterates $x_n$ 
from gradient descent have a limit, for each $x_0 \sim \mu_0$.

We have just proved that
$$\mathrm{P}_{\mu_0}\left\{ (x_n)_n \text{ is Cauchy} \right\} = 1~,$$
or, equivalently, that $X_n \sim \mu_n$ is almost surely Cauchy, which 
implies \cite{advanced_proba} that $\mu_n$ converges almost surely to a 
certain measure $\mu_\infty$. Moreover, since 
$\lim_{n \to \infty} \| \nabla \En(x_n) \| = 0$, the strict 
saddle condition implies that $x_n$ does not converge to saddle points, so we conclude that necessarily 
$$\mu_\infty \left[\Phi_\NN^{-1}(y) \right] = \mathrm{P}_{\mu_0}\left\{ \lim_{n \to \infty} x_n \in \Phi_\NN^{-1}(y) \right\} = 1~,$$
therefore that $\mu_\infty$ is supported in the microcanonical ensemble $\Phi_\NN^{-1}(y)$, which finishes the proof. $\square$




\subsection{Proof of part (ii)}

We first compute how the entropy is modified at each gradient step. 
By definition of the pushforward measure, for any diffeomorphism $\varphi$ 
and any measurable $g$
$$\E_{x \sim \varphi_{\#}\mu} g(x) = \E_{x \sim \mu} g(\varphi(x))~. $$ 
Also, from a change of variables we have, by denoting $\tilde{\mu}= \varphi_{\#} \mu$, 
$\tilde{\mu}(x) = |J \varphi^{-1}(x) | \mu(\varphi^{-1}(x))~,$
and thus 
$$\log \tilde{\mu}(x) = \log \mu(\varphi^{-1}(x)) - \log | J \varphi( \varphi^{-1}(x))| ~.$$
It follows that
$$-\E_{x \sim \tilde{\mu}} \log \tilde{\mu}(x) = -\E_{x\sim \mu}  \log {\mu}(x) + \E_{x\sim \mu} \log | J \varphi(x) |$$
and hence
\begin{equation}
\label{flam1}
H( \varphi_{\#} \mu) = H(\mu) - \E_{\mu} \log | J \varphi(x) | ~.
\end{equation}

The change in entropy by applying the diffeomorphism is thus given by the term $\E_{\mu} \log | J \varphi(x) |$, 
and thus the entropy of $\mu_n$ is given by 
$$H(\mu_n) = H(\mu_0) - \sum_{n' \leq n} \E_{\mu_{n'}} \log | J \varphi_n(x) | $$ 
By definition, the Jacobian of $\varphi_n$ is 
\begin{equation}
\label{flam2}
J \varphi_n(x) = {\bf 1} - \gamma_n \left( \sum_{k\leq K} \nabla^2 \phi_k(x) (\phi_k(x) - y_k) + J \Phi_\NN(x)^T J \Phi_\NN(x) \right)~.
\end{equation}

We know that $\Phi$ is Lipschitz, which implies that $\| J \Phi(x) \| \leq \beta$, and
 that $\nabla \Phi$ is also Lipschitz, meaning that $\| \nabla^2 \phi_k(x) \| \leq \eta$ for all $k$. 
Applying the Cauchy-Schwartz inequality, it follows that 
$$\left \| \sum_{k\leq K} \nabla^2 \phi_k(x) (\phi_k(x) - y_k)  \right \| \leq \eta K \sqrt{\En(x)}~.$$
We abuse notation and redefine $\eta := \eta K$ since $K$ is a constant. 
 Also, the term $J \Phi(x)^T J \Phi(x)$ is of rank at most $K$. 
 We can thus write $J \varphi_n(x)$ as
 \begin{equation}
\label{flam3}
 J \varphi_n(x) = A_n(x) + B_n(x)~,
 \end{equation}
  with $A_n(x)$ full rank $\NN$ and with singular values within the interval $(1 - \gamma_n \eta \sqrt{\En(x)} , 1 + \gamma_n \eta \sqrt{\En(x)} )$; 
 and $-B_n(x)$ positive semidefinite of rank $K$, with singular values bounded by $\gamma_n \beta^2$. 
It follows that the singular values of $J \varphi_n(x)$, called $\lambda_1, \dots, \lambda_\NN$, satisfy
\begin{eqnarray*}
| \log | J \varphi_n(x) | | & \leq & \sum_{i=1}^\NN | \log \lambda_i | \\ 
&\leq & \sum_{i=1}^{\NN-K} \max( |\log (1 + \gamma_n \eta \sqrt{\En(x)} )| , |\log( 1 - \gamma_n \eta \sqrt{\En(x)} )| ) \\
& & + \sum_{i=1}^K  |\log( 1 - \gamma_n \beta^2 )|  \\
&\leq & (\NN-K) \log( 1 + \gamma_n \eta \sqrt{\En(x)} ) + K \log ( 1 + \gamma_n \beta^2) + o(\gamma_n^2)
\end{eqnarray*}
and thus up to second order terms we have
\begin{eqnarray}
\label{flam4}
 \E_{\mu_n} \log |J \varphi_n(x)|  &\leq&  (\NN-K) \log \left(1 + \gamma_n \eta \E_{\mu_n} \sqrt{\En(x)} \right) + K \log \left(1 + \gamma_n \beta^2\right) ~, \nonumber \\
&\leq&  (\NN-K) \gamma_n \eta \E_{\mu_n} \sqrt{\En(x)} + K \gamma_n \beta^2 ~,
\end{eqnarray}
where we have used Jensen's inequality on the concave function $\log(1+x)$ and $\log(1+x) \leq x$ for $x\geq 0$
to obtain the inequality $\E \log(1 + X) \leq \log ( 1 + \E X)$. 
Denoting by $r_n =  \E_{\mu_{n}} \sqrt{\En(x)}$ the average distance to the microcanonical ensemble at iteration $n$,
it results from (\ref{flam4}) that after $n$ steps of gradient descent the entropy rate has decreased at most
$$\left(1 - \frac{K}{\NN}\right) \eta \sum_{n'\leq n}  \gamma_{n'} r_{n'}   + \frac{K}{\NN}  \beta^2 \sum_{n' \leq n} \gamma_{n'}~.$$

\section{Proof of Corollary \ref{convergencecoro}}
\label{convergencecoroproof}

The proof is a direct application of Theorem \ref{convergencetheo} 
and Sard's theorem, that states that if $\Phi_\NN$ is a ${\bf C}^\infty$ Lipschitz function, then the image of its critical points 
$\{ x ~,;\, |J \Phi_\NN (x) | = 0\}$ has zero measure. We can thus 
apply Theorem \ref{theogradconv} from part (ii) of the proof of 
Theorem \ref{convergencetheo} for almost every $y$ $\square$.

\section{Proof of Theorem \ref{convergencetheo2}}
\label{globalconvtheoremi}
We show  that $\Phi_\NN(x) = \{\NN^{-1} \| x \star h_k\|_2^2 \}_k$ satisfies the 
strict saddle condition. 
Here $x \in \R^d$, and we recall that the Fourier transform 
is defined as $\hat{x}(\omega) = \sum_u x(u) e^{-i \omega u 2\pi/d}$, with $\omega \in (-d/2,d/2]$.
The gradient of the loss function $\En(x) = \frac{1}{2}\| \Phi(x) - y \|^2$ 
is 
$$\nabla \En(x) = J \Phi_\NN(x)^T ( \Phi_\NN(x) - y)~,$$
and its Hessian is 
$$\nabla^2 \En(x) = \sum_{k} \nabla^2 \phi_k(x) v_k + J \Phi_\NN(x)^\top J \Phi_\NN(x)~,$$
where $v_k = \phi_k(x) - y_k$.
Expressing the gradient and the Hessian in the Fourier domain yields
\begin{eqnarray}
\label{cosa}
\nabla \En(\hat{x}) &=& \hat{x} \cdot (\sum_k v_k | \hat{h_k}|^2) \\
\nabla^2 \En(\hat{x})(\om, \om') &=& \sum_k v_k | \hat{h_k}(\om)|^2 \delta(\om-\om') + \hat{x}(\om) | \hat{h_k}(\om)|^2  \hat{x}(\om')^* | \hat{h_k}(\om')|^2 ~.
\end{eqnarray}
The Hessian thus contains a diagonal term and a rank-$K$ term. 
We need to show that a critical point $x$ satisfying $\nabla \En(\hat{x}) = 0$ with $\| v \| >0$ has a Hessian 
matrix with at least one negative eigenvalue. 
From (\ref{cosa}), it follows that a critical point satisfies
\begin{equation}
\label{cosa0}
\forall \om~,~ \hat{x}(\om)  \cdot (\sum_k v_k | \hat{h_k}(\om)|^2)  = 0~.
\end{equation}
Let $C = \{ \om~;~\hat{x}(\om) \neq 0 \}$. The Hessian is expressed in terms of block matrices 
regrouping the frequencies in $C$ as
$$\nabla^2 \En( \hat{x}) = \left(\begin{array}{c|c}
\mathrm{M} & 0 \\
\hline
0 & \nabla^2_{C,C} 
\end{array} \right)~,
$$
where $\mathrm{M}$ is the diagonal matrix of size $(d - |C|) \times (d - |C|)$ given by 
the frequencies outside $C$, such that $\hat{x}(\om) = 0$:
$$\mathrm{M}_{\omega,\omega} = \sum_k v_k | \hat{h_k}(\om)|^2~,~\omega \notin C~.$$
We examine the diagonal block corresponding to $\mathrm{M}$.
The image of $\Phi_\NN$ is the convex cone $\mathcal{C}$ in $\mathbb{R}_{+}^K$ determined by the directions 
$o_\om = (| \hat{h_1}(\om)|^2, \dots, | \hat{h_K}(\om_j)|^2) \in \R^K$, $\om=1\dots \NN$. 
Without loss of generality, we assume here that $\| o_\om \| > 0$ for all $\om$, since frequencies that are invisible 
to all the filters do not play any role in the gradient descent.
The target $y$ is by
hypothesis in the interior of $\mathcal{C}$. Further, any two directions $o,o'$ in $\mathcal{C}$ satisfy 
\begin{eqnarray*}
\langle o, o' \rangle &=& \sum_k | \hat{h_k}(\om)|^2 | \hat{h_k}(\om')|^2 > 0~,
\end{eqnarray*}
since the filters have compact spatial support. 

If $C$ is empty, then $x=0$, which implies that $v = \Phi(x) - y = -y$ 
has all its entries negative, and therefore $\text{diag}(\sum_k v_k | \hat{h_k}(\om)|^2) < 0$. 
We shall thus assume in the following that $C$ is non-empty.
Similarly, we verify that the space spanned by $o_\omega$, $\omega \in C$, cannot have full rank $K$.
Indeed, if this was the case, the first order optimality condition (\ref{cosa0}) reveals that $v$ should be orthogonal to 
all directions $o_\omega$, $\omega \in C$. Since this system has rank $K$, this contradicts the fact that $v \neq 0$.

We can thus write $\mathcal{C}$ as generated by directions $\mathcal{O}_C = \{ o_\omega; \omega \in C\}$ 
and $\mathcal{O}_{\overline{C}} = \{ o_\omega ; \omega \notin C \}$, with $|\mathcal{O}_{\overline{C}}|>0$, $|\mathcal{O}_C|>0$. Since $y$ is in the interior, it follows that 
\begin{equation}
\label{grit}
y = \sum_{\omega \in C} \beta_\omega o_\omega + \sum_{\omega \notin C} \gamma_\omega o_\omega~,~\beta_\omega, \gamma_\omega >0 \forall~\omega~. \end{equation}
We need to show that there exists at least one $\omega \notin C$ such that $\langle v, o_\omega \rangle < 0$. 
Suppose otherwise, i.e. that for all $\omega \notin C$, $\langle \Phi_d(x), o_\omega \rangle \geq \langle y, o_\omega \rangle$. Since $o_\omega \in \mathcal{O}_C \Rightarrow \langle \Phi_d(x), o_\omega \rangle = \langle y, o_\omega \rangle$ by the first order critical conditions, we have
\begin{eqnarray}
\label{grit2}
\langle y, y \rangle &=& \sum_{\omega \in C} \beta_\omega \langle o_\omega, y \rangle + \sum_{\omega \notin C} \gamma_\omega \langle o_\omega,y \rangle \nonumber \\
&\leq& \sum_{\omega \in C} \beta_\omega \langle o_\omega, \Phi_d(x) \rangle + \sum_{\omega \notin C} \gamma_\omega \langle  o_\omega, \Phi_d(x) \rangle~.
\end{eqnarray}
On the other hand, from (\ref{grit}) we also have 
\begin{equation}
    \langle y, \Phi_d(x) \rangle = \sum_{\omega \in C} \beta_\omega \langle o_\omega, \Phi_d(x) \rangle + \sum_{\omega \notin C} \gamma_\omega \langle  o_\omega, \Phi_d(x) \rangle~,
\end{equation}
and since $\Phi(x) = \sum_{\omega \in C} \alpha_\omega o_\omega$ is a linear combination of vectors in $\mathcal{O}_{\overline{C}}$, we also have $\langle \Phi(x), y \rangle = \langle \Phi(x), \Phi(x) \rangle$. This implies from (\ref{grit2}) that
\begin{equation}
  \langle y, y \rangle \leq \langle y, \Phi_d(x) \rangle = \langle \Phi_d(x), \Phi_d(x) \rangle~,
\end{equation}
which leads to $y = \Phi(x)$ and therefore $v=0$, which is a contradiction.

Finally, if $x \in \Phi_\NN^{-1}(y)$ 
for $y \in \Phi_\NN(\domainr)^\circ$, 
then $y$ falls necessarily inside the convex hull of $\mathcal{C}$, which implies that 
$\{\nabla \phi_k(x) = \hat{x}(\om) \cdot |\hat{h}_k|^2(\om) \}_{k \leq K}$ have rank $K$. 
This concludes the proof $\square$. 

\section{Proof of Proposition \ref{enenprop}}
\label{enenpropproof}
If $\gamma = 0$ then (\ref{nsdf8sydf01332230}) proves that
\[
\|x\|_2^2 = \|x \star \psi_{J,0}\|_2^2 + 
\sum_{j'=1}^{\log_2 d} \sum_q \|x \star \psi_{j',q'}\|_2^ 2 .
\]
If $J = \log_2 d$ then $\psi_{J,0} (u)= d^{-1} 1_{\Lambda_d}$ 
and $x \star \psi_{J,0}(u)$ is the average of $x$ over $\Lambda_d$. 
We thus get
\begin{equation}
\label{enerconsdfsd}
\|x\|_2^2 = d^{-1} \Big( \sum_{u} x(u)\Big)^2 + 
\sum_{j'=1}^{\log_2 d} \sum_q \|x \star \psi_{j',q'}\|_2^ 2 .
\end{equation}
Replacing $x$ by $|x \star \psi_{j,q}|$ gives
\[
\|x \star \psi_{j,q}\|_2^2 = d^{-1} \|x \star \psi_{j,q} \|_1^2 +
\sum_{j'=1}^{\log_2 d} \sum_q \| |x \star \psi_{j,q}| \star \psi_{j',q'}\|_2^ 2 .
\]
We finally prove (\ref{nsdf8sydf0133223}) by 
decomposing each term 
$\| |x \star \psi_{j,q}| \star \psi_{j',q'}\|_2^ 2$ into an
$\bf l^1$ norm plus a sum of $\bf l^2$ norms, obtained
replacing $x$
by $||x \star \psi_{j,q}| \star \psi_{j',q'}|$
in (\ref{enerconsdfsd}).

\section{Proof of Theorem \ref{theorexmsparse}}
\label{sparsewavetheo}

Let us first prove property (i). 
Young inequality is proved by observing that
\[
\| x \star \psi_{j,q} \|_1 = \sum_{n \in \LaN} \Big| \sum_{u \in \LaN} x(u)\, \psi_{j,q} (n-u)\Big|
\leq \sum_{n \in \LaN} \sum_{u \in \LaN} |x(u)\, \psi_{j,q}(n-u)| = \|x\|_1 \, \|\psi_{j,q} \|_1~.
\]
The inequality is an equality if and only if for any fixed $n$, 
the product $x(u)\, \psi_{j,q}(n-u)$ has a constant phase when $u$ varies.
Since 
$x(u)$ 
is real, its phase is either $0$ or $\pi$.
It implies that $\psi_{j,q}(n-u)$ has a phase modulo $\pi$ which does
not depend upon $u$ when $x(u) \psi_{j,q} (n-u)\neq 0$ and
hence $x(u) \neq 0$. Since
the phase of $\psi$ is $\varphi(\xi.u)$, 
the phase of $\psi_{j,q} (u) = 2^{-\ell j} \psi (2^{-j} r^{-1}_q u)$ is 
$\varphi( 2^{-j} \xi_q.u)$ with $\xi_q = r_q \xi$ so 
\begin{equation}
\label{insdfnsdf}
\forall u \in \LaN~~,~~\varphi (2^{-j}\xi_q .(n-u))  = a(2^{-j} n) + k \pi~~\mbox{if}~~
x(u)\,\psi_{j,q}(n-u) \neq 0~\mbox{with}~k \in \Z~.
\end{equation}
Since $\varphi$ is bi-Lipschitz, there exists $\beta > 0$ such that
\begin{equation}
\label{bilipsh2}
\beta^{-1} |a - a'| \leq |\varphi_q (a) - \varphi_q (a')| \leq 
\beta |a - a'|~.
\end{equation}
Since $\psi_q(0) \neq 0$ and $\psi_q$ is continuous, there exists
$\alpha > 0$ such that $|\psi_q (u)| > 0$ for $u \in [-\alpha,\alpha]^\ell$. 
If $2^{-j} |u-u'| \leq 2 \alpha$ then for $n = (u+u')/2$ we have
$2^{-j} |n -u| \leq \alpha$ and $ 2^{-j} |n -u'| \leq  \alpha$,
so $\psi_{j,q}(n-u) \neq 0$ and $\psi_{j,q}(n-u') \neq 0$. 
If the inner product $\xi_q.(u-u')$ is not zero then (\ref{bilipsh2}) implies
that $|\varphi(2^{-j}\xi_1.(n-u)) - \varphi_q(2^{-j}\xi_q.(n-u'))| > 0$.
So if $x(u)$ and $x(u')$ are non-zero (\ref{insdfnsdf}) implies that
\[
|\varphi (2^{-j}\xi_1.(n-u))  - \varphi (2^{-j} \xi_q.(n-u'))|  \geq  \pi .
\]
It follows from (\ref{bilipsh2}) that if $2^{-j} |u-u'| \leq 2 \alpha$ then 
\[
2^{-j} \beta |\xi_q .(u - u')| \geq  \pi ,
\]
which proves $|\xi_q . (u - u')| \geq  C\,2^{j}~$  for $C = \min(\pi \beta^{-1},2 \alpha |\xi_q|)$, and hence part (i). 

Let us now prove property (ii). 
Since $\psi_q$ has a compact support it is included in
$[-\gamma,\gamma]^\ell$ for $\gamma$ large enough. Since the support
of $x$ are points of distance at least $\Delta$ it results that for
any $n \in \Z^\ell$ and $2^j \leq \Delta \,\gamma^{-1}$,  the product 
$x(u) \psi_{j,q} (n-u)$ is non-zero for at most one $u \in \Z^\ell$.
It results that
\[
\| x \star \psi_{j,q} \|_1 = \sum_{n \in \LaN} \Big| \sum_{u \in \LaN} x(u)\, \psi_{j,q} (n-u)\Big|
= \sum_{n \in \LaN} \sum_{u \in \LaN} |x(u)|\, |\psi_{j,q}(n-u)| = \|x\|_1 \, \|\psi_{j,q} \|_1~.
\]
The hypothesis (\ref{corcond}) implies that
$\|x'\|_1 = \|x' \star \psi_{j,q} \|_1$ for 
all $q \leq Q$ and $2^j \leq \Delta\,\min(1,\gamma^{-1})$.
Applying Theorem \ref{theorexmsparse} for
$2^j \geq 2^{-1} \Delta\,\min(1,\gamma^{-1})$
proves that $x'(u)$ and $x'(u')$ are
non-zero only for all $q \leq Q$ we have $\xi_q .(u-u') = 0$ or 
$|\xi_q . (u - u')| \geq  C'\,\Delta$, 
where $C'$ does not depend upon $x$ and $x'$. 

Since the $\{ \xi_q \}_{q \leq Q}$ are $Q \geq \ell$ different
rotations of a non-zero $\xi \in \R^\ell$,
they define a frame of $\R^\ell$. It results 
that there exists $A$ and $B$ such that for any $v \in \R^\ell$
\begin{equation}
\label{framein}
A \,|v| \leq \sum_{q \leq Q} |v.\xi_q| \leq B\, |v|~.
\end{equation}
This inequality applied to $v = u-u' \neq 0$ proves that 
there exists $q \leq Q$ such that $\xi_q .(u-u') \neq 0$.
If $x(u) \neq 0$ and $x(u') \neq 0$ then we proved that
if $\xi_q .(u-u') \neq 0$ then
$|\xi_q . (u - u')| \geq  C'\,\Delta$. The frame inequality (\ref{framein})
implies that $|u - u'| \geq B^{-1}\,C'\,\Delta$ which shows that
any two points in the support of $x'$ have a distance at least $C\,\Delta$
with $C = C' B^{-1}$.